\documentclass[reprint,superscriptaddress,amsmath,amssymb,aps,pra,floatfix]{revtex4-2}
\usepackage{graphicx}
\usepackage[usenames,dvipsnames]{xcolor}

\usepackage[colorlinks=true,citecolor=blue,linkcolor=blue,breaklinks=true]{hyperref}
\usepackage{changes}
\usepackage{soul}
\usepackage{url}

\usepackage{changes}

\newcommand{\fig}{Fig.~}

\begin{document}

\title{Mechanically Induced Correlated Errors on Superconducting Qubits \\ with Relaxation Times Exceeding 0.4 Milliseconds}

\author{Shingo Kono$^\dagger$} 
\thanks{These authors contributed equally.\\ Corresponding authors: \\ \noindent $^\dagger$ shingo.kono@epfl.ch\\ \noindent $^\ddagger$ tobias.kippenberg@epfl.ch} 

\author{Jiahe Pan}
\thanks{These authors contributed equally.\\ Corresponding authors: \\ \noindent $^\dagger$ shingo.kono@epfl.ch\\ \noindent $^\ddagger$ tobias.kippenberg@epfl.ch} 

\author{Mahdi Chegnizadeh}
\thanks{These authors contributed equally.\\ Corresponding authors: \\ \noindent $^\dagger$ shingo.kono@epfl.ch\\ \noindent $^\ddagger$ tobias.kippenberg@epfl.ch} 

\author{Xuxin Wang}

\author{Amir Youssefi}

\author{Marco Scigliuzzo}

\author{Tobias~J.~Kippenberg$^\ddagger$} 

\affiliation{Institute of Physics, Swiss Federal Institute of Technology Lausanne (EPFL), Lausanne, Switzerland}
\affiliation{Center for Quantum Science and Engineering, EPFL, Lausanne, Switzerland}

\begin{abstract}
Superconducting qubits are one of the most advanced candidates to realize scalable and fault-tolerant quantum computing. 
Despite recent significant advancements in the qubit lifetimes, the origin of the loss mechanism for state-of-the-art qubits is still subject to investigation. 
Moreover, successful implementation of quantum error correction requires negligible correlated errors among qubits.
Here, we realize ultra-coherent superconducting transmon qubits based on niobium capacitor electrodes, with lifetimes exceeding 0.4 ms. 
By employing a nearly quantum-limited readout chain based on a Josephson traveling wave parametric amplifier, we are able to simultaneously record bit-flip errors occurring in a multiple-qubit device, revealing that the bit-flip errors in two highly coherent qubits are strongly correlated.
By introducing a novel time-resolved analysis synchronized with the operation of the pulse tube cooler in a dilution refrigerator, we find that a pulse tube mechanical shock causes nonequilibrium dynamics of the qubits, leading to correlated bit-flip errors as well as transitions outside of the computational state space.
Our observations confirm that coherence improvements are still attainable in transmon qubits based on the superconducting material that has been commonly used in the field.
In addition, our findings are consistent with qubit dynamics induced by two-level systems and quasiparticles, deepening our understanding of the qubit error mechanisms.
Finally, these results inform possible new error-mitigation strategies by decoupling superconducting qubits from their mechanical environments. 
\end{abstract}

\maketitle

\section{Introduction}
\vspace{-10pt}
Superconducting qubits have become a viable platform both for scientific and technological applications, ranging from fundamental quantum optical experiments~\cite{gu2017microwave}, hybrid quantum systems~\cite{clerk2020hybrid} to quantum information science~\cite{kjaergaard2020superconducting}.
In particular, they have attracted much attention in fault-tolerant quantum computing, achieving important milestones, including the realization of high-fidelity quantum gate and readout on a multiple-qubit system~\cite{barends2014superconducting,heinsoo2018rapid}, the reports of quantum supremacy~\cite{arute2019quantum,wu2021strong}, and the demonstrations of surface codes~\cite{krinner2022realizing,zhao2022realization,Google_Quantum_AI2023-wv}.
Despite such encouraging progress, realizing large-scale superconducting quantum computing is still an outstanding challenge~\cite{gambetta2017building}. 
Although quantum error correction promises reliable and scalable quantum computing, it strictly requires the physical errors of a large number of qubits to be sufficiently smaller than a certain threshold, and, more importantly, to be uncorrelated~\cite{lidar2013quantum,fowler2012surface}. 
Since the physical errors in superconducting qubits are dominantly limited by their coherence~\cite{barends2014superconducting,heinsoo2018rapid,xu2020high}, considerable efforts towards improvements in the qubit lifetimes~\cite{siddiqi2021engineering} have been made using insights from diverse fields, ranging from classical and quantum circuit engineering~\cite{reed2010fast,paik2011observation,yan2016flux,nguyen2019high,gordon2022environmental} to material science~\cite{Murray2021-fv,place2021new,wang2022towards,deng2023titanium}.
However, it still remains unclear whether the qubit lifetimes can be enhanced steadily.
In addition, more coherent superconducting qubits are more sensitive to small changes in their environments, imposing a challenge on their scalability. 
For instance, when a superconducting qubit is dominantly coupled to a few two-level systems (TLSs), its relaxation time is largely fluctuating~\cite{klimov2018fluctuations,burnett2019decoherence}.
More recently, it has been reported that the absorption of ionizing radiation generates high-energy phonons in a qubit substrate, which leads to nonequilibrium quasiparticles, causing correlated charge-parity switchings~\cite{wilen2021correlated} and energy relaxations~\cite{mcewen2022resolving} of superconducting qubits.
To verify fault tolerance, it is, therefore, more and more important to characterize a highly coherent multiple-qubit system more carefully, i.e., not only reporting their averaged coherence but also studying the time and frequency dependence of their coherence, as well as confirming the absence of correlated errors.
Indeed, such characterizations have revealed dominant loss mechanisms and sources of fluctuations in superconducting qubits, such as surface dielectric loss~\cite{wang2015surface}, TLSs~\cite{klimov2018fluctuations,martinis2005decoherence,grabovskij2012strain,lisenfeld2019electric}, nonequilibrium quasiparticles~\cite{wang2014measurement,serniak2018hot}, and ionizing radiation~\cite{cardani2021reducing,vepsalainen2020impact,thorbeck_tls_2022}.

\begin{figure*}[t]
\includegraphics[width=17cm]{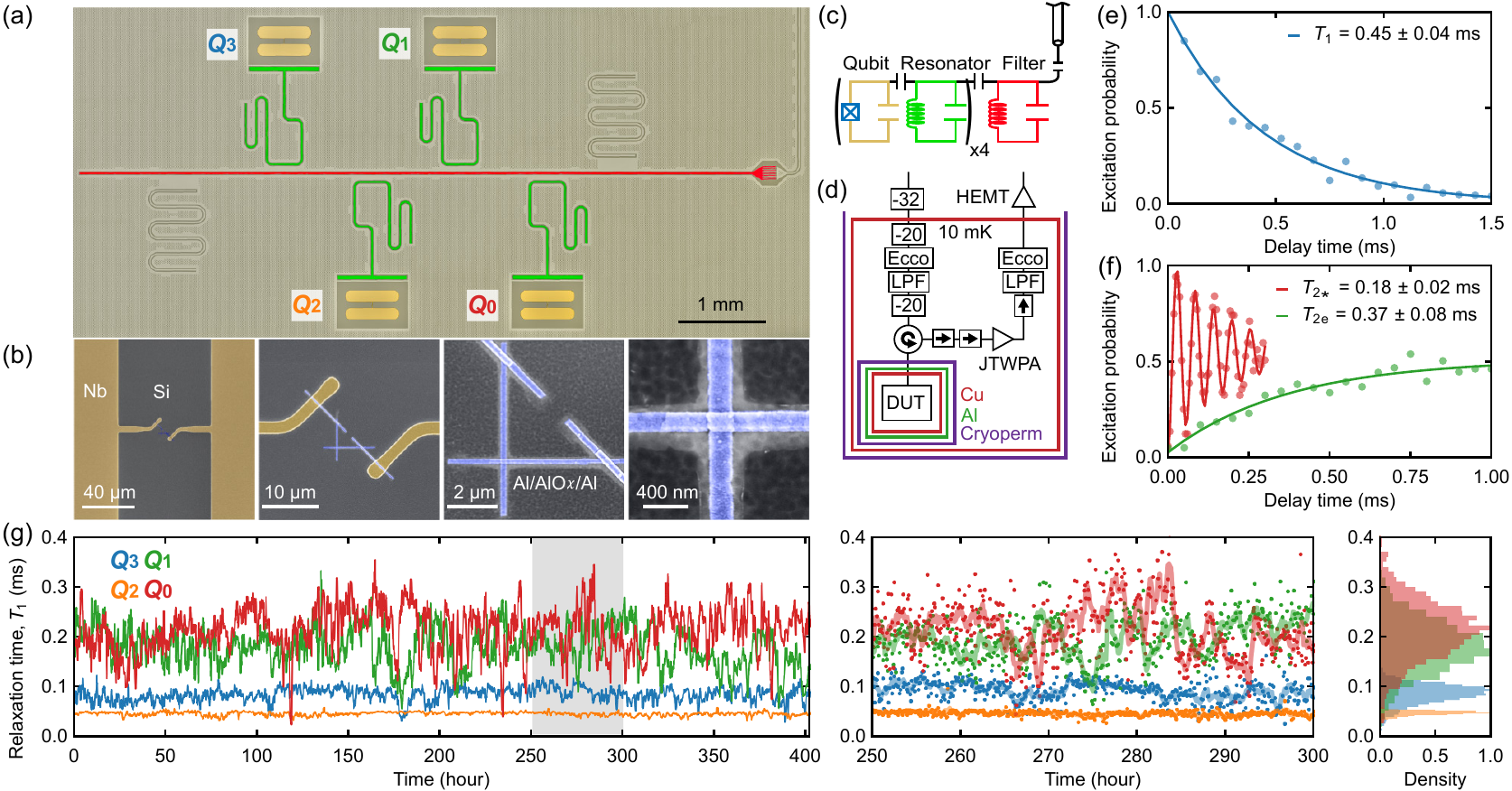}
\vspace{-5pt}
\caption{
\textbf{Ultra-coherent multiple superconducting transmon qubit device based on Nb electrodes.}
(a)~False-colored optical micrograph of a superconducting device, consisting of four transmon qubits (yellow, $Q_0$ -- $Q_3$) with individual readout resonators~(green) coupled to a shared Purcell filter~(red).
(b)~False-colored scanning electron microscope images of an $\mathrm{Al}/\mathrm{AlO}_x/\mathrm{Al}$ Josephson junction~(blue) shunted by a Nb capacitor~(yellow) on a silicon substrate~(gray).
(c)~Equivalent circuit of the device.
(d)~Simplified cryogenic wiring. 
(e),(f)~Time traces of the excitation probability of qubit $Q_0$, showing the best relaxation times and the Ramsey and Hahn-echo dephasing times. 
(g)~Long-term stability of the relaxation times of the four qubits. 
The middle panel shows the magnified plot for the gray region of the left panel, where the dots are the results obtained from the individual time traces, while the solid lines are their smoothed data with a 5-hour time window.
The right panel shows the height-wise normalized histograms.
}
\label{fig:1}
\vspace{-10pt}
\end{figure*}

Here, we realize ultra-coherent superconducting transmon qubits based on niobium electrodes with relaxation times exceeding 0.4~ms and report a new source of correlated bit-flip errors, caused by mechanical bursts generated by the pulse tube cooler of a dry dilution refrigerator~\cite{olivieri2017vibrations}.
This is revealed by a novel time-resolved analysis of the residual excitation probabilities and quantum jumps of multiple qubits, which are synchronized with the operation of the pulse tube cooler. 
Although the origin of the mechanical sensitivity of long-lived superconducting qubits could not be determined unambiguously in this work, our observations are consistent with TLS- and quasiparticle-mediated qubit decays to phononic baths~\cite{muller2019towards,glazman2021bogoliubov}. 
Moreover, these findings suggest future strategies for fault-tolerant superconducting quantum computing, including the development of acoustically shielded superconducting devices~\cite{rosen2019protecting}, mechanical shock-resilient sample packaging~\cite{PIRRO2000331,maisonobe2018vibration}, and a vibration-free dilution  refrigerator~\cite{Kalra2016-ey,cao2022vibration,uhlig2023dry}.

\begin{figure*}[t]
\includegraphics[width=17cm]{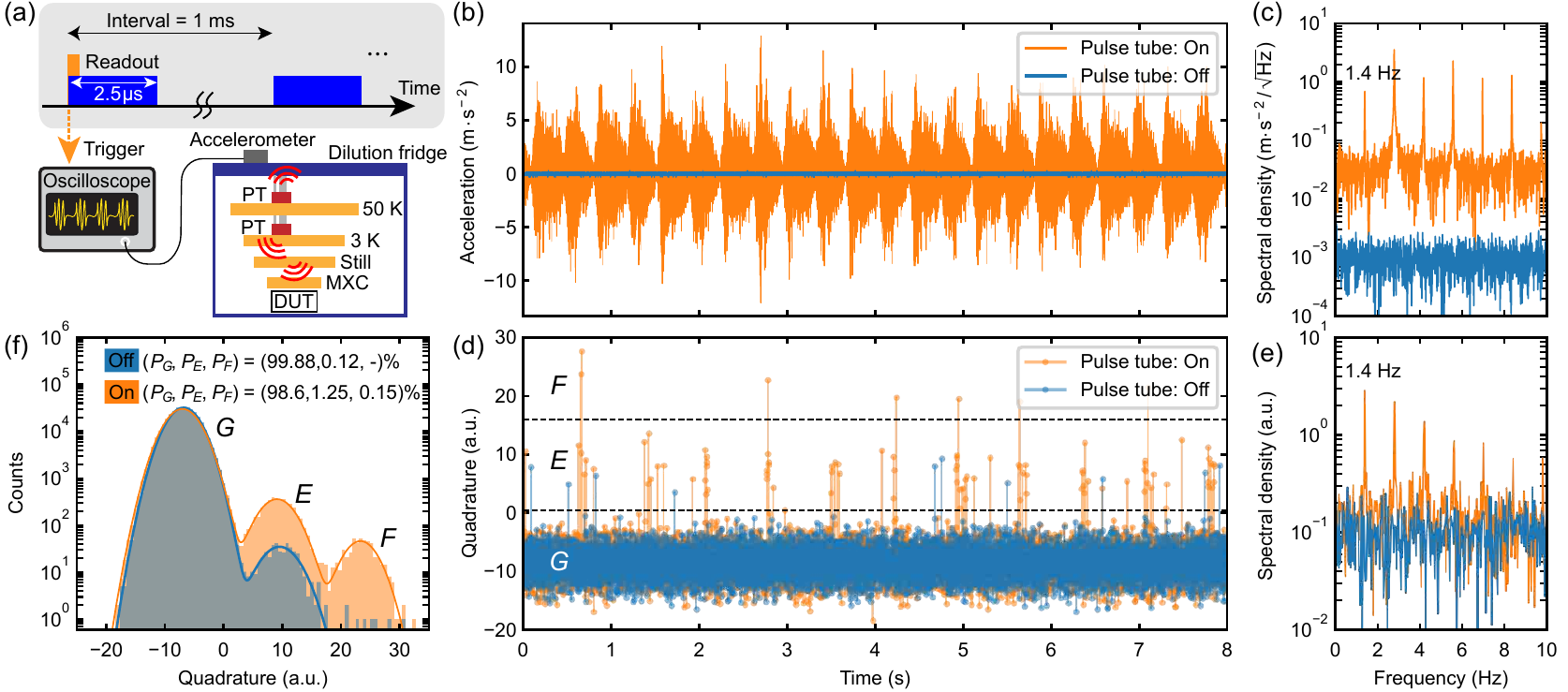}
\vspace{-5pt}
\caption{
\textbf{Out of equilibrium dynamics of a transmon qubit, induced by mechanical shockwaves from the pulse tube cooler.}
(a)~Pulse scheme and simplified experimental setup for simultaneously recording a vibrational noise generated by the pulse tube cooler~(PT) and a time trace of the single-shot readout outcomes. 
(b)~Time trace of the acceleration of the vibrational noise and (c)~its amplitude spectral density~(ASD) with the pulse tube cooler on~(orange) and off~(blue). The fundamental frequency of the harmonics is approximately 1.4~Hz.
(d)~Time trace of the readout quadrature amplitude and (e)~its ASD. The black dashed lines show a threshold to distinguish between the $G$ and $E$ states and one between the $E$ and $F$ states, respectively.
(f)~Histograms of the qubit readout quadrature amplitudes when the pulse tube cooler is on and off.
The solid lines are weighted multiple Gaussian distributions fitted to the histograms in order to obtain the occupation probabilities. 
}
\label{fig:2}
\vspace{-10pt}
\end{figure*}

\vspace{-10pt}
\section{Results}
\vspace{-10pt}
\subsection{Ultra-coherent transmon qubits}
\vspace{-10pt}
We develop ultra-coherent superconducting transmon qubits, formed by a single $\mathrm{Al}/\mathrm{AlO}_x/\mathrm{Al}$ Josephson junction shunted by a Nb capacitor, fabricated on a high-resistivity silicon substrate.
Figures~\ref{fig:1}(a) and (c) show an optical micrograph of a fabricated multiple superconducting qubit device and its equivalent circuit model, respectively, including four frequency-fixed transmon qubits with resonance frequencies ranging from 4.8~GHz to 6.2~GHz and anharmonicities of 0.26~GHz on average.
As the metal-substrate interface of the Al film fabricated by a lift-off process can not be as clean as that of the Nb film directly sputtered on the silicon substrate, we minimize the area of the Al electrodes and bandage patches~\cite{osman2021simplified} to reduce the energy participation ratio in the interface~[see Fig.~\ref{fig:1}(b)].
To realize multiplexed dispersive readout, all the qubits are individually coupled to $\lambda/4$ readout resonators with different resonance frequencies around 7~GHz, sharing a $\lambda/2$ Purcell filter~\cite{jeffrey2014fast}.
The filter is connected to a feed line, along which frequency-multiplexed control and readout signals are sent. 
The filter is designed to have a 7-GHz resonance frequency and a 300-MHz bandwidth, suppressing the qubit radiative decay rates to a level of $\mathcal{O}(10~\mathrm{Hz})$.
The state-dependent dispersive shifts and the readout resonator bandwidths are designed to be $\mathcal{O}(1~\rm{MHz})$. 
Note that the different dispersive shifts of the readout resonators for the first and second excited states enable us to distinguish the readout signals corresponding to the first three states ($G$, $E$, and $F$) in a single shot. 
See Table~\ref{Table:S1} in Appendix~A for the full characterization of the system parameters.

\begin{figure}[t]
\includegraphics[width=8.5cm]{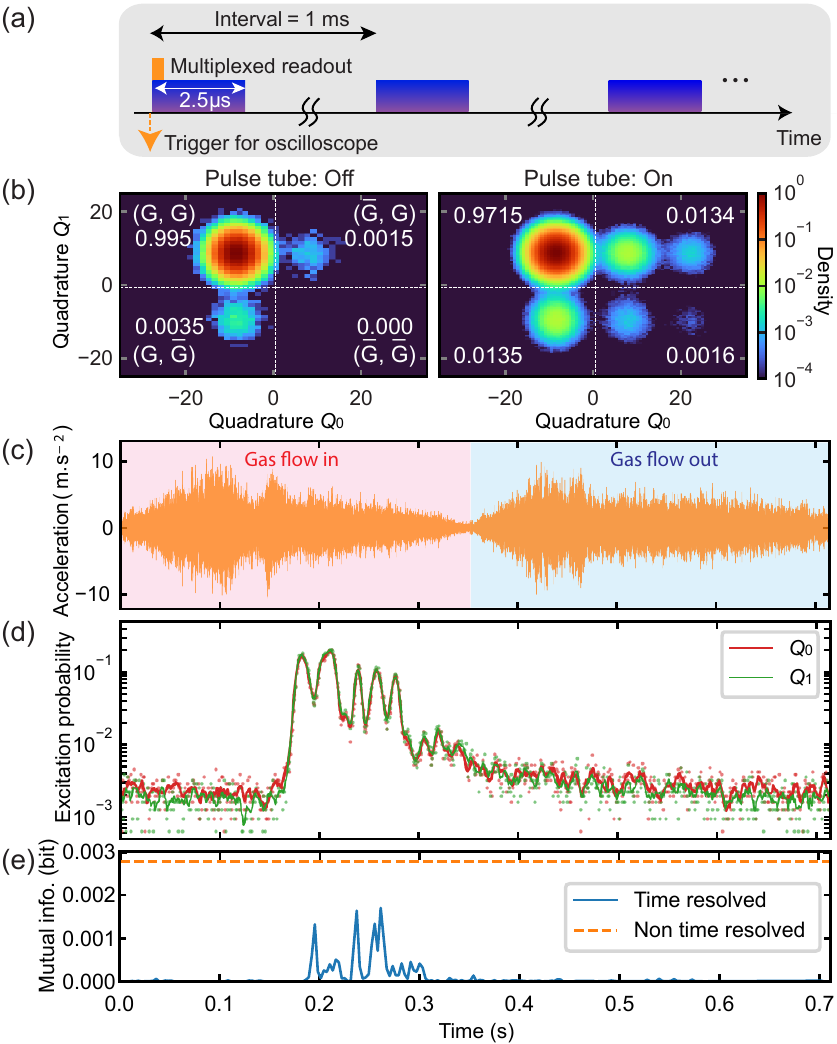}
\vspace{-5pt}
\caption{
\textbf{Mechanically induced correlated qubit excitations.}
(a)~Pulse scheme for measuring the qubit excitation probabilities in a time-resolved manner, synchronized with the periodic vibrational noise generated by the pulse tube cooler.
(b)~Height-wise normalized 2D histograms of the readout quadrature amplitudes for qubits $Q_0$ and $Q_1$ with the pulse tube cooler off and on, respectively.
(c)~Time trace of the acceleration of the referential vibrational noise within one period.
There are two phases in one cycle of the pulse tube operation: helium gas flow in~(red shaded) and out~(red shaded).
(d)~Time-resolved qubit residual excitation probabilities of $Q_0$ and $Q_1$, time-aligned to the referential periodic vibrational noise. 
The dots and lines are the raw data and their smoothed data with a 5-ms time window, respectively.
(e)~Mutual information as a function of time within the one vibrational period, obtained from the time-resolved simultaneous single-shot data for $Q_0$ and $Q_1$.
The blue solid line is the time-resolved MI, while the orange dashed line is the non-time-resolved one.
}
\label{fig:3}
\vspace{-10pt}
\end{figure}

As schematically shown in Fig.~\ref{fig:1}(d), the fabricated device is mounted at the mixing chamber stage ($\sim$10~mK) of a dry dilution refrigerator, enclosed in a multilayer shielding: copper radiation shields and magnetic shields of aluminum and cryoperm.
The transmon qubits are characterized using a nearly quantum-limited broadband Josephson traveling wave parametric amplifier~\cite{macklin2015near}, allowing us to perform the simultaneous single-shot readout of the qubits by frequency-multiplexing~\cite{heinsoo2018rapid}.
The readout error probabilities for the $G$ and $E$ states are characterized to be $<0.001$ and $<0.03$, which are limited by separation errors and readout-induced state-flip errors~\cite{boissonneault2009dispersive}, respectively (see Appendix E2).
To suppress thermal and backward amplifier noises, the input and output lines are heavily attenuated and isolated, respectively, while both are equipped with low-pass filters and eccosorb filters~(see Appendix~B).

Figures~\ref{fig:1}(e) and (f) show the time traces of the excitation probability of the transmon qubit with a resonance frequency of 4.8~GHz~($Q_0$), showing the best relaxation times ($T_1 = 0.45\pm0.04$~ms) and the Ramsey and Hahn-echo dephasing times ($T_{2*} = 0.18\pm0.02$~ms and $T_{2\mathrm{e}} = 0.37\pm0.08$~ms), respectively.
Our observations confirm that coherence improvements are still possible with widely employed superconducting material systems involving $\mathrm{Al}/\mathrm{AlO}_x/\mathrm{Al}$ Josephson junctions and Nb electrodes fabricated on a silicon substrate~\cite{gordon2022environmental}.

As shown in Fig.~\ref{fig:1}(g), we measure the long-term stability of the relaxation times of the four qubits, showing significantly large fluctuations, especially for the longer-lived qubits ($Q_0$ and $Q_1$) with averaged $T_1$ of approximately $0.2$~ms and relative standard deviations of 30~\%, while $Q_2$ and $Q_3$ with averaged $T_1$ of 0.04-0.08~ms exhibit smaller relative deviations of 10-20~\%. 
The Allan deviation analysis of the fluctuations implies that the relaxation times of our long-lived transmon qubits are mainly limited by TLSs~\cite{burnett2019decoherence}~(see Appendix~E1).

\vspace{-10pt}
\subsection{Effect of pulse tube cooler on qubit excitation}
\vspace{-10pt}
We perform the single-shot readout for the longest lifetime transmon qubit~($Q_0$).
As shown in \fig\ref{fig:2}(a), we apply 2.5-$\mu$s long readout pulses repeatedly with an interval of 1~ms ($\gg$ averaged $T_1$), which is sufficiently long to prepare the qubit in the equilibrium.
In addition to the conventional qubit measurement setup, an accelerometer, mechanically anchored to the top plate of the dilution refrigerator, converts the vibrational noise to a voltage signal. 
The converted signal is acquired by an oscilloscope that is operated synchronously with the qubit readout sequence via a trigger signal generated by the qubit measurement setup, enabling us to simultaneously record both the vibrational noise and the single-shot readout outcomes.

Figures~\ref{fig:2}(b) and (d) show the synchronized time trace records of the vibrational noise and the qubit single-shot readout quadrature amplitude, respectively.
When the pulse tube cooler is on~(orange), the qubit is more frequently excited to the first excited state ($E$), even to the second excited state ($F$), while with the cooler off~(blue), it mostly remains in the ground state ($G$).
Here, we temporarily switch off the pulse tube cooler without affecting the base temperature ($\approx 5$~minutes).
Note that the readout signals for the $E$ and $F$ states of $Q_0$ can be distinguished well in one quadrature projected in an optimal phase.
More importantly, when the pulse tube cooler is on, the qubit becomes excited periodically in time, synchronized with the periodic vibrational noise.
Figures~\ref{fig:2}(c) and (e) show the amplitude spectral densities of the vibrational noise and the qubit readout quadrature, respectively, showing both have harmonics with exactly the same fundamental frequency of approximately 1.4~Hz when the pulse tube cooler is on.

Figure~\ref{fig:2}(f) shows the histograms of the readout quadratures (the number of each data $\approx 3\times 10^5$) with the pulse tube cooler on and off. 
To mitigate the readout separation errors, we obtain the occupation probability in each qubit state by fitting the histogram to weighted multiple Gaussian distributions.
When the pulse tube cooler is switched off, we achieve a high initialization fidelity of 99.88~\% by passive cooling, where the corresponding effective temperature is $T_\mathrm{eff}=34$~mK, although the base temperature is approximately $10$~mK.
Importantly note that the excitation probability due to the readout backaction is $<$~6e-5 (see Appendix E2).
In contrast, when the pulse tube is on, the qubit is excited to the $E$ state with a higher probability ($P_E=1.25$~\%), and even the occupation probability in the $F$ state is not negligible ($P_F=0.15$~\%).
More interestingly, the occupation probability distribution is not in the thermal equilibrium, i.e., the effective temperature is $T_\mathrm{eff}=53$~mK for the $G$-$E$ transition, while $T_\mathrm{eff}=102$~mK for the $E$-$F$ transition. 
This implies that the qubit is not simply excited by a local thermal heating of the phononic environment, but it is excited by nonequilibrium dynamics of the mechanical shock generated by the pulse tube cooler.

\vspace{-10pt}
\subsection{Mechanically induced correlated excitations}
\vspace{-10pt}
We study the existence of correlated excitations for two of the long-lived qubits, $Q_0$ and $Q_1$ (averaged $T_1 \approx 0.2$~ms).
Figure~\ref{fig:3}(a) shows the pulse scheme for the multiplexed single-shot readout of the two qubits, while Figure~\ref{fig:3}(b) shows the 2D histograms of the readout quadrature outcomes of approximately $3\times 10^5$ and $10^7$ when the pulse tube cooler is off and on, respectively. 
The excitation probability, or the probability when the state is found in the $E$ or $F$ states ($\bar{G}$), can be characterized for both the qubits, as shown in Figure~\ref{fig:3}(b).
To quantitatively study the correlated excitations, we use mutual information~(MI) in the unit of bit, which is defined as
\begin{equation}
\label{eq:1}
I = \sum_{\mathcal{X},\mathcal{Y}} \: P(\mathcal{X},\mathcal{Y}) \log_2\left[\frac{P(\mathcal{X},\mathcal{Y})}{P(\mathcal{X})\cdot P(\mathcal{Y})}\right].
\end{equation}
Here, $P(\mathcal{X},\mathcal{Y})$ is the joint probability of qubit $Q_0$ in the $\mathcal{X}$ state and qubit $Q_1$ in the $\mathcal{Y}$ state, while $P(\mathcal{X})$ and $P(\mathcal{Y})$ are the marginal probabilities of $Q_0$ in $\mathcal{X}$ and $Q_1$ in $\mathcal{Y}$, respectively, where $\mathcal{X}$ and $\mathcal{Y}$ can be $G$ and $\bar{G}=E$ or $F$.
This quantifies how much information about the excitation of one qubit we obtain from the other qubit~($0\leq I \leq1$~bit). 
For the case when the pulse tube cooler is off, the MI is $I<$~1e-5~bit, while that with the pulse tube on is found to be $I=0.0028$~bit~[orange dashed line in Fig.~\ref{fig:3}(e)], showing there is a significant correlation in their excitations that are induced by the mechanical shocks. 


\begin{figure*}[t]
\includegraphics[width=17cm]{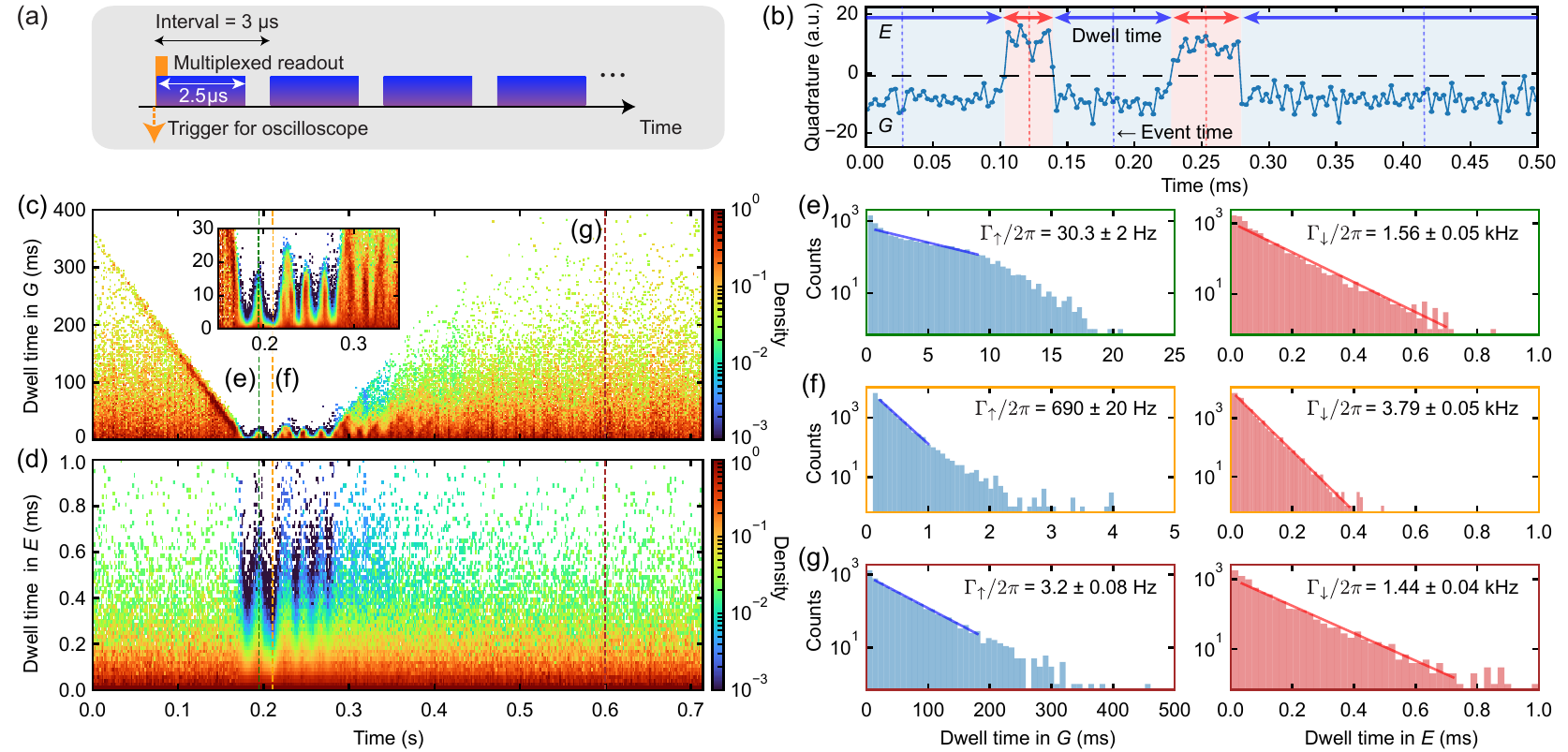}
\vspace{-5pt}
\caption{
\textbf{Mechanically induced quantum jumps.}
(a)~Pulse scheme for continuously monitoring the states of $Q_0$ and $Q_1$, where 2.5-$\mu$s long multiplexed readout pulses are repeated successively with an interval of 3~$\mu$s, synchronized with the measurement of the vibrational noise from the pulse tube cooler.
(b)~An example of a time trace of the single-shot readout quadrature for qubit $Q_0$, showing several quantum jumps. The shaded blue and red regions correspond to the dwell events in the $G$ and $E$ states, respectively. The event time, centered at each dwell event, is shown with a vertical dashed line, while the dwell time is shown with a double-sided horizontal arrow.
(c),(d)~2D histograms of the event time and the dwell times in the $G$ and $E$ states, respectively, where the event time is time-aligned to the referential periodic vibrational noise, shown in Fig.~\ref{fig:3}(c).
The histograms are normalized by the height at the first dwell time bin for every event time.
The inset in (c) shows the same in a magnified region.
(e)-(g)~Time-resolved histograms of the dwell times in the $G$ and $E$ states for different event times specified with dashed lines in (c) and (d).
The solid lines are the fitting results, shown within the fitting regions, resulting in the corresponding transition rates.
}
\label{fig:4-1}
\vspace{-10pt}
\end{figure*}

To investigate the origin of the correlated excitations, we develop a time-resolved analysis of the residual qubit excitation probabilities, synchronized with the periodic vibrational noise.
As shown in \fig\ref{fig:3}(a), we repeat a sequence consisting of $2^{12}$ readout pulses (approximately 4 seconds in total) while simultaneously recording the vibrational noise.
Since the starting time of each sequence is not synchronized with the phase of the periodic vibrational noise, we need to time-align every time trace of the single-shot data with respect to the periodic vibrational noise. 
To this end, we first specify one period of the vibrational noise as a reference, as shown in Fig.~\ref{fig:3}(c). 
Then, we time-align every trace of the single-shot outcomes by maximizing the cross-correlation of the simultaneously recorded vibrational noise with the referential one. 
Consequently, we can accumulate a sufficient number of the single-shot outcomes at an arbitrary time of interest within the vibrational period to obtain the time-resolved residual excitation probabilities of the qubits.
Figure~\ref{fig:3}(d) shows the results of the time-resolved measurement of the excitation probabilities obtained from the sequences repeated about 3000 times.
Although the two operational phases of the pulse tube cooler exhibit similar acceleration of vibrations, the qubits are only frequently excited during the gas-flow-in phase.
Furthermore, we find that both the excitation probabilities are increased synchronously from the lowest values ($\approx 0.002$) by a factor of more than one hundred.
This implies that the two qubits are dominantly excited by the common mechanical shocks via their phononic baths.


Figure~\ref{fig:3}(e) shows the time-resolved MI as a function of time within the one vibrational period, which is obtained from the time-resolved multiplexed single-shot readout outcomes for the two qubits using Eq.~(\ref{eq:1}). 
The time-resolved MI (blue solid line) is always smaller than the non-time-resolved value (orange dashed line), implying the time-resolved events are not strongly correlated, but macroscopic parameters, such as the intrinsic decay rates or bath occupations, are changing synchronously for both the qubits, causing the correlated errors in the non-time-resolved analysis.

\vspace{-10pt}
\subsection{Mechanically induced correlated \\ quantum jumps}
\vspace{-10pt}

Next, we study the mechanical effect on quantum jumps of the transmon qubits, i.e., the effect on their transition rates between the $G$ and $E$ states.
As shown in Fig.~\ref{fig:4-1}(a), we repeat a sequence to continuously monitor the states of $Q_0$ and $Q_1$ about 3000 times, while simultaneously recording the vibrational noise from the pulse tube cooler.
Each sequence consists of $2^{19}$ successive multiplexed readout pulses with an interval of 3~$\mu$s, corresponding to a length of $\approx 1.6$~seconds. 
We apply a 2-point moving average to the raw successive readout outcomes to reduce the separation readout errors for the following quantum jump analysis.
Here, we will first focus on the time-resolved measurement of the transition rates of qubit $Q_0$, and then study the existence of a correlation in quantum jumps between qubits $Q_0$ and $Q_1$.

\begin{figure}[t]
\includegraphics[width=8.5cm]{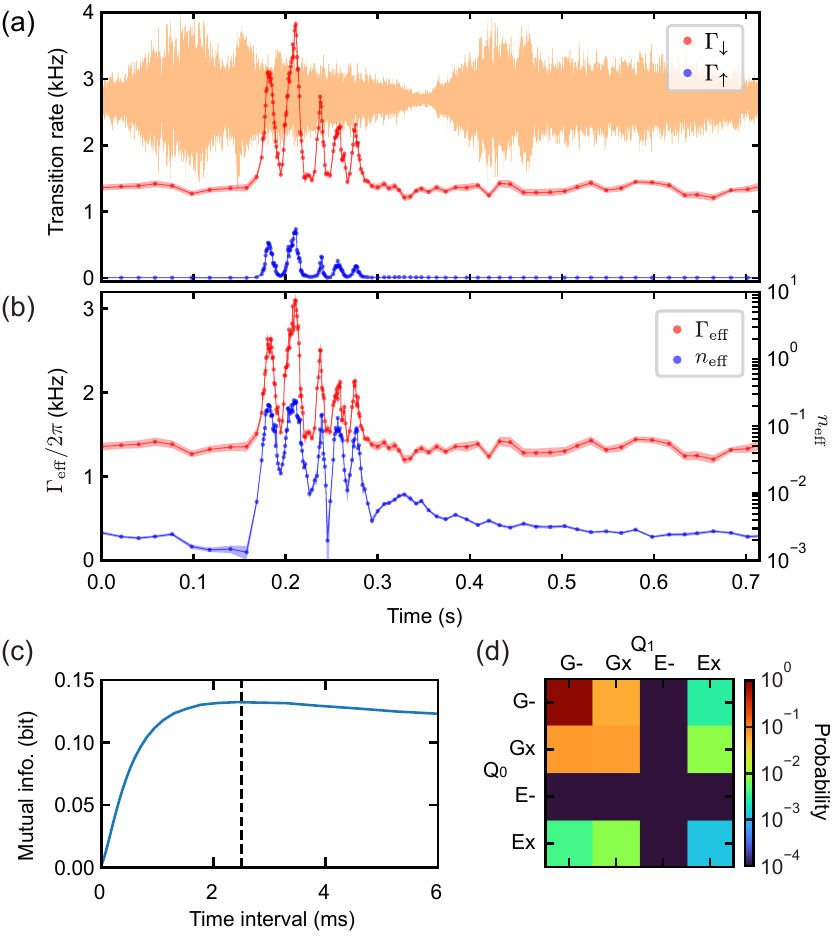}
\vspace{-5pt}
\caption{
\textbf{Time-resolved transition rates and correlated quantum jumps.}
(a)~Time-resolved excitation and decay rates of $Q_0$ ($\Gamma_\uparrow$ and $\Gamma_\downarrow$), time-aligned to the referential periodic vibrational noise (orange). 
The transition rates are obtained by fitting the time-resolved histograms of the dwell times to exponential curves.
(b)~Time-resolved effective intrinsic decay rate and bath occupation of $Q_0$ ($\Gamma_\mathrm{eff}$ and $n_\mathrm{eff}$), which are calculated from $\Gamma_\uparrow$ and $\Gamma_\downarrow$.
The shaded regions in (a) and (b) depict the fitting errors.
(c)~Mutual information~(MI) between the error probabilities of qubits $Q_0$ and $Q_1$, as a function of the time interval. The dashed line marks the time interval showing the maximal MI.
(d)~Table of the total 16 error probabilities of $Q_0$ and $Q_1$, taken at the time interval which maximizes the MI, where the data is clipped to the minimum of $10^{-4}$.
}
\label{fig:4-2}
\vspace{-10pt}
\end{figure}

In a similar manner to the previous time-resolved measurement, we can time-align every continuous monitoring trace with respect to the referential periodic vibrational noise, shown in Fig.~\ref{fig:3}(c).
As shown in Fig.~\ref{fig:4-1}(b), every time-aligned trace contains several quantum jump events, from which we sample the event time (vertical dashed line) and the dwell time (a double-sided horizontal arrow) for the $G$ and $E$ states, individually.
Figures~\ref{fig:4-1}(c) and (d) show the 2D histograms of the event time within the one period of the vibrational noise and the dwell times in the $G$ and $E$ states, respectively, where each histogram contains approximately $10^6$ dwell events.

To determine the transition rates between the $G$ and $E$ states in a time-resolved fashion, we calculate the histogram of the dwell time at a time of interest within the vibrational period. 
The time-resolved histograms for both the $E$ and $G$ dwell times, shown in Figs.~4(e)--(g), are obtained by using the dwell events [Figs.~4(c) and (d)] within a bin centered at the chosen time.
The bin width is set to the average $G$ dwell times for obtaining both the $G$ and $E$ dwell-time histograms.
This is because the time resolution of the measurement of the transition rate from the $G$ to $E$ states is limited by the inverse of the rate, approximately corresponding to the average of the $G$ dwell time around the time of interest. 
Furthermore, the number of the $E$ dwell events within a bin width of the average $E$ dwell times (fundamental time resolution) is not sufficiently large, requiring a wider bin width to accumulate more data points, and eventually limiting the time resolution.
Then, we fit each histogram to an exponential curve, resulting in the corresponding transition rate~\cite{vijay2011observation}. 
Note that we optimize the fitting region for the $G$ dwell time data by minimizing the fitting error, in order to exclude any mechanical shock effects that could occur at the tail of the $G$ dwell-time histogram and cause deviations from a standard exponential distribution, as seen in Fig.~\ref{fig:4-2}(c).

Figure~\ref{fig:4-2}(a) shows the time-resolved transition rates of $\Gamma_\downarrow$ and $\Gamma_\uparrow$ for $Q_0$ as a function of time within the one period of the referential vibrational noise, while Fig.~\ref{fig:4-2}(b) shows the effective intrinsic decay rate and bath occupation ($\Gamma_\mathrm{eff}$ and $n_\mathrm{eff}$), which can be calculated using simple relations, $\Gamma_\downarrow = \Gamma_\mathrm{eff}(n_\mathrm{eff}+1)$ and $\Gamma_\uparrow = \Gamma_\mathrm{eff} n_\mathrm{eff}$.
Note that the time steps are adaptively chosen according to the average $G$ dwell time due to the time resolution limitations as discussed above.
The effective bath occupation ($n_\mathrm{eff}$) is increased by several orders of magnitude at several points, while the effective decay rate ($\Gamma_\mathrm{eff}$) is moderately perturbed.

We further study whether there is a correlation in quantum jumps between qubits $Q_0$ and $Q_1$.
When the qubits are continuously monitored for a certain time interval, an error, corresponding to a state transition, will occur with a finite probability. We classify continuous 2-qubit records into 16 possible events: the initial 2-qubit states \{$GG,GE,EG,GG$\} and the error occurrence within the time interval \{\texttt{--},\texttt{-×},\texttt{×-},\texttt{××}\}.
Here, ``\texttt{-}" stands for the occurrence of no bit-flip event, and ``\texttt{×}" stands for the occurrence of at least one bit-flip event.
As shown with an example in Fig.~\ref{fig:4-2}(d), we can, therefore, calculate a $4\times4$ matrix of the error probabilities at a specific time interval. 

To quantitatively study the existence of correlated bit-flip errors during continuous monitoring, we use mutual information~(MI), defined in Eq.~(\ref{eq:1}), using the $4\times4$ error probability matrix in this case.
The MI corresponds to a quantitative value of the amount of the correlation in the quantum jumps, i.e., how much information about the occurrence of a bit-flip error in one qubit we can obtain from the other one~($0\leq I \leq2$~bit). 
Figure~\ref{fig:4-2}(c) presents the MI of the bit-flip error probabilities between $Q_0$ and $Q_1$ as a function of the time interval, showing the maximum of $I=0.13$ at 2.5~ms.
It clearly shows that there is a correlation in the quantum jump events between the two qubits, which corresponds to the existence of a correlated error in their gates.

\vspace{-10pt}
\subsection{Effect of a controlled mechanical shock}
\begin{figure}[t]
\includegraphics[width=8.5cm]{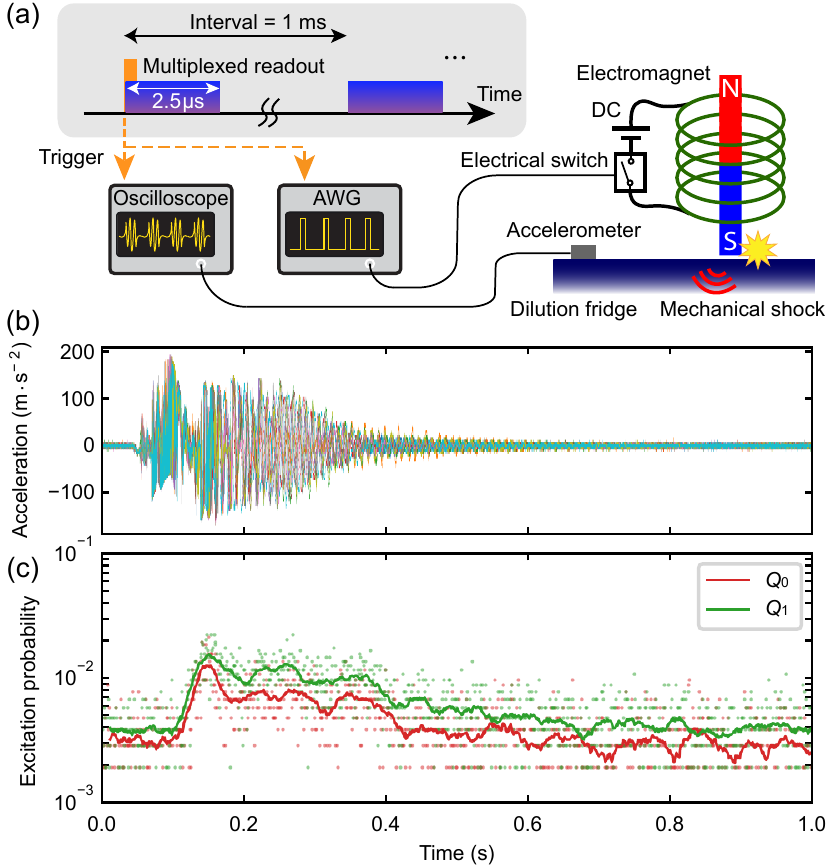}
\vspace{-5pt}
\caption{
\textbf{Qubit excitations induced by a controlled pure mechanical shock.}
(a)~Pulse scheme and simplified experimental setup for simultaneously recording a time trace of the qubit single-shot readout outcomes and a vibrational signal generated by a controlled mechanical shock.
In addition to the oscilloscope setup for the vibration measurement, an AWG sends a synchronized switching signal to trigger a current in an electromagnet, leading to a mechanical shock.
(b)~Multiple time traces of the acceleration signal generated by the controlled mechanical shock.
(c)~Time-resolved qubit excitation probabilities, synchronized with the controlled mechanical shock while the pulse tube cooler is deactivated.
The dots and lines are the raw data and their smoothed data with a 5-ms time window, respectively.
}
\label{fig:5}
\vspace{-10pt}
\end{figure}

\vspace{-10pt}
To rule out that the qubit excitations are due to possible electrical noise produced by the pulse tube cooler, we mount on the top plate of the dilution refrigerator an electric hammer based on an electromagnet~[see Fig.~\ref{fig:5}(a)]. 
In this manner, we generate a purely mechanical shock by a pulsed DC current in the electromagnet, synchronized with the qubit readout sequence via a trigger signal.
We simultaneously record both the vibrational signal and the qubit single-shot readout outcomes while the pulse tube cooler is deactivated.

Figure~\ref{fig:5}(b) presents the time trace of the acceleration generated by the controlled mechanical shock, showing an impulse shock followed by a damped oscillation at eigenfrequencies of the refrigerator.
Figure~\ref{fig:5}(c) shows the time-resolved excitation probabilities of qubits $Q_0$ and $Q_1$, measured synchronously with the mechanical shock.
The excitation probabilities are obtained from the single-shot outcomes of approximately 1000 for each time of interest.
We observe that both the qubits are similarly excited by the purely mechanical shock, which implies that any possible electrical noise generated from the pulse tube cooler is not dominantly involved in the qubit excitations.

\vspace{-10pt}
\subsection{Time-resolved measurement of microwave bath}
\vspace{-10pt}
One possible process to excite the qubits mechanically is that a mechanical shock would change the microwave property of the feed line, leading to an increase in the microwave background noise.
For example, a change in the transmission through the filters and attenuators may alter the amount of thermal noise coming from the higher temperature stages of the refrigerator.
Moreover, as shown in Fig.~\ref{fig:SI_Setup} of Appendix~B, we use an attenuator and a terminator made of crystalline quartz in order to acoustically thermalize the feed line to the base temperature.
Such microwave components would convert a mechanical shock to microwave noise, eventually exciting the qubits.
This concern motivates us to perform the time-resolved analysis of the microwave transmission ratio ($S_{21}$) and background noise of the measurement chain, synchronized with the measurement of the periodic vibrational noise generated by the pulse tube cooler.

\begin{figure}[t]
\includegraphics[width=8.5cm]{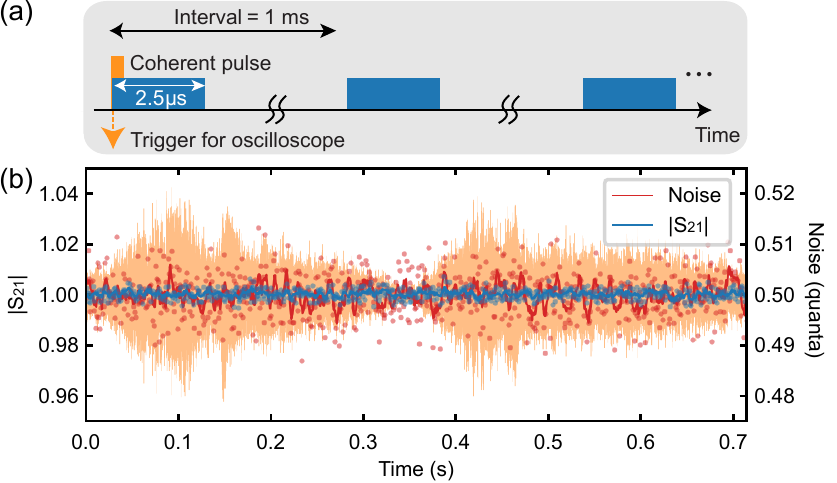}
\vspace{-5pt}
\caption{
\textbf{Time-resolved measurement of microwave bath.}
(a)~Pulse scheme for characterizing microwave bath, where 2.5-$\mu$s long coherent pulses are repeated with an interval of 1~ms, synchronized with the measurement of the vibrational noise from the pulse tube cooler.
(b)~Transmission and background noise of the full measurement chain, measured in a time-resolved fashion and synchronized with the referential periodic vibrational noise~(orange).
The transmission and background noise are normalized by the non-time-resolved values, respectively. 
Moreover, the background noise is rescaled to the vacuum noise level (1/2), to be in the photon unit.
The dots and lines are the raw data and their smoothed data with a 5-ms time window, respectively.
}
\label{fig:6}
\vspace{-10pt}
\end{figure}

As shown in Fig.~\ref{fig:6}(a), we apply coherent pulses at around 7~GHz, which is off-resonant with the resonators and qubits.
By measuring the average and variance of the coherent amplitude of the pulse in a time-resolved manner (similar to the experiment for Fig.~\ref{fig:3}), we can characterize the microwave transmission ratio and background noise at a time of interest within the period of the vibrational noise.
Figure~\ref{fig:6}(b) shows the microwave transmission amplitude ratio ($|S_{21}|)$ and background noise normalized by the non-time-resolved values, respectively.
No significant variation is observed in the time-resolved transmission ratio and background noise.

Here, we can estimate at least how much background noise increase is required to excite the qubits to the extent observed in Fig.~\ref{fig:3}(d).
From numerical simulations based on the finite-element method, the external coupling rate of qubits $Q_0$ and $Q_1$ to the feed line is of the order of $\Gamma_\mathrm{ex}/2\pi \approx 10$~Hz, showing strong suppression of the radiation loss by the Purcell filter.
In contrast, the experimentally measured relaxation rate is of the order of 1~kHz, confirming that qubits are strongly under-coupled to the feed line ($\Gamma_{\rm{in}}/2\pi \approx 1$~kHz, where $\Gamma_{\rm{in}}$ is the internal loss rate of the qubit).
Assuming the intrinsic qubit bath temperature is negligible, we can estimate the microwave bath photon occupation ($n_\mathrm{ex}$) required to have the residual excitation probability of $P_E \approx 0.2$ by using 
\begin{equation}
	P_E = \frac{\Gamma_\mathrm{ex}n_\mathrm{ex}}{\Gamma_\mathrm{ex}(2n_\mathrm{ex}+1) + \Gamma_\mathrm{in}}.
\end{equation}
This results in $n_\mathrm{ex} \approx 30$.
Furthermore, when the thermal photon occupation is measured with a measurement chain of a quantum efficiency $\eta$, the measured photon occupation is calculated as $\tilde{n}_\mathrm{ex}=\eta n_\mathrm{ex}$.
Even though the quantum efficiency can not be reliably calibrated using our experimental data, we can safely assume that the efficiency is at least of the order of 0.1, given that the nearly quantum-limited JTWPA is used as a preamplifier~\cite{youssefi2022squeezed}.
If it dominantly heats up the qubits, the background noise needs to be increased to be of the order of $\tilde{n}_\mathrm{ex}\approx 3$.
However, there is no increase observed at such a level in the background noise, as shown in Fig.~\ref{fig:6}(b).
This confirms that the microwave bath can not play any role in the excitation of the qubits.

\vspace{-10pt}
\section{Discussion}
\vspace{-10pt}

While we clearly observe that the transmon qubits suffer from correlated bit-flip errors due to the global nature of mechanical shocks, the physical origin still remains uncertain.
Here, we propose and discuss possible explanations for the mechanically-induced qubit excitations and quantum jumps of the qubits.

First, we can rule out the possibility that the qubits are excited by an abrupt thermal heating of the mixing chamber plate caused by a mechanical shock and not time-resolved by a conventional thermometer.
In fact, this is not consistent with the nonthermal probability distribution of the qubit produced by the pulse tube shock [see Fig.~\ref{fig:2}(f)].

In addition, we verify that the mechanical vibrations do not affect the microwave environment around the qubit frequencies, although they can cause low-frequency electrical noise in cables, degrading the coherence of a spin qubit~\cite{Kalra2016-ey}. 
Indeed, we record no fluctuation in the transmission ratio and background noise of the feed line while the mechanical bursts excite the qubits (see Fig.~\ref{fig:6}).
Furthermore, the absence of variation in the microwave background noise also supports the lack of heating of the mixing chamber plate (thermally coupled to the microwave environment via the attenuators). 
In addition, this is also consistent with the qubit excitations caused by a pure mechanical shock, as shown in Fig.~\ref{fig:5}, where any possible electrical noise generated by the pulse tube cooler does not play an important role in the qubit excitations.
Nevertheless, our characterization does not cover the frequency range for infrared photons, which are more sensitive to a small mechanical displacement in microwave components and can induce a qubit decay by breaking Cooper pairs~\cite{corcoles2011protecting}.

We associate the mechanical sensitivity of the qubits with TLS- and quasiparticle-mediated interactions to their phononic baths. 
On one hand, high-energy phonons break Cooper pairs, resulting in nonequilibrium quasiparticles in the qubit electrodes~\cite{martinis2021saving,glazman2021bogoliubov}. 
The quasiparticles are quickly cooled down to the lowest energy level above the superconducting energy gap, remaining there for several tens of milliseconds before recombining into Cooper pairs~\cite{wang2014measurement,mcewen2022resolving}. 
This nonequilibrium quasiparticle bath could be heated by scattering with the phonons produced by the pulse tube shock, resulting in excitations or relaxations of the qubits. 
On the other hand, TLSs couple to a qubit via their electric dipole-dipole interactions while coupling to a phononic bath via their strain potential~\cite{muller2019towards}. Therefore, they can mediate between the qubit and the phononic bath, resulting in the mechanical sensitivity of the qubit.
Finally, both a change in the strain~\cite{lisenfeld2015observation} or the saturation of the TLS bath~\cite{andersson2021acoustic} can alter the coupling between the qubit and the TLSs, leading to a fluctuation in the qubit decay rate to the phononic bath.  
Additionally, the TLS model also explains the long-term fluctuations that are observed in the lifetimes of our qubits~\cite{klimov2018fluctuations,burnett2019decoherence}, as shown in Fig~\ref{fig:1}(g). 


\vspace{-10pt}
\section{Conclusion and outlook}
\vspace{-10pt}
In summary, in this work, we presented a novel time-resolved measurement technique to study the mechanical sensitivity of ultra-coherent transmon qubits, synchronized with the operation of the pulse tube cooler of a dilution refrigerator. 
Our results demonstrated that the vibrations generated by the pulse tube cooler induce dominant bit-flip errors in the qubits. 
Moreover, the global nature of the mechanical bursts on the multi-qubit device causes correlated errors among the qubits, which are detrimental to realizing large-scale quantum computing based on quantum error correction. 
While the origin of mechanical sensitivity of the qubits could not be established unequivocally, our observations are consistent with TLS- and qusiparticle-induced qubit decay models~\cite{muller2019towards,glazman2021bogoliubov}, and provide valuable insights into the loss mechanisms that limit the state-of-the-art qubit coherence.

Our findings suggest several strategies to mitigate the mechanical sensitivity of superconducting qubits, including the use of a suspended qubit substrate with phononic crystal structures~\cite{rosen2019protecting} and sample packages that employ both mechanical isolation and thermal conductivity~\cite{PIRRO2000331,maisonobe2018vibration}, in order to isolate superconducting qubits from mechanical vibrations. 
Furthermore, our results would emphasize the importance of vibration-free dilution refrigerators~\cite{Kalra2016-ey,cao2022vibration,uhlig2023dry} in achieving long and stable coherence times in superconducting devices. 
We believe that these insights will be valuable for the development of next-generation superconducting-qubit technologies that pave the way for realizing fault-tolerant quantum computing.

\vspace{-10pt}
\section*{Acknowledgment}
\vspace{-10pt}
We thank Sebastian Cozma and Pasquale Scarlino for helping with designing the sample package and Yang Xu for developing the Nb deposition and etching, respectively.
Moreover, we thank MIT Lincoln Laboratory and William D.~Oliver for providing the JTWPA.
This work was supported by the EU H2020 research and innovation programme under the Marie Sklodowska-Curie grant agreement No.~101033361~(QuPhon), and from the European Research Council~(ERC) grant No.~835329~(ExCOM-cCEO). 
This work was also supported by the Swiss National Science Foundation~(SNSF) under grant No.~204927 and the NCCR QSIT grant No.~51NF40-185902.  
All devices were fabricated in the Center of MicroNanoTechnology~(CMi) at EPFL.

\vspace{-10pt}
\section*{Data and materials availability}
\vspace{-10pt}
The code and data used to produce the plots within this paper will be available at a Zenodo open-access repository. 
All other data used in this study are available from the corresponding authors upon reasonable request.

\appendix

\section{System parameters}
\vspace{-10pt}
The system parameters of the multi-qubit device are summarized in Table~\ref{Table:S1}.

The qubit frequencies ($\omega_\mathrm{q}$), relaxation times ($T_1$), and dephasing times ($T_{2*}$ and $T_{2\mathrm{e}}$) are obtained as the averages of the long-term stability measurement data collected over 400 hours, respectively, shown in Fig.~\ref{fig:SI_ADEV}, while the error bars are calculated as the standard deviations.
The best relaxation and dephasing times, shown in Figs.~\ref{fig:1}(e) and (f), are observed in the first cooling down for the multi-qubit device, while the long-stability measurement is conducted in the following cooling down, where the relaxation and dephasing times are slightly degraded possibly due to additional oxidation of the Nb and Si surfaces.
The anharmonicities ($\alpha$) are characterized by the $E$-$F$ control in the time domain ($Q_0$ and $Q_1$) and the two-photon transition in the qubit excitation spectra ($Q_2$ and $Q_3$).

The frequencies, external-coupling rates, and internal-loss rates of the readout resonators are characterized from the reflection spectra with the qubits in the $G$ states in continuous-wave (CW) measurement.
Since the qubits are well cooled down to their ground states in our experimental setup, the effective intrinsic losses due to the qubit excitations are negligible~\cite{kono2020breaking}.
The dispersive shifts for $Q_0$ and $Q_1$ are determined by the resonance frequency difference of the qubit-state-dependent reflection spectra of the resonators in the time-domain protocol [see Fig.~\ref{fig:SI_RO}(a)], while those for $Q_2$ and $Q_3$ are obtained from the photon-number resolved qubit excitation spectra.

\begin{table*}[bt]
	\caption{\textbf{System parameters.} The dispersive shifts, denoted by $\chi_{GE}$ ($\chi_{GF}$), are the frequency difference between the resonators with the qubits in the $G$ and $E$ ($F$) states.}
	\vspace{8pt}
	\begin{tabular}{l c c c c} \hline \hline
		\multicolumn{1}{c}{\textbf{Parameter}} & \qquad \qquad $Q_0$ \qquad \qquad & \qquad \qquad $Q_1$ \qquad \qquad & \qquad \qquad $Q_2$ \qquad \qquad & \qquad \qquad $Q_3$ \qquad \qquad \\ \hline
		Qubit frequency, $\omega_\mathrm{q}/2\pi$ (GHz) & 4.794064 $\pm$ 8e-6 & 5.20603 $\pm$ 20e-6 & 5.721 $\pm$ 100e-6 & 6.23127 $\pm$ 30e-6 \\
		Anharmonicity, $\alpha/2\pi$ (GHz) & 0.272 & 0.266 & 0.263 & 0.250 \\
		Relaxation time, $T_1$ (ms) & 0.21 $\pm$ 0.06 & 0.18 $\pm$ 0.05 & 0.04 $\pm$ 0.005 & 0.08 $\pm$ 0.02\\
		Ramsay dephasing time, $T_{2*}$ (ms) & 0.1 $\pm$ 0.05 & 0.06 $\pm$ 0.03 & 0.02 $\pm$ 0.01 & 0.06 $\pm$ 0.03\\
		Echo depfasing time, $T_{2\mathrm{e}}$ (ms) & 0.29 $\pm$ 0.09 & 0.22 $\pm$ 0.07 & 0.08 $\pm$ 0.01 & 0.08 $\pm$ 0.03\\
		Simulated purcell limit (ms) & 127 & 35 & 26 & 0.2 \\ \hline
		Resonator frequency, $\omega_\mathrm{r}/2\pi$ (GHz) & 7.07605 & 6.97984 & 6.885998 & 6.797376 \\
		External coupling rate, $\kappa_\mathrm{ex}/2\pi$ (MHz) & 1.85 & 1.06 & 0.102 & 0.52 \\
		Internal loss rate, $\kappa_\mathrm{in}/2\pi$ (MHz) & 0.11 & $<$ 0.01 & 0.02  & 0.01 \\
		Dispersive shift for $E$, $\chi_{GE}/2\pi$ (MHz) & 0.65 & 0.95 & 1.7 & 6.6 \\
		Dispersive shift for $F$, $\chi_{GF}/2\pi$ (MHz) & 1.04 & 1.55 & - & - \\
		\hline \hline
	\end{tabular}
	\label{Table:S1}
\end{table*}

\begin{figure}[t]
\includegraphics[width=8.5cm]{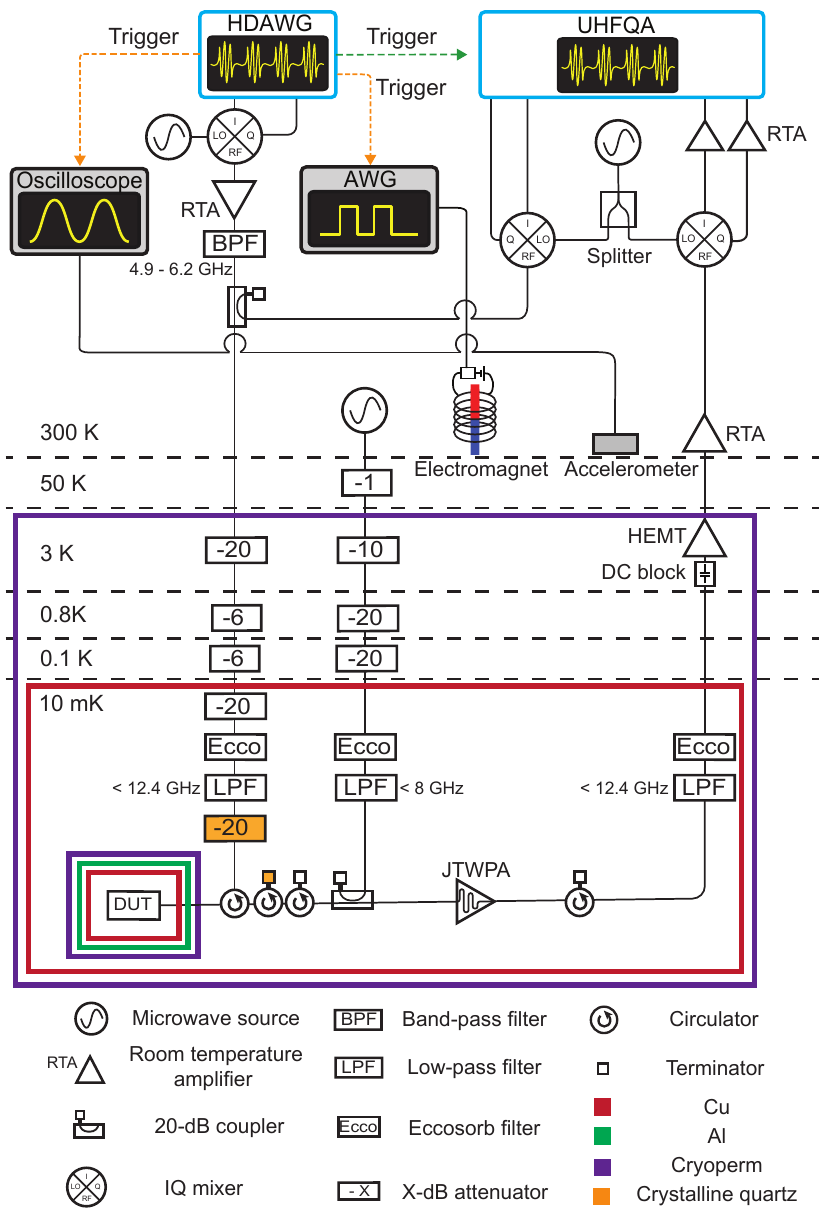}
\vspace{-5pt}
\caption{
\textbf{Cryogenic wiring and room temperature measurement setup.}
}
\label{fig:SI_Setup}
\vspace{-10pt}
\end{figure}

\vspace{-10pt}
\section{Cryogenic wiring and room temperature measurement setup}
\begin{figure}[t]
\includegraphics[width=8.5cm]{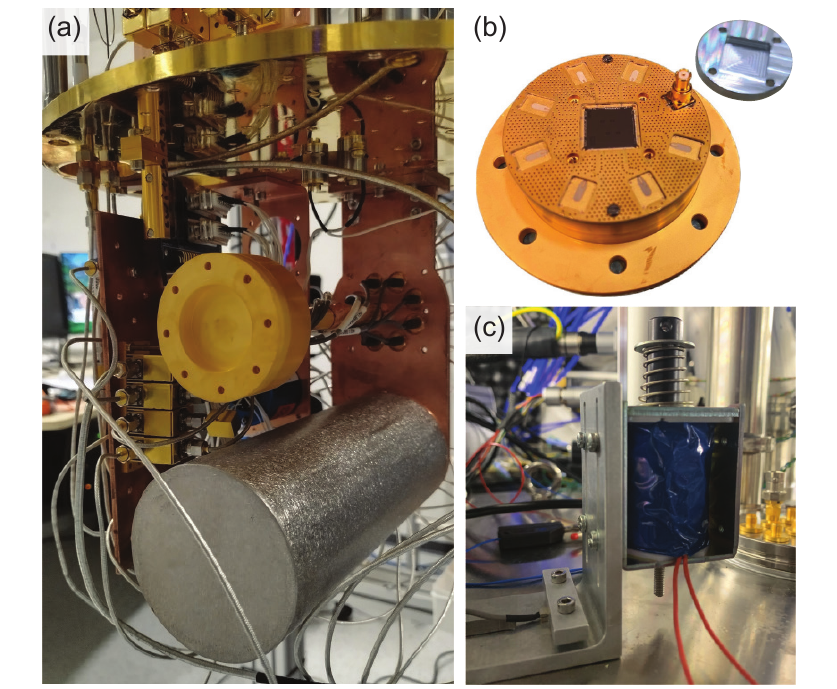}
\vspace{-5pt}
\caption{
\textbf{Sample holder and package, electric hammer, and accelerometer.}
(a)~Two slots for sample holders. The DUT is mounted inside the multilayer shielding at the bottom slot.
(b)~Multi-qubit device wire-bonded to a PCB on a sample table. 
An Al lid is covered to suppress qubit radiation loss.
(c)~Electric hammer based on an electromagnet.
An accelerometer is attached to the jig for the hammer.
}
\label{fig:SI_SetupPhoto}
\vspace{-10pt}
\end{figure}

\vspace{-10pt}
The experimental setup is shown in Fig.~\ref{fig:SI_Setup}.
The chip is fixed on a sample table made of gold-plated Cu, wire-bonded to a PCB based on coplanar waveguides sandwiched by double ground planes, and covered by an Al lid~[see Fig.~\ref{fig:SI_SetupPhoto}(b)].
There is an air gap designed under the chip on the sample table to increase the frequency of spurious chip modes well above our working frequencies~\cite{lienhard2019microwave}.
We thermally and mechanically anchor the device under test (DUT) to the mixing chamber stage ($\approx 10$~mK) of a dilution refrigerator (BlueFors BF-LD250). 
We isolate the DUT from environment fluctuations with multiple shielding: Al and cryoperm shields are used for reducing magnetic noise, while Cu shields are used for thermalizing the qubit radiation field to the base temperature~[see Fig.~\ref{fig:SI_SetupPhoto}(a)].
The input line to the DUT is equipped with a series of cryogenic attenuators ($-72$~dB in total) to suppress thermal noise from the higher temperature stages, while the output line is equipped with several isolators to prevent back heating from amplifiers. 
The readout signals are amplified by a Josephson traveling wave parametric amplifier (JTWPA) and a high electron mobility transistor (HEMT) amplifier in the dilution refrigerator, allowing us to realize a nearly quantum-limited microwave measurement.
The continuous pump signal to operate the JTWPA is added to the readout chain after the DUT via a directional coupler. 
We optimize the pump power and frequency to maximize the signal-to-noise ratio around the readout frequencies, resulting in about a 20~dB gain.
In addition, all the input, output, and pump lines are equipped with low-pass filters~(LPFs) and eccosorb filters~(Eccos) to reduce the contamination of high-frequency photons. 

For multiplexed control and readout of the transmon qubits, we employ an arbitrary waveform generator (Zurich Instruments HDAWG) and a quantum analyzer (Zurich Instruments UHFQA) to generate, acquire, and analyze intermediate frequency (IF) pulse sequence. 
We up- and down-convert frequency-multiplexed IF signals using IQ mixers operated with continuous waves generated from a multi-channel microwave source (AnaPico APMS12G). 
The up-converted control signal is amplified, filtered to cut the output amplifier noise around the readout frequencies, and combined with the readout signal via a directional coupler. 
The frequency-multiplexed readout microwave signals are down-converted, amplified by room temperature amplifiers (RTA), digitized, and, demodulated by the UHFQA, leading to the $I$ and $Q$ quadratures for each readout frequency. 
The operation of the UHFQA, generating and digitizing the readout signals, is synchronized with the HDAWG via a trigger signal.

For monitoring the vibrations of the top plate of the dilution refrigerator, we use an oscilloscope (Keysight 1000X), operated synchronously with the HDAWG via a trigger signal. 
The acceleration of the top plate is continuously converted to a voltage signal by an accelerometer (KEMET VS-BV203-B) mounted on it [see Fig.~\ref{fig:SI_SetupPhoto}(c)].
The accelerometer can be activated with a 5-V DC voltage bias (not shown in Fig. \ref{fig:SI_Setup}).
Upon a trigger signal from the HDAWG, the oscilloscope starts to record the converted voltage signal.
The recorded voltage signals are expressed in the unit of acceleration by using the sensitivity of 20~$\mathrm{mV/m/s}^2$. 

For artificially generating a pure mechanical shock on the top plate of the refrigerator, we use an electric hammer based on a circuit consisting of an electromagnet, a 15-V DC voltage bias, and an electrical switch [see Fig. \ref{fig:SI_SetupPhoto}(c)].
When the electrical switch is on, a current flows in the electromagnet, accelerating the magnet core and resulting in a mechanical shock on the top plate.
When the electrical switch is off, the magnet core is detached from the top plate by the elastic force of a spring. 
We use an AWG (Tektronix AFG3252C), operated synchronously with the HDAWG, to control the electrical switch by a pulsed voltage signal.
In the experiment for Fig.~\ref{fig:5}, the electric hammer is activated periodically with 50-ms pulsed signals with a period of 1.5~seconds.

\vspace{-10pt}
\section{Fabrication details}
\vspace{-10pt}
In this section, we provide the detailed fabrication process flow for ultra-coherent superconducting qubit devices, including $\mathrm{Al}/\mathrm{Al}\mathrm{O}_x/\mathrm{Al}$ junctions and Nb electrodes.

We use (100)-oriented intrinsic float zone double-side polished silicon wafers with a high resistivity ($>10\rm{\;k\Omega}$-cm), supplied from Siegert. 
The thickness of the wafers is 525~$\mu$m, while the diameter is 4 inches.

\textbf{Substrate preparation:} 
Prior to the Nb deposition, the wafers are first cleaned by Piranha solution, which is dedicated to removing the organic contamination on the substrate. 
This is done by dipping the wafer into two identical and successive baths of sulfuric acid ($\rm{H_2SO_4}$ 96~\% at 100°C) activated by hydrogen peroxide ($\rm{H_2O_2}$ 30~\%) for 5 minutes each, followed by dipping in two DI water baths for 4 minutes each and spin-drying. 
Afterward, the wafers are dipped in buffered HF solution ($\rm{NH_4F}$ 40~\% and $\rm{HF}$ 50~\% with a volume dilution ratio of 7:1) for 5 minutes to remove the silicon oxide on the surface, followed by DI water washing and spin-drying. 
The cleaned wafers are then quickly transferred to the load-lock chamber of the sputtering tool (in less than 3 minutes) to avoid re-oxidation.

\textbf{Deposition of niobium:} 
Immediately after native oxide removal, we deposit 150-nm Nb thin film using DC sputtering (Pfeiffer SPIDER Sputtering system). 
The argon flow rate is optimized such that the thin film has around $-100$~MPa stress~\cite{kreikebaum2020superconducting}, which is measured by comparing the wafer bow before and after the sputtering.
The film stress is $-130$~MPa stress for the multi-qubit device used in the main experiment.

\textbf{Niobium patterning and etching:} 
To pattern large structures including transmon pads, resonators, and coplanar waveguides, the wafer undergoes dehydration, followed by coating with 1.5-$\mu$m AZ ECI 3007 photo-resist and baking at $100^\circ$C for 2 minutes (automatic coater, ACS200 Gen3).
The resist is patterned using a direct laser writer (MLA 150) and developed with AZ 726 MIF (an organic solution based on TMAH) after post-exposure baking at 110$^{\circ}$C for 1 minute (ACS200 Gen3).
Prior to the Nb etching, we perform 10-s oxygen plasma at 200~W power (Tepla GiGAbatch) to remove the resist residues.

The Nb film is dry-etched with $\rm{SF_6}$ plasma (SPTS APS). The pressure of the chamber is 2~mTorr, the gas flow is 40~SCCM, and the RF source power is 250~W. 
The etching time is determined every time by monitoring the intensity of the 440-nm spectral line which corresponds to SiF (the by-product of Si etching), normally resulting in about 70 seconds. 
The etching is manually interrupted 5 seconds after a sharp increase in the intensity, i.e., the signature of the end of the Nb etching.
The additional etching time of 5 seconds ensures that the Nb films are fully etched, leading to an approximately 150-nm over-etching of the silicon substrate [see Fig.~\ref{fig:SI_cross}(b)].
Before starting the Nb etching for the actual wafer, we clean the chamber with oxygen plasma for 5 minutes. Then, we run an identical Nb etching process with a dummy silicon wafer for 2 minutes.

The photo-resist is removed with the following procedures.
First, the surface of the resist denatured by the Nb etching is removed by a low-power oxygen plasma (200 W) for 2 minutes (Tepla GiGAbatch). The time is well calibrated so that not all the resist is removed with the oxygen plasma.
The wafer is then dipped in a clean 1165 remover for 5 minutes at $60^{\circ}$C with high-power sonication to remove most of the resist.
Successively, the wafer is dipped in another clean 1165 remover solution and kept overnight.
The beaker containing the wafer and remover is heated up to $60^{\circ}$C with strong sonication for 5 minutes.
The wafer is cleaned in acetone and IPA for 3 minutes each at $60^{\circ}$C with high-power sonication sequentially, followed by drying with a nitrogen gun.

\textbf{Electron-beam lithography:} 
Prior to the e-beam resist coating, we use low-power oxygen plasma  (200 W) for 1 minute to remove any possible resist residues.
The wafer is then dipped into HF acid (1\% diluted) for 5 minutes to remove the oxide layers on the Si and Nb surfaces, followed by dipping in two DI water baths and spin-drying.

Afterward, we immediately coat the wafer with bilayer e-beam resist using the following procedures.
First, a 500-nm MMA EL9 resist layer is coated, followed by baking at $180^\circ$C for 5 minutes. 
Then, a 500-nm PMMA 495 A8 resist layer is coated, followed by baking at $180^\circ$C for 5 minutes. 

The e-beam lithography is done with a beam diameter of approximately 4 nm, enabled with a 100-kV acceleration voltage and a 200-pA current (Raith EBPG5000+). 
During the exposure, two doses are used for making the resist structures for the Manhattan process with in-situ bandage pads~\cite{osman2021simplified}. 
First, a high dose (1600~$\mu\rm{C}/\rm{cm}^2$) is used to expose PMMA for defining the junction structure, while the proximity effect also exposes MMA, resulting in undercuts around the exposed area.
Second, a low dose (350~$\mu\rm{C}/\rm{cm}^2$) is used to expose MMA at the tips of all the line structures for making undercuts to refrain Al films from being deposited on the MMA side walls.
After the exposure, the resist is developed using a solution of MIBK:IPA (1:3) for 2 minutes, followed by dipping in IPA for 1 minute and drying with a nitrogen gun.

\textbf{Shadow evaporation:} 
We use Plassys MEB550SL3, which is an ultra-high vacuum 3-chamber system dedicated to shadow evaporation for fabricating Josephson junctions. This system has a separate load-lock chamber equipped with argon ion milling and UV-lamp to generate ozone for ashing, while it has an evaporation chamber and an oxidation chamber, separately.
It can transfer a wafer between the three chambers without breaking the vacuum. 

The wafer with the bilayer resist coating is loaded in the load-lock chamber and is pumped it for 10 hours (pressure $<10^{-7}$~Torr). 
The recipe begins with generating ozone for 1 minute to remove the resist residues on the Si surface.
To ensure no resist residues, the ashing time is calibrated such that e-beam resists are etched by approximately 5 nm.

The wafer is then transferred to the evaporation chamber for fabricating Josephson junctions. 
Prior to every Al evaporation, we evaporate titanium at a rate of 0.2 nm/min for 2 minutes (with the closed shutter) and wait for 4 minutes such that the chamber pressure becomes below $2\times10^{-9}$~Torr. 
In our e-beam lithographic design, we can deposit aluminum lines selectively by changing the in-plane rotation angle according to the angle of the line patterns to be used.
To deposit the bottom layer of the junction, the wafer is tilted by $\theta =  45^\circ$ without an in-plane rotation ($\phi=0^\circ$), and Al is evaporated by 40 nm at a deposition rate of 0.5~nm/sec. 

\begin{figure}[t]
\includegraphics[scale=1]{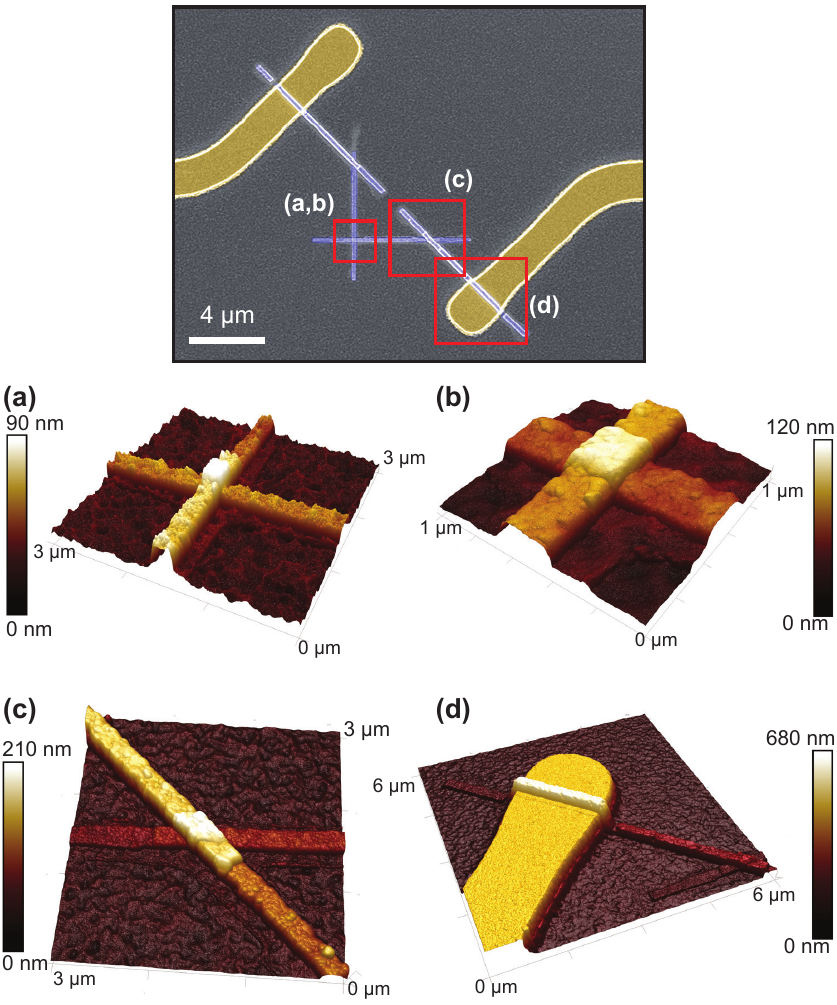}
\vspace{-5pt}
\caption{
\textbf{Surface investigation on the device.} Atomic force microscopy (AFM) characterizes the surfaces of different parts of the device. (a),(b)~AFM images of the Josephson junction. (c),(d)~AFM images of the Al-Al and Al-Nb connections, respectively.
}
\label{fig:SI_AFM}
\vspace{-10pt}
\end{figure}

The wafer is then transferred to the oxidation chamber to make the oxide layer. The oxidation time and pressure are calibrated as 10 minutes and 0.07~Torr, respectively, to obtain the target junction resistance of $\approx 7$~k$\Omega$. 

Afterward, the wafer is returned to the evaporation chamber and tilted by $\theta = 45^\circ$, where Al is evaporated by 30 nm two times at $\phi=\pm 90^\circ$, respectively.
The two evaporations are useful to cover all the faces of the oxidized bottom layer to stabilize the junction resistance.

Before connecting the junction to the Nb pads, we have to remove the niobium oxide from the connection parts by using argon ion milling in the load-lock chamber. 
To this end, the wafer is transferred to the load-lock chamber and tilted by $45^\circ$,  where argon milling is performed for 4 minutes each at two in-plane rotation angles of $\phi = -45^\circ$ and $135^\circ$, respectively. 
The time and power of the milling are calibrated to assure removing all the niobium oxide layers and connecting the Al leads to the Nb pads without a resistance layer in between. 

Next, the wafer is transferred to the evaporation chamber and tilted by $45^\circ$, where 30/100/100-nm Al evaporations are performed at $\phi =135^\circ/-45^\circ/135^\circ$, respectively. 
The first 30-nm Al evaporation is required to avoid the disconnectivity in Al leads due to the shadowing effect, while two 100-nm Al evaporation is used for connecting the Al leads to the Nb pads above the over-etched Si substrate. 

The final step is to transfer the wafer to the load-lock chamber and oxidize it with high-purity oxygen at 15~Torr for 10~minutes for making a clean oxide layer as a protection barrier before taking out the wafer from the chamber. 

\textbf{Lift-off:}
After the evaporation, the wafer is dipped and kept in remover 1165 for 6 hours at room temperature.
High power sonication at $60^\circ$C is then used for 45 minutes to further assist the lift-off process.
Afterward, the wafer is dipped in acetone and IPA at $80 ^\circ$C with high power sonication for 10 minutes each sequentially, followed by drying using a nitrogen gun.

\textbf{Dicing:} 
To protect the wafer from possible contamination in dicing, the wafer is coated with 1.5~$\mu m$ photo-resist (AZ ECI 3007), where a monolayer of HMDS vapor is deposited before the photo-resist coating, which might remain on the Nb and Si surfaces after the final cleaning, that would prevent re-oxidation of the Nb and Si surface~\cite{nersisyan2019manufacturing}.
It is then diced into chips (Disco DAD321) with specific care about the electrostatic discharge potential damages during the dicing.

\textbf{Final cleaning:}
After dicing, chips are dipped in remover 1165 for 30 minutes at $50^{\circ} $C without sonication. 
Then, chips are dipped in acetone and IPA solutions for 10 minutes each at $50^{\circ} $C without sonication, sequentially.
Finally, chips are dried with a nitrogen gun.

\textbf{Packaging:}
We fix a chip on a sample table using diluted GE varnish with acetone.
The chip is then wire-bonded to a PCB with 20-$\mu$m diameter Al wires (F\&S Bondtec 56i) and covered with an Al lid.

\begin{figure}[t]
\includegraphics[scale=1]{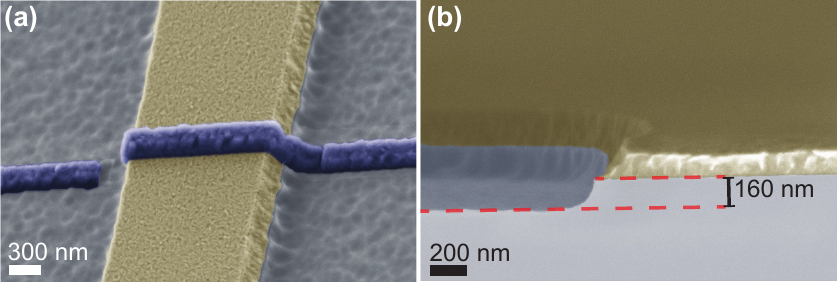}
\vspace{-5pt}
\caption{
\textbf{Cross section of the etched Nb film on the Si substrate.} 
(a)~Tilted SEM image of the Nb-Al connection. 
(b)~Cross-sectional SEM image of the Nb-Si interface. Plasma etching of the Nb layer is controlled with time and results in over-etching of the silicon substrate by $\approx 160$~nm.
}
\label{fig:SI_cross}
\vspace{-10pt}
\end{figure}

\vspace{-10pt}
\section{Sample investigation}
\vspace{-10pt}
The scanning electron microscope (SEM) images of the Josephson junction in the final devices are shown in the main text. 
We further investigate the surface properties and topography of the junction, as well as those for the Al-Al and Al-Nb connections using atomic force microscopy (AFM) (see Fig.~\ref{fig:SI_AFM}). 
We measure an average roughness for the silicon surface of $R_\mathrm{a}^\text{(Si)} = 2.6$~nm, which we attribute to the DRIE over-etching of the silicon in the Nb etching process. 
The roughness of the Nb layer is measured as $R_\mathrm{a}^\text{(Nb)} = 1.1$~nm (Fig.~\ref{fig:SI_AFM}(d)). 
The step size from the Si surface to the Nb top surface is found to be $\sim325$~nm, which includes the Nb layer thickness and the Si over-etched thickness.
Figure~\ref{fig:SI_cross} shows a tilted SEM image of the Nb-Al connection as well as a cross-sectional SEM of the Nb-Si interface. 
Note that the thickness of the Nb layer in the cross-sectional SEM becomes thinner by the cleaving process than the actual one.
Nevertheless, we estimate the depth of the over-etched silicon to be $\approx 160$~nm from the cross-sectional SEM. Subtracting this depth from the total step size measured with the AFM, we estimate the actual thickness of the Nb layer to be $\approx 165$~nm. 
Furthermore, the cross-sectional SEM shows that the Nb etching process does not result in silicon undercuts, but realizes a tapered profile, which is useful for making the Nb-Al connection easier and improving the relaxation times~\cite{nersisyan2019manufacturing}.

In Figs.~\ref{fig:1}(b) and \ref{fig:SI_AFM}(b), there are unknown thin layers visible on the silicon surface around the Al structures. 
Since the thin layers remain within the undercut of the e-beam resist, this may be due to the re-deposition of the e-beam resist during the argon milling in the shadow evaporation step.

\begin{figure*}[t]
\includegraphics[width=17cm]{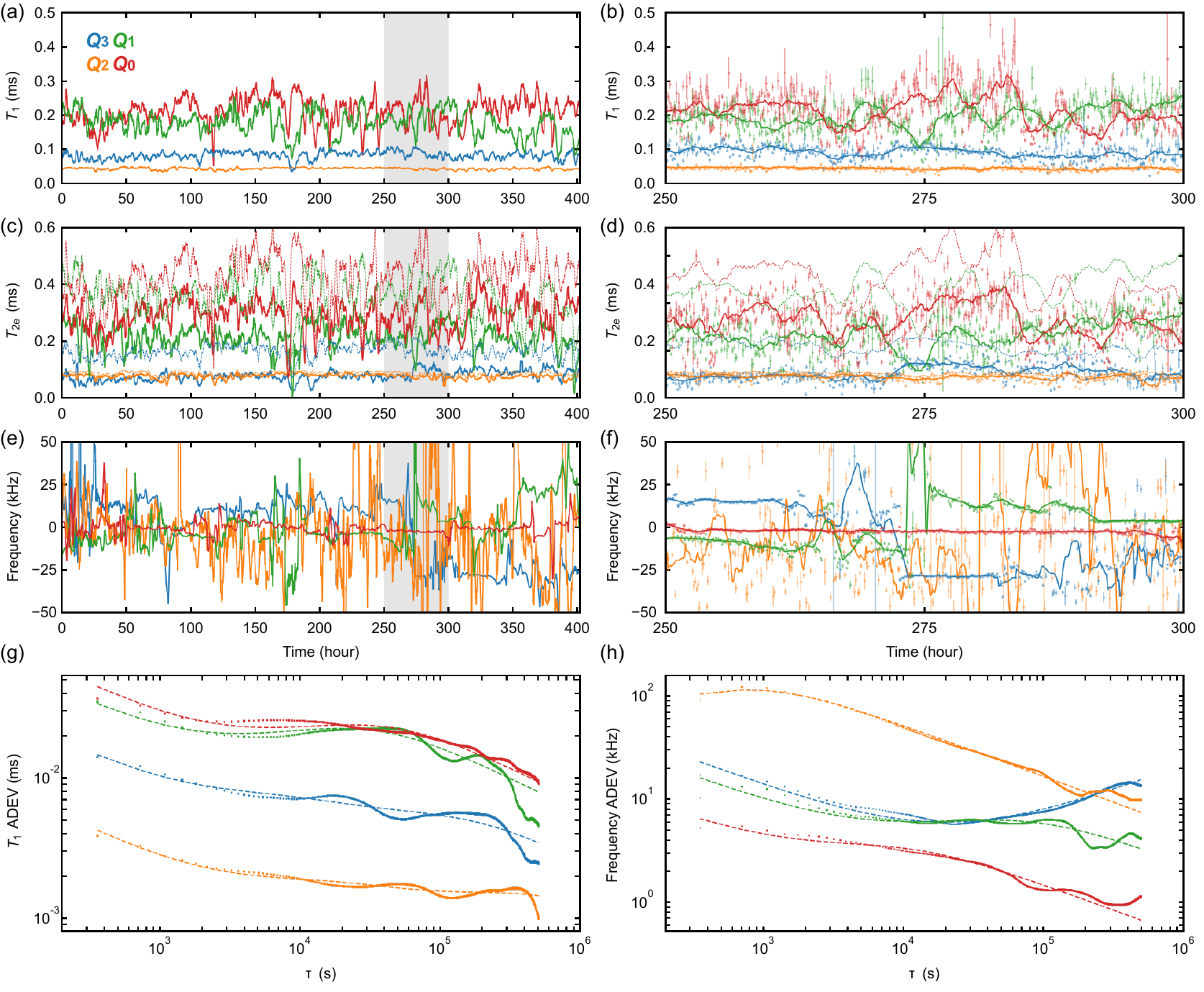}
\vspace{-5pt}
\caption{
\textbf{Long-term stability of qubit parameters and Allan deviation analysis.}
(a)--(f)~Relaxation time ($T_1$), Hahn-echo dephasing time ($T_{2\mathrm{e}}$), and qubit frequency of four qubits $Q_0$--$Q_3$ as a function of time.
(b),(d) and (f) are the magnified plots for the gray regions in the left panels, respectively.
The dots in the magnified plots are the results obtained from the individual time traces, while the solid lines in both plots are their smoothed data with a 5-hour time window.
The dashed lines for $T_{2\mathrm{e}}$ shows the $T_1$ limit, i.e., $2T_1$.
(g),(h)~ Allan deviation of relaxation time and qubit frequency, respectively. The dots are the data, while the dashed lines are the fitting results based on the TLS model.
}
\label{fig:SI_ADEV}
\vspace{-10pt}
\end{figure*}

\vspace{-10pt}
\section{Extended data}
\vspace{-10pt}
\subsection{Allan deviation (ADEV) analysis}
\vspace{-10pt}
Allan deviation analysis is an useful technique to identify the source of random fluctuations. 
In this section, we will first show that two kind of random process mediated by TLS will have identical signature on ADEV. 
Next, the ADEV of the long-term qubit parameter fluctuations data are calculated and are fitted by the analytical model.

The first noise process we study is called random telegraph noise, which means a random variable $x$ jumps between two possible values (0 and 1 for simplicity). 
The dynamics is governed by a set of rate equation,
\begin{equation}
	\begin{aligned}
		& P_1(t+d t)=P_0(t) \gamma_0 d t+P_1(t)\left(1-\gamma_1d t\right), \\
		& P_0(t+d t)=P_1(t) \gamma_1 d t+P_0(t)\left(1-\gamma_0d t\right),
	\end{aligned}
\end{equation}
where $\gamma_0$ and $\gamma_1$ are the transition rates, while $P_0(t)$ and $P_1(t)$ are the probability for random binary variable $x$ to be 0 and 1 in time $t$, respectively. 
The process could be understood as a Markovian process between the two values, where the average dwell time in value 0 is $\frac{\gamma_1}{\gamma_0+\gamma_1}$ of the total time, and $\frac{\gamma_0}{\gamma_0+\gamma_1}$ for value 1. 
In the interacting defect model, the state transitions of a low-frequency TLS with excitation rate $\gamma_0$ and relaxation rate $\gamma_1$ could change the frequency of a high-frequency TLS coupled to a qubit, resulting in the fluctuation in the qubit relaxation time by random telegraph noise~\cite{muller2019towards}.

The second noise process we will consider is a Poisson reset process triggered by events with rate $\gamma$. 
After each event, a random continuous variable $x$ is drawn from a probability distribution, $P(x)$.
The random fluctuations caused by gamma and cosmic ray~\cite{mcewen2022resolving,thorbeck_tls_2022,wilen2021correlated} could be modeled with this process, where the absorption of ionizing radiation may cause charge rearrangement in a substrate, resulting in the random reset of the frequency of a TLS that induces a qubit decay.

To begin with, we will calculate the auto-correlation for these two process. 
For the random telegraph noise, where $x = 0$ or $1$, the only non-zero contribution in $c_{xx}(\tau)$ appears when $x(t)=x(t+\tau)=1$.
Therefore, the auto-correlation is written as
\begin{equation}
	\begin{aligned}
		c_{xx}(\tau) & =  \langle x(t)x(t+\tau) \rangle                                   \\
		& = P\left(x(t)=1\right)P\left(x(t+\tau)=1|x(t)=1\right)                                   \\
		& = \frac{\gamma_0}{\gamma_1+\gamma_0}P_{11}(\tau), \label{selfcorr}
	\end{aligned}
\end{equation}
where $P\left(x(t)=1\right)=\gamma_0/(\gamma_1+\gamma_0)$ is valid due to the detailed balance condition.
For simplicity, we denote $P_{11}(\tau) = P\left(x(t+\tau)=1|x(t)=1\right)$ and $P_{01}(\tau) = P\left(x(t+\tau)=0|x(t)=1\right)$. 
Furthermore, they should satisfy a set of equations:
\begin{equation}
	\begin{aligned}
		& P_{11}(\tau) +P_{01}(\tau) = 1,                                              \\
		P_{11}(\tau+d\tau) = & P_{11}(\tau) (1-\gamma_1d\tau) + P_{01}(\tau)\gamma_0 d\tau, 
	\label{diffeqn}
	\end{aligned}
\end{equation}
where we take the limit of $d\tau\rightarrow0$ so the chance of having more than 1 switching during $d\tau$ vanishes. Eq.~(\ref{diffeqn}) is then transformed in to a differential equation of $P_{11}(\tau)$ as
\begin{equation}
\frac{dP_{11}(\tau)}{d\tau}=-P_{11}(\tau)(\gamma_1+\gamma_0)+\gamma_0,
\end{equation}
which could readily be solved with the initial condition $P_{11}(0)=1$ as
\begin{equation}
P_{11}(\tau)=\frac{\gamma_0}{\gamma_0+\gamma_1}+\frac{\gamma_1}{\gamma_0+\gamma_1}e^{-(\gamma_0+\gamma_1)|\tau|}.
\end{equation}
Substituting the solution into Eq.~(\ref{selfcorr}), we obtain $c_{xx}(\tau)$ as
\begin{equation}
	c_{xx}(\tau) = \frac{\gamma_0\gamma_1}{(\gamma_0+\gamma_1)^2}e^{-(\gamma_0+\gamma_1)|\tau|},
\end{equation}
where we omit the constant part.

For the Poisson reset process, $x(t)$ and $x(t+\tau)$ would be the same value if there are no switching event within time $\tau$, and would be drawn from 2 i.i.d distributions if there are at least 1 switching event within $\tau$.
Then, the auto-correlation is obtained as
\begin{equation}
	\begin{aligned}
		c_{xx}(\tau) & = \langle x(t)x(t+\tau) \rangle \\
		& = \langle x_0x_1 \rangle P(\text{switched within $\tau$}) \\
		&\qquad + \langle x^2_0 \rangle P(\text{not switched within $\tau$}) \\
		& = \langle x_0x_1 \rangle (1-e^{-\gamma |\tau|}) + \langle x^2_0 \rangle e^{-\gamma |\tau|},
	\end{aligned}
\end{equation}
where $x_0$ and $x_1$ are random variables drawn from 2 i.i.d distributions.
For simplicity, we consider $P(x)$ to be a normal distribution of zero mean and $\sigma^2$ variance.
Hence, the auto-correlation of $x$ for Poisson reset process is calculated as 
\begin{equation}
c_{xx}(\tau)=\sigma^2e^{-\gamma|\tau|}.
\end{equation}

To summarize, the two noise processes we discussed have the same form in the auto-correlation function, i.e., a double-exponential function. 
By applying the Wiener–Khinchin theorem, we obtain a Lorentzian-type power spectral density~(PSD) of random variable $x$ as $S_{xx}(\omega) = 2A\frac{\alpha}{\alpha^2+\omega^2}$ for both the noise process, with $\{A=\frac{\gamma_0\gamma_1}{(\gamma_0+\gamma_1)^2},\alpha=\gamma_0+\gamma_1\}$ for the random telegraph noise and $\{A=\sigma^2,\alpha=\gamma\}$ for the Poisson reset process.

To continue, we calculate the ADEV by following the reference~\cite{VANVLIET1982261}. From the definition of ADEV, we write
\begin{equation}
	\begin{aligned}
		\sigma^2_x(\tau) & = \langle[\bar x(t) - \bar x(t+\tau)]^2\rangle/2                          \\
		& =\frac{1}{2\tau^2}\langle[q(t+\tau)-q(t)-q(t+2\tau)+q(t+\tau)]^2\rangle   \\
		& =\frac{1}{2\tau^2}[2c_{qq}(2\tau)-8c_{qq}(\tau)+6c_{qq}(0)], 
	\label{adev}
	\end{aligned}
\end{equation}
where we assume the process is stationary. Here $q(t)$ is the phase variable defined as $q(t)=\int_0^tx(\tau)d\tau$, in the sense that we treat $x(t)$ as the frequency variable in the conventional setup for ADEV, i.e., $\frac{dq(t)}{dt}=x(t)$ and $\bar x(t) =\int_t^{t+\tau}x(t)dt/\tau= [q(t+\tau)-q(t)]/\tau$.

The PSDs of $q(t)$ and $x(t)$ is related as $S_{qq}(\omega) = S_{xx}(\omega)/\omega^2 = 2A\frac{\alpha}{\omega^2(\alpha^2+\omega^2)}$. 
Then, we perform an inverse Fourier transform for $S_{qq}(\omega)$ to obtain the auto-correlation for $q$ as
\begin{equation}
c_{qq}(\tau) = -A\frac{e^{-\alpha|t|}+\alpha|t|}{\alpha^2}. 
\label{phaseautocorr}
\end{equation}
Substituting $c_{qq}(\tau)$ into Eq.~(\ref{adev}), we obtain the ADEV as
\begin{equation}
	\sigma_x(\tau) = \frac{\sqrt{A(4e^{-\alpha t}-e^{-2\alpha t}+2\alpha t-3)}}{\tau\alpha}.
	\label{eq:s_x}
\end{equation}


To characterize the long-term stability of the transmon parameters (frequency, $T_1$, $T_{2\mathrm{e}}$, etc.), we record the fluctuations of the qubit coherences and frequency over 400 hours, which are shown in Fig.~\ref{fig:SI_ADEV}(a)-(f). 
The one cycle consisting of the relaxation, Ramsay, and Hahn-echo sequences is repeated with an interval of about 6 minutes, where the four qubits are simultaneously controlled and read out by frequency multiplexing. 
Note that the qubit frequencies are obtained by the Ramsey measurements.

To confirm whether these fluctuations are mediated by TLSs with the noise processes we discussed in the last section, we fit the ADEV of qubit $T_1$ and frequency with the following model:
\begin{equation}
	\sigma_x(\tau) =  \sigma_{x_1}(\tau) + \sigma_{x_2}(\tau) + k \sqrt{\frac{1}{\tau}},
\end{equation}
where $\sigma_{x_1}(\tau)$ and $\sigma_{x_2}(\tau)$ are two ADEV with independent parameters ($A$ and $\alpha$) based on Eq.~(\ref{eq:s_x}), while the term $\sqrt{1/\tau}$ is a white noise contribution with a coefficient $k$. 
The fitting results are shown in Figs.~\ref{fig:SI_ADEV}(g) and (h) in dashed lines. The fitted lines are consistent with the ADEV data we observed, which implies that the long-term fluctuations are likely to be related to the TLS environment. 
However, as discussed in the last section, we can not distinguish whether the TLSs are fluctuated by a global Poisson process, e.g. gamma and cosmic rays, or they are fluctuated by the low-frequency TLSs they are coupled to.

\begin{figure}[t]
\includegraphics[width=8.5cm]{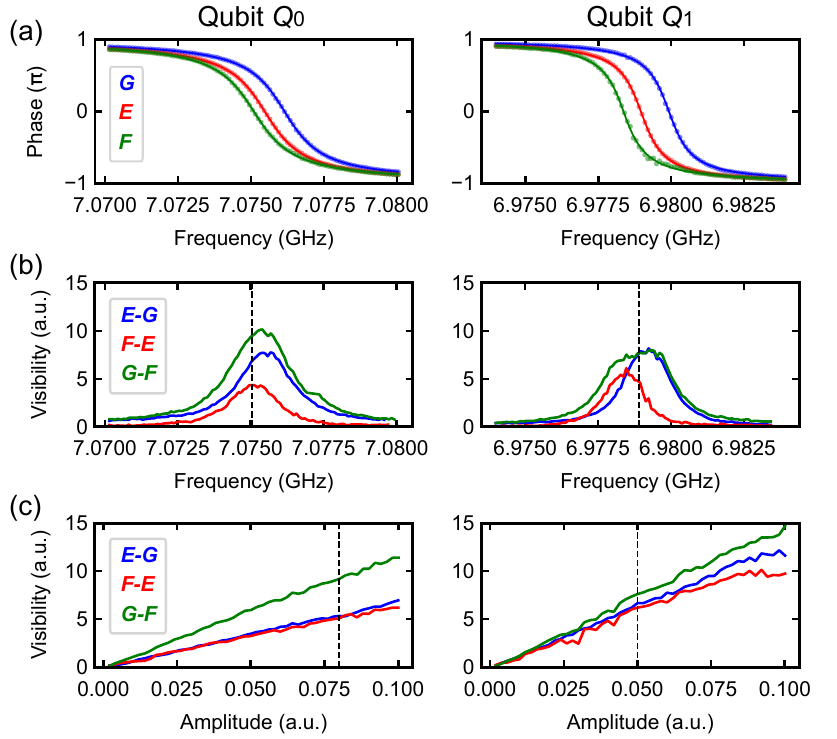}
\vspace{-5pt}
\caption{
\textbf{Readout optimization for qubits $Q_0$~(left) and $Q_1$~(right).}
(a)~Reflection phase spectra of readout resonators with each qubit prepared in the $G$, $E$, and $F$ states, respectively.
(b)~Visibilities between $E$ and $G$, $F$ and $E$, and $G$ and $F$ as a function of readout frequency.
(c)~Visibilities as a function of readout amplitude.
The vertical dashed lines in (b) and (c) depict the optimized values.
}
\label{fig:SI_RO}
\vspace{-10pt}
\end{figure}

\vspace{-10pt}
\subsection{Single-shot readout calibration and characterization}
\vspace{-10pt}
Here, we discuss how to calibrate and characterize the multiplexed single-shot readout of two of the transmon qubits that are used in the main experiment.

Figure~\ref{fig:SI_RO}(a) shows the qubit-state-dependent reflection phase spectra of the readout resonators for qubits $Q_0$ and $Q_1$, respectively.
After the preparation of the qubits in the $G$, $E$, and $F$ states, we send and measure the multiplexed readout pulse, sweeping the readout frequencies.
By fitting the phase of the reflection spectra, we obtain the resonator frequency with the qubit in each state, resulting in the qubit-state dependent frequency shifts.
The qubit-state dependent resonator frequency shifts for qubits $Q_0$ and $Q_1$ are summarized in Table~\ref{Table:S1}.

Figure~\ref{fig:SI_RO}(b) show the visibilities as a function of the readout frequency, where the visibility is defined as the distance of the complex amplitudes of the readout signals for the different qubit states.
Note that the readout amplitude is set to the one showing the linear resonator response.
To distinguish between all the three different states, the readout frequency is optimized such that the minimal one among the three different visibilities is maximized.
The optimized readout frequencies for $Q_0$ and $Q_1$ are shown with the dashed lines in Fig.~\ref{fig:SI_RO}(b), respectively.

To optimize the readout amplitude, we measure the readout signals for the three different states as a function of the readout amplitude with the optimized readout frequency.
Figure~\ref{fig:SI_RO}(c) show the visibilities as a function of the readout amplitudes for $Q_0$ and $Q_1$, respectively.
As the readout amplitude increases, the visibilities increase linearly due to the signal-to-noise ratio improvements.
However, when the readout amplitude further increases, the visibilities start to show saturations slightly.
This is due to the qubit state flip during the readout, caused by the off-resonant drive of the qubit~\cite{boissonneault2009dispersive}.
We choose a readout amplitude that does not show significant saturations but realizes sufficiently large separations between the different states in a single-shot measurement.
The optimized readout amplitudes are shown with the dashed lines in Fig.~\ref{fig:SI_RO}(c).

\begin{figure}[t]
\includegraphics[width=8.5cm]{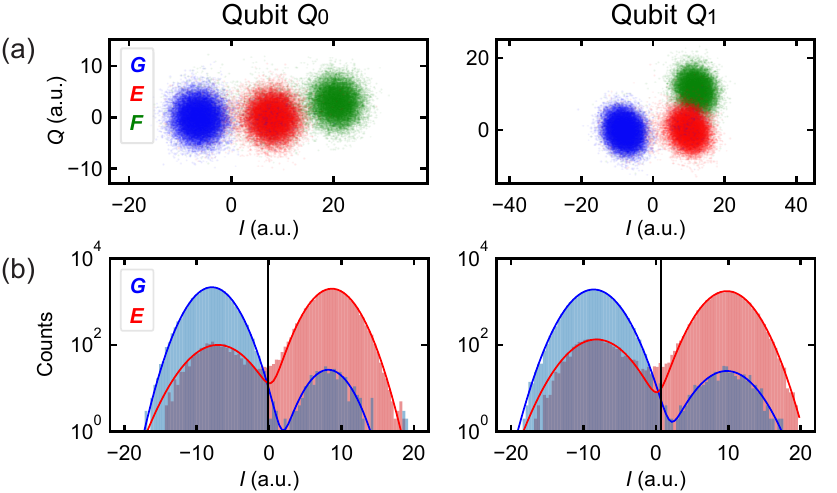}
\vspace{-5pt}
\caption{
\textbf{Readout characterization for qubits $Q_0$~(left) and $Q_1$~(right).}
(a)~Scatter plots of the single-shot readout signals in the complex plane when each qubit is prepared in the $G$, $E$, and $F$ states.
(b)~Histograms of readout quadrature $I$ for the $G$ and $E$ states. The solid lines are the fitting results to obtain the mean and standard deviation of each Gaussian peak.
}
\label{fig:SI_RC}
\vspace{-10pt}
\end{figure}

Figure~\ref{fig:SI_RC}(a) show the scatter plots of the complex amplitudes of the optimized single-shot readout pulses for the three different states of qubits $Q_0$ and $Q_1$, respectively.
We apply principal component analysis~(PCA) to the raw complex amplitudes in order to maximize the separation between the signals for the $G$ and $E$ states in the real axis ($I$).
Note that this process corresponds to an ordinary linear operation consisting of phase rotation and displacement in the complex plane.
For simplicity, we use only the real part ($I$) of the readout signals to distinguish between $G$ and $\bar{G}=$ $E$ or $F$ for the main analysis.
Notably, the real part ($I$) can distinguish well between the $G$, $E$, and $F$ states for $Q_0$, enabling us to characterize the occupation probability of the three states using the histogram of quadrature amplitude $I$~[see Fig.~\ref{fig:2}(f)].

Figure~\ref{fig:SI_RC}(b) shows the histograms of the single-shot readout quadratures ($I$) for qubits $Q_0$ and $Q_1$, respectively, prepared in the $G$ and $E$ states.
Due to the residual excitation probability and the infidelity of the $\pi$ control, there is a finite probability in the non-target states.
Nevertheless, each histogram can be fitted with the sum of weighted Gaussian distributions, extracting the mean and standard deviation of each Gaussian distribution. 

From the fitting results, we can estimate the separation readout errors. 
In our analysis, the threshold to distinguish between the two states is set to the center of the two Gaussian distributions, as shown with the black vertical lines in Fig.~\ref{fig:SI_RC}(b).
Thus, the distance between each peak and the threshold is calculated as $\bar{\mu} = |\mu_E-\mu_G|/2$, where $\mu_G$ and $\mu_E$ are the means of the distributions for the $G$ and $E$ states, respectively.
The separation error probability ($\varepsilon_\uparrow^\mathrm{s}$ and $\varepsilon_\downarrow^\mathrm{s}$), i.e., the probability that one state is detected as the other due to the inefficient signal-to-noise ratio, can be calculated as 
$\varepsilon^\mathrm{s}_\uparrow = \frac{1}{2}\mathrm{erfc}\left(\frac{\bar{\mu}}{\sqrt{2}\:\sigma_G}\right)$ and $\varepsilon^\mathrm{s}_\downarrow = \frac{1}{2}\mathrm{erfc}\left(\frac{\bar{\mu}}{\sqrt{2}\:\sigma_E}\right)$, where $\sigma_G$ and $\sigma_E$ are the standard deviations for the $G$ and $E$ states, respectively.
From the fitting results, we find that $\bar{\mu}/\sigma_G\approx \bar{\mu}/\sigma_E > 6.5$, which results in the separation readout error probabilities of $\varepsilon^\mathrm{s}_\uparrow \approx \varepsilon^\mathrm{s}_\downarrow < 0.001$ for both qubits $Q_0$ and $Q_1$.

\begin{figure}[t]
\includegraphics[width=8.5cm]{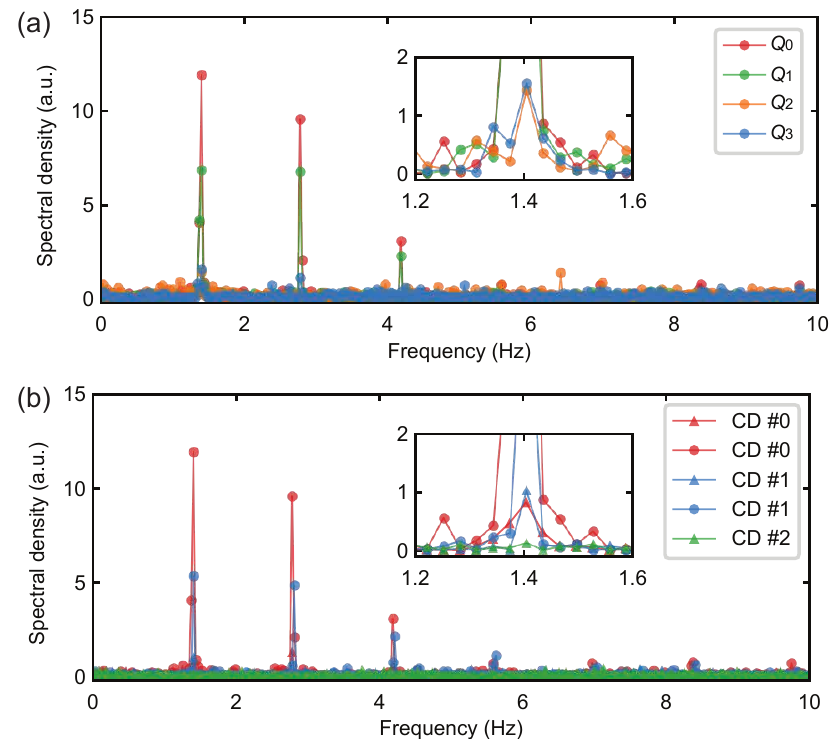}
\vspace{-5pt}
\caption{
\textbf{Mechanical sensitivities for different qubits and for different cooling downs.}
(a)~Amplitude spectral densities~(ASD) of the single-shot quadrature time traces for four qubits $Q_0$--$Q_3$.
(b)~ASDs of the single-shot quadrature time traces for $Q_0$, measured in different cooling downs (CDs \#0--2) with the same cryogenic wiring for the DUT.
The data in the same color, but with different symbols, show the ones measured in the same cooling down, but at different timing.
To be fair, the separations between the readout signals for the $G$ and $E$ states are always normalized to be 1 before obtaining the ASDs in this analysis.
}
\label{fig:SI_PSD}
\vspace{-10pt}
\end{figure}

To estimate the state-flip errors during the readout, we perform continuous monitoring of the qubit states, with a sequence in which 2.5~$\mu$s-long readout pulses are applied successively with an interval of 3~$\mu$s, as shown in Fig.~\ref{fig:4-1}(a).
To avoid the pulse tube effect on the state-flip error analysis, we determine the transition rates by fitting the dwell-time histograms obtained in a time-resolved fashion, as shown for the case of qubit $Q_0$ in Fig.~\ref{fig:4-2}(a).
Here, we use the transition rates at a time far from the periodic pulse tube shock (at a time of 0.6~seconds within the one period).
This results in the excitation rates of $\Gamma_\uparrow/2\pi = 3.23 \pm 0.08$~Hz and $2.4 \pm 0.06 $~Hz for $Q_0$ and $Q_1$, and the decay rates of $\Gamma_\downarrow/2\pi = 1.44 \pm 0.04$~kHz and $1.17 \pm 0.04$~kHz, respectively.

Using the transition rates during the readout ($\Gamma_\downarrow$ and $\Gamma_\uparrow$), we can estimate the upper bound on the state-flip readout error probabilities ($\varepsilon_\uparrow^\mathrm{f}$ and $\varepsilon_\downarrow^\mathrm{f}$).
For simplicity, a state-flip readout error can be considered a transition event occurring within the total readout time ($T_\mathrm{readout}=3$~$\mu$).
Therefore, the upper bounds for the $G$ and $E$ states can be estimated as
$
\varepsilon_\uparrow^\mathrm{f} < \Gamma_\uparrow T_\mathrm{readout}
$
and 
$
\varepsilon_\downarrow^\mathrm{f} < \Gamma_\downarrow T_\mathrm{readout}
$, respectively.
Using the transition rates during the readout, the upper bounds on the state-flip readout error probabilities are determined to be
$
\varepsilon_\uparrow^\mathrm{f} \lesssim 6$e-5
and 
$
\varepsilon_\downarrow^\mathrm{f} < 0.03
$ for both qubits $Q_0$ and $Q_1$.

On the other hand, we can characterize the transition rates in the free evolution using $\Gamma_\uparrow^\mathrm{free} = P_\mathrm{th}/T_1$ and $\Gamma_\downarrow^\mathrm{free} = (1-P_\mathrm{th})/T_1$, where $P_\mathrm{th}$ and $T_1$ is the residual thermal excitation probability and relaxation time of the qubit, respectively.
The experimentally obtained values for both $Q_0$ and $Q_1$ ($T_1 \approx 0.2$~ms and $P_\mathrm{th} \approx 0.015$) result in $\Gamma_\uparrow^\mathrm{free}/2\pi \approx 10$~Hz and $\Gamma_\downarrow^\mathrm{free} \approx 0.8$~kHz.
Note that these values are obtained in a non-time-resolved fashion, where they are affected by the pulse tube shocks on average. 
From this analysis, we find that the excitation rate during the readout is smaller than that in the free evolution. 
This is because the excitation rate during the readout is characterized in a time-resolved manner, minimizing the pulse tube effect.
In contrast, the decay rate during the readout is larger than that in the free evolution, which is due to a finite readout backaction, i.e., off-resonant-drive-induced decay~\cite{boissonneault2009dispersive}.

\vspace{-10pt}
\subsection{Mechanical sensitivities for different qubits and for different cooling downs}
\vspace{-10pt}
Here, we study the dependence of the mechanical sensitivity on different qubits and different cooling downs~(CDs). 
To see the mechanical sensitivity of the qubits, we show the amplitude spectral density~(ASD) of the time trance of single-shot qubit readout outcomes while the pulse tube cooler is activated, as shown in Fig.~\ref{fig:2}(e).
If the ASD shows larger harmonics with a fundamental frequency of about 1.4~Hz, the qubit is more sensitive to the mechanical shock generated by the pulse tube cooler.

Figure~\ref{fig:SI_PSD}(a) shows the ASDs of the single-shot readout time traces for the four different qubits.
It clearly shows that the longer-lived qubits ($Q_0$ and $Q_1$) are significantly more sensitive to the mechanical shock than the shorter-lived qubits ($Q_2$ and $Q_3$).
This implies that the relaxation times of the longer-lived qubits are limited by mechanical sensitive baths, while those of the shorter-lived qubits are limited by other loss mechanisms.

Figure~\ref{fig:SI_PSD}(b) shows the ASDs of the single-shot readout time trace of qubit $Q_0$ for three different cooling downs~(CDs) of the dilution refrigerator. 
Importantly, the cryogenic wiring for the four-qubit device is kept exactly the same, while other cryogenic wirings that are completely isolated from the main experiment are changed every cooling down.
Nevertheless, the mechanical sensitivity of the qubit is varying depending on cooling downs.
Interestingly, the sensitivity changes even within the same cooling down~(see the results in CDs \#0 and \#1).
This is consistent with the qubit parameter fluctuations due to the TLS baths (see Figs~\ref{fig:SI_ADEV}).
In particular, the ASD for CD \#2 shows no mechanical sensitivity, although we don't see any crucial changes in the cryogenic wiring.

One unique setup for our dilution refrigerator is a fiber base optical setup at the still stage, where we are testing electro-optical converter chips via optical fiber coupling with epoxy glue.
However, even though the optical setup is kept exactly the same between CDs \#1 and \#2, the mechanical sensitivity are completely different between these two cooling downs.
This may imply that the optical setup does not play an important role in the mechanical sensitivity of the qubit.
This is also consistent with that we observe a finite mechanical sensitivity of the qubits even during a cooling down in the absence of samples connected to the optical fibers.


\bibliography{Refs}

\begin{thebibliography}{61}%
\makeatletter
\providecommand \@ifxundefined [1]{%
 \@ifx{#1\undefined}
}%
\providecommand \@ifnum [1]{%
 \ifnum #1\expandafter \@firstoftwo
 \else \expandafter \@secondoftwo
 \fi
}%
\providecommand \@ifx [1]{%
 \ifx #1\expandafter \@firstoftwo
 \else \expandafter \@secondoftwo
 \fi
}%
\providecommand \natexlab [1]{#1}%
\providecommand \enquote  [1]{``#1''}%
\providecommand \bibnamefont  [1]{#1}%
\providecommand \bibfnamefont [1]{#1}%
\providecommand \citenamefont [1]{#1}%
\providecommand \href@noop [0]{\@secondoftwo}%
\providecommand \href [0]{\begingroup \@sanitize@url \@href}%
\providecommand \@href[1]{\@@startlink{#1}\@@href}%
\providecommand \@@href[1]{\endgroup#1\@@endlink}%
\providecommand \@sanitize@url [0]{\catcode `\\12\catcode `\$12\catcode
  `\&12\catcode `\#12\catcode `\^12\catcode `\_12\catcode `\%12\relax}%
\providecommand \@@startlink[1]{}%
\providecommand \@@endlink[0]{}%
\providecommand \url  [0]{\begingroup\@sanitize@url \@url }%
\providecommand \@url [1]{\endgroup\@href {#1}{\urlprefix }}%
\providecommand \urlprefix  [0]{URL }%
\providecommand \Eprint [0]{\href }%
\providecommand \doibase [0]{https://doi.org/}%
\providecommand \selectlanguage [0]{\@gobble}%
\providecommand \bibinfo  [0]{\@secondoftwo}%
\providecommand \bibfield  [0]{\@secondoftwo}%
\providecommand \translation [1]{[#1]}%
\providecommand \BibitemOpen [0]{}%
\providecommand \bibitemStop [0]{}%
\providecommand \bibitemNoStop [0]{.\EOS\space}%
\providecommand \EOS [0]{\spacefactor3000\relax}%
\providecommand \BibitemShut  [1]{\csname bibitem#1\endcsname}%
\let\auto@bib@innerbib\@empty
\bibitem [{\citenamefont {Gu}\ \emph {et~al.}(2017)\citenamefont {Gu},
  \citenamefont {Kockum}, \citenamefont {Miranowicz}, \citenamefont {Liu},\
  and\ \citenamefont {Nori}}]{gu2017microwave}%
  \BibitemOpen
  \bibfield  {author} {\bibinfo {author} {\bibfnamefont {X.}~\bibnamefont
  {Gu}}, \bibinfo {author} {\bibfnamefont {A.~F.}\ \bibnamefont {Kockum}},
  \bibinfo {author} {\bibfnamefont {A.}~\bibnamefont {Miranowicz}}, \bibinfo
  {author} {\bibfnamefont {Y.-x.}\ \bibnamefont {Liu}},\ and\ \bibinfo {author}
  {\bibfnamefont {F.}~\bibnamefont {Nori}},\ }\bibfield  {title} {\bibinfo
  {title} {Microwave photonics with superconducting quantum circuits},\ }\href
  {https://www.sciencedirect.com/science/article/pii/S0370157317303290}
  {\bibfield  {journal} {\bibinfo  {journal} {Physics Reports}\ }\textbf
  {\bibinfo {volume} {718}},\ \bibinfo {pages} {1} (\bibinfo {year}
  {2017})}\BibitemShut {NoStop}%
\bibitem [{\citenamefont {Clerk}\ \emph {et~al.}(2020)\citenamefont {Clerk},
  \citenamefont {Lehnert}, \citenamefont {Bertet}, \citenamefont {Petta},\ and\
  \citenamefont {Nakamura}}]{clerk2020hybrid}%
  \BibitemOpen
  \bibfield  {author} {\bibinfo {author} {\bibfnamefont {A.}~\bibnamefont
  {Clerk}}, \bibinfo {author} {\bibfnamefont {K.}~\bibnamefont {Lehnert}},
  \bibinfo {author} {\bibfnamefont {P.}~\bibnamefont {Bertet}}, \bibinfo
  {author} {\bibfnamefont {J.}~\bibnamefont {Petta}},\ and\ \bibinfo {author}
  {\bibfnamefont {Y.}~\bibnamefont {Nakamura}},\ }\bibfield  {title} {\bibinfo
  {title} {Hybrid quantum systems with circuit quantum electrodynamics},\
  }\href {https://www.nature.com/articles/s41567-020-0797-9} {\bibfield
  {journal} {\bibinfo  {journal} {Nature Physics}\ }\textbf {\bibinfo {volume}
  {16}},\ \bibinfo {pages} {257} (\bibinfo {year} {2020})}\BibitemShut
  {NoStop}%
\bibitem [{\citenamefont {Kjaergaard}\ \emph {et~al.}(2020)\citenamefont
  {Kjaergaard}, \citenamefont {Schwartz}, \citenamefont {Braum{\"u}ller},
  \citenamefont {Krantz}, \citenamefont {Wang}, \citenamefont {Gustavsson},\
  and\ \citenamefont {Oliver}}]{kjaergaard2020superconducting}%
  \BibitemOpen
  \bibfield  {author} {\bibinfo {author} {\bibfnamefont {M.}~\bibnamefont
  {Kjaergaard}}, \bibinfo {author} {\bibfnamefont {M.~E.}\ \bibnamefont
  {Schwartz}}, \bibinfo {author} {\bibfnamefont {J.}~\bibnamefont
  {Braum{\"u}ller}}, \bibinfo {author} {\bibfnamefont {P.}~\bibnamefont
  {Krantz}}, \bibinfo {author} {\bibfnamefont {J.~I.-J.}\ \bibnamefont {Wang}},
  \bibinfo {author} {\bibfnamefont {S.}~\bibnamefont {Gustavsson}},\ and\
  \bibinfo {author} {\bibfnamefont {W.~D.}\ \bibnamefont {Oliver}},\ }\bibfield
   {title} {\bibinfo {title} {Superconducting qubits: Current state of play},\
  }\href
  {https://www.annualreviews.org/doi/abs/10.1146/annurev-conmatphys-031119-050605}
  {\bibfield  {journal} {\bibinfo  {journal} {Annual Review of Condensed Matter
  Physics}\ }\textbf {\bibinfo {volume} {11}},\ \bibinfo {pages} {369}
  (\bibinfo {year} {2020})}\BibitemShut {NoStop}%
\bibitem [{\citenamefont {Barends}\ \emph {et~al.}(2014)\citenamefont
  {Barends}, \citenamefont {Kelly}, \citenamefont {Megrant}, \citenamefont
  {Veitia}, \citenamefont {Sank}, \citenamefont {Jeffrey}, \citenamefont
  {White}, \citenamefont {Mutus}, \citenamefont {Fowler}, \citenamefont
  {Campbell} \emph {et~al.}}]{barends2014superconducting}%
  \BibitemOpen
  \bibfield  {author} {\bibinfo {author} {\bibfnamefont {R.}~\bibnamefont
  {Barends}}, \bibinfo {author} {\bibfnamefont {J.}~\bibnamefont {Kelly}},
  \bibinfo {author} {\bibfnamefont {A.}~\bibnamefont {Megrant}}, \bibinfo
  {author} {\bibfnamefont {A.}~\bibnamefont {Veitia}}, \bibinfo {author}
  {\bibfnamefont {D.}~\bibnamefont {Sank}}, \bibinfo {author} {\bibfnamefont
  {E.}~\bibnamefont {Jeffrey}}, \bibinfo {author} {\bibfnamefont {T.~C.}\
  \bibnamefont {White}}, \bibinfo {author} {\bibfnamefont {J.}~\bibnamefont
  {Mutus}}, \bibinfo {author} {\bibfnamefont {A.~G.}\ \bibnamefont {Fowler}},
  \bibinfo {author} {\bibfnamefont {B.}~\bibnamefont {Campbell}}, \emph
  {et~al.},\ }\bibfield  {title} {\bibinfo {title} {Superconducting quantum
  circuits at the surface code threshold for fault tolerance},\ }\href
  {https://www.nature.com/articles/nature13171} {\bibfield  {journal} {\bibinfo
   {journal} {Nature}\ }\textbf {\bibinfo {volume} {508}},\ \bibinfo {pages}
  {500} (\bibinfo {year} {2014})}\BibitemShut {NoStop}%
\bibitem [{\citenamefont {Heinsoo}\ \emph {et~al.}(2018)\citenamefont
  {Heinsoo}, \citenamefont {Andersen}, \citenamefont {Remm}, \citenamefont
  {Krinner}, \citenamefont {Walter}, \citenamefont {Salath{\'e}}, \citenamefont
  {Gasparinetti}, \citenamefont {Besse}, \citenamefont {Poto{\v{c}}nik},
  \citenamefont {Wallraff} \emph {et~al.}}]{heinsoo2018rapid}%
  \BibitemOpen
  \bibfield  {author} {\bibinfo {author} {\bibfnamefont {J.}~\bibnamefont
  {Heinsoo}}, \bibinfo {author} {\bibfnamefont {C.~K.}\ \bibnamefont
  {Andersen}}, \bibinfo {author} {\bibfnamefont {A.}~\bibnamefont {Remm}},
  \bibinfo {author} {\bibfnamefont {S.}~\bibnamefont {Krinner}}, \bibinfo
  {author} {\bibfnamefont {T.}~\bibnamefont {Walter}}, \bibinfo {author}
  {\bibfnamefont {Y.}~\bibnamefont {Salath{\'e}}}, \bibinfo {author}
  {\bibfnamefont {S.}~\bibnamefont {Gasparinetti}}, \bibinfo {author}
  {\bibfnamefont {J.-C.}\ \bibnamefont {Besse}}, \bibinfo {author}
  {\bibfnamefont {A.}~\bibnamefont {Poto{\v{c}}nik}}, \bibinfo {author}
  {\bibfnamefont {A.}~\bibnamefont {Wallraff}}, \emph {et~al.},\ }\bibfield
  {title} {\bibinfo {title} {Rapid high-fidelity multiplexed readout of
  superconducting qubits},\ }\href
  {https://journals.aps.org/prapplied/abstract/10.1103/PhysRevApplied.10.034040}
  {\bibfield  {journal} {\bibinfo  {journal} {Physical Review Applied}\
  }\textbf {\bibinfo {volume} {10}},\ \bibinfo {pages} {034040} (\bibinfo
  {year} {2018})}\BibitemShut {NoStop}%
\bibitem [{\citenamefont {Arute}\ \emph {et~al.}(2019)\citenamefont {Arute},
  \citenamefont {Arya}, \citenamefont {Babbush}, \citenamefont {Bacon},
  \citenamefont {Bardin}, \citenamefont {Barends}, \citenamefont {Biswas},
  \citenamefont {Boixo}, \citenamefont {Brandao}, \citenamefont {Buell} \emph
  {et~al.}}]{arute2019quantum}%
  \BibitemOpen
  \bibfield  {author} {\bibinfo {author} {\bibfnamefont {F.}~\bibnamefont
  {Arute}}, \bibinfo {author} {\bibfnamefont {K.}~\bibnamefont {Arya}},
  \bibinfo {author} {\bibfnamefont {R.}~\bibnamefont {Babbush}}, \bibinfo
  {author} {\bibfnamefont {D.}~\bibnamefont {Bacon}}, \bibinfo {author}
  {\bibfnamefont {J.~C.}\ \bibnamefont {Bardin}}, \bibinfo {author}
  {\bibfnamefont {R.}~\bibnamefont {Barends}}, \bibinfo {author} {\bibfnamefont
  {R.}~\bibnamefont {Biswas}}, \bibinfo {author} {\bibfnamefont
  {S.}~\bibnamefont {Boixo}}, \bibinfo {author} {\bibfnamefont {F.~G.}\
  \bibnamefont {Brandao}}, \bibinfo {author} {\bibfnamefont {D.~A.}\
  \bibnamefont {Buell}}, \emph {et~al.},\ }\bibfield  {title} {\bibinfo {title}
  {Quantum supremacy using a programmable superconducting processor},\ }\href
  {https://www.nature.com/articles/s41586\%20019\%201666\%205} {\bibfield
  {journal} {\bibinfo  {journal} {Nature}\ }\textbf {\bibinfo {volume} {574}},\
  \bibinfo {pages} {505} (\bibinfo {year} {2019})}\BibitemShut {NoStop}%
\bibitem [{\citenamefont {Wu}\ \emph {et~al.}(2021)\citenamefont {Wu},
  \citenamefont {Bao}, \citenamefont {Cao}, \citenamefont {Chen}, \citenamefont
  {Chen}, \citenamefont {Chen}, \citenamefont {Chung}, \citenamefont {Deng},
  \citenamefont {Du}, \citenamefont {Fan} \emph {et~al.}}]{wu2021strong}%
  \BibitemOpen
  \bibfield  {author} {\bibinfo {author} {\bibfnamefont {Y.}~\bibnamefont
  {Wu}}, \bibinfo {author} {\bibfnamefont {W.-S.}\ \bibnamefont {Bao}},
  \bibinfo {author} {\bibfnamefont {S.}~\bibnamefont {Cao}}, \bibinfo {author}
  {\bibfnamefont {F.}~\bibnamefont {Chen}}, \bibinfo {author} {\bibfnamefont
  {M.-C.}\ \bibnamefont {Chen}}, \bibinfo {author} {\bibfnamefont
  {X.}~\bibnamefont {Chen}}, \bibinfo {author} {\bibfnamefont {T.-H.}\
  \bibnamefont {Chung}}, \bibinfo {author} {\bibfnamefont {H.}~\bibnamefont
  {Deng}}, \bibinfo {author} {\bibfnamefont {Y.}~\bibnamefont {Du}}, \bibinfo
  {author} {\bibfnamefont {D.}~\bibnamefont {Fan}}, \emph {et~al.},\ }\bibfield
   {title} {\bibinfo {title} {Strong quantum computational advantage using a
  superconducting quantum processor},\ }\href
  {https://journals.aps.org/prl/abstract/10.1103/PhysRevLett.127.180501}
  {\bibfield  {journal} {\bibinfo  {journal} {Physical review letters}\
  }\textbf {\bibinfo {volume} {127}},\ \bibinfo {pages} {180501} (\bibinfo
  {year} {2021})}\BibitemShut {NoStop}%
\bibitem [{\citenamefont {Krinner}\ \emph {et~al.}(2022)\citenamefont
  {Krinner}, \citenamefont {Lacroix}, \citenamefont {Remm}, \citenamefont
  {Di~Paolo}, \citenamefont {Genois}, \citenamefont {Leroux}, \citenamefont
  {Hellings}, \citenamefont {Lazar}, \citenamefont {Swiadek}, \citenamefont
  {Herrmann} \emph {et~al.}}]{krinner2022realizing}%
  \BibitemOpen
  \bibfield  {author} {\bibinfo {author} {\bibfnamefont {S.}~\bibnamefont
  {Krinner}}, \bibinfo {author} {\bibfnamefont {N.}~\bibnamefont {Lacroix}},
  \bibinfo {author} {\bibfnamefont {A.}~\bibnamefont {Remm}}, \bibinfo {author}
  {\bibfnamefont {A.}~\bibnamefont {Di~Paolo}}, \bibinfo {author}
  {\bibfnamefont {E.}~\bibnamefont {Genois}}, \bibinfo {author} {\bibfnamefont
  {C.}~\bibnamefont {Leroux}}, \bibinfo {author} {\bibfnamefont
  {C.}~\bibnamefont {Hellings}}, \bibinfo {author} {\bibfnamefont
  {S.}~\bibnamefont {Lazar}}, \bibinfo {author} {\bibfnamefont
  {F.}~\bibnamefont {Swiadek}}, \bibinfo {author} {\bibfnamefont
  {J.}~\bibnamefont {Herrmann}}, \emph {et~al.},\ }\bibfield  {title} {\bibinfo
  {title} {Realizing repeated quantum error correction in a distance-three
  surface code},\ }\href {https://www.nature.com/articles/s41586-022-04566-8}
  {\bibfield  {journal} {\bibinfo  {journal} {Nature}\ }\textbf {\bibinfo
  {volume} {605}},\ \bibinfo {pages} {669} (\bibinfo {year}
  {2022})}\BibitemShut {NoStop}%
\bibitem [{\citenamefont {Zhao}\ \emph {et~al.}(2022)\citenamefont {Zhao},
  \citenamefont {Ye}, \citenamefont {Huang}, \citenamefont {Zhang},
  \citenamefont {Wu}, \citenamefont {Guan}, \citenamefont {Zhu}, \citenamefont
  {Wei}, \citenamefont {He}, \citenamefont {Cao} \emph
  {et~al.}}]{zhao2022realization}%
  \BibitemOpen
  \bibfield  {author} {\bibinfo {author} {\bibfnamefont {Y.}~\bibnamefont
  {Zhao}}, \bibinfo {author} {\bibfnamefont {Y.}~\bibnamefont {Ye}}, \bibinfo
  {author} {\bibfnamefont {H.-L.}\ \bibnamefont {Huang}}, \bibinfo {author}
  {\bibfnamefont {Y.}~\bibnamefont {Zhang}}, \bibinfo {author} {\bibfnamefont
  {D.}~\bibnamefont {Wu}}, \bibinfo {author} {\bibfnamefont {H.}~\bibnamefont
  {Guan}}, \bibinfo {author} {\bibfnamefont {Q.}~\bibnamefont {Zhu}}, \bibinfo
  {author} {\bibfnamefont {Z.}~\bibnamefont {Wei}}, \bibinfo {author}
  {\bibfnamefont {T.}~\bibnamefont {He}}, \bibinfo {author} {\bibfnamefont
  {S.}~\bibnamefont {Cao}}, \emph {et~al.},\ }\bibfield  {title} {\bibinfo
  {title} {Realization of an error-correcting surface code with superconducting
  qubits},\ }\href
  {https://journals.aps.org/prl/abstract/10.1103/PhysRevLett.129.030501}
  {\bibfield  {journal} {\bibinfo  {journal} {Physical Review Letters}\
  }\textbf {\bibinfo {volume} {129}},\ \bibinfo {pages} {030501} (\bibinfo
  {year} {2022})}\BibitemShut {NoStop}%
\bibitem [{\citenamefont {{Google Quantum
  AI}}(2023)}]{Google_Quantum_AI2023-wv}%
  \BibitemOpen
  \bibfield  {author} {\bibinfo {author} {\bibnamefont {{Google Quantum AI}}},\
  }\bibfield  {title} {\bibinfo {title} {Suppressing quantum errors by scaling
  a surface code logical qubit},\ }\href
  {https://www.nature.com/articles/s41586-022-05434-1} {\bibfield  {journal}
  {\bibinfo  {journal} {Nature}\ }\textbf {\bibinfo {volume} {614}},\ \bibinfo
  {pages} {676} (\bibinfo {year} {2023})}\BibitemShut {NoStop}%
\bibitem [{\citenamefont {Gambetta}\ \emph {et~al.}(2017)\citenamefont
  {Gambetta}, \citenamefont {Chow},\ and\ \citenamefont
  {Steffen}}]{gambetta2017building}%
  \BibitemOpen
  \bibfield  {author} {\bibinfo {author} {\bibfnamefont {J.~M.}\ \bibnamefont
  {Gambetta}}, \bibinfo {author} {\bibfnamefont {J.~M.}\ \bibnamefont {Chow}},\
  and\ \bibinfo {author} {\bibfnamefont {M.}~\bibnamefont {Steffen}},\
  }\bibfield  {title} {\bibinfo {title} {Building logical qubits in a
  superconducting quantum computing system},\ }\href
  {https://www.nature.com/articles/s41534-016-0004-0} {\bibfield  {journal}
  {\bibinfo  {journal} {npj quantum information}\ }\textbf {\bibinfo {volume}
  {3}},\ \bibinfo {pages} {2} (\bibinfo {year} {2017})}\BibitemShut {NoStop}%
\bibitem [{\citenamefont {Lidar}\ and\ \citenamefont
  {Brun}(2013)}]{lidar2013quantum}%
  \BibitemOpen
  \bibfield  {author} {\bibinfo {author} {\bibfnamefont {D.~A.}\ \bibnamefont
  {Lidar}}\ and\ \bibinfo {author} {\bibfnamefont {T.~A.}\ \bibnamefont
  {Brun}},\ }\href@noop {} {\emph {\bibinfo {title} {Quantum error
  correction}}}\ (\bibinfo  {publisher} {Cambridge university press},\ \bibinfo
  {year} {2013})\BibitemShut {NoStop}%
\bibitem [{\citenamefont {Fowler}\ \emph {et~al.}(2012)\citenamefont {Fowler},
  \citenamefont {Mariantoni}, \citenamefont {Martinis},\ and\ \citenamefont
  {Cleland}}]{fowler2012surface}%
  \BibitemOpen
  \bibfield  {author} {\bibinfo {author} {\bibfnamefont {A.~G.}\ \bibnamefont
  {Fowler}}, \bibinfo {author} {\bibfnamefont {M.}~\bibnamefont {Mariantoni}},
  \bibinfo {author} {\bibfnamefont {J.~M.}\ \bibnamefont {Martinis}},\ and\
  \bibinfo {author} {\bibfnamefont {A.~N.}\ \bibnamefont {Cleland}},\
  }\bibfield  {title} {\bibinfo {title} {Surface codes: {Towards} practical
  large-scale quantum computation},\ }\href
  {https://journals.aps.org/pra/abstract/10.1103/PhysRevA.86.032324} {\bibfield
   {journal} {\bibinfo  {journal} {Physical Review A}\ }\textbf {\bibinfo
  {volume} {86}},\ \bibinfo {pages} {032324} (\bibinfo {year}
  {2012})}\BibitemShut {NoStop}%
\bibitem [{\citenamefont {Xu}\ \emph {et~al.}(2020)\citenamefont {Xu},
  \citenamefont {Chu}, \citenamefont {Yuan}, \citenamefont {Qiu}, \citenamefont
  {Zhou}, \citenamefont {Zhang}, \citenamefont {Tan}, \citenamefont {Yu},
  \citenamefont {Liu}, \citenamefont {Li} \emph {et~al.}}]{xu2020high}%
  \BibitemOpen
  \bibfield  {author} {\bibinfo {author} {\bibfnamefont {Y.}~\bibnamefont
  {Xu}}, \bibinfo {author} {\bibfnamefont {J.}~\bibnamefont {Chu}}, \bibinfo
  {author} {\bibfnamefont {J.}~\bibnamefont {Yuan}}, \bibinfo {author}
  {\bibfnamefont {J.}~\bibnamefont {Qiu}}, \bibinfo {author} {\bibfnamefont
  {Y.}~\bibnamefont {Zhou}}, \bibinfo {author} {\bibfnamefont {L.}~\bibnamefont
  {Zhang}}, \bibinfo {author} {\bibfnamefont {X.}~\bibnamefont {Tan}}, \bibinfo
  {author} {\bibfnamefont {Y.}~\bibnamefont {Yu}}, \bibinfo {author}
  {\bibfnamefont {S.}~\bibnamefont {Liu}}, \bibinfo {author} {\bibfnamefont
  {J.}~\bibnamefont {Li}}, \emph {et~al.},\ }\bibfield  {title} {\bibinfo
  {title} {High-fidelity, high-scalability two-qubit gate scheme for
  superconducting qubits},\ }\href
  {https://journals.aps.org/prl/abstract/10.1103/PhysRevLett.125.240503}
  {\bibfield  {journal} {\bibinfo  {journal} {Physical Review Letters}\
  }\textbf {\bibinfo {volume} {125}},\ \bibinfo {pages} {240503} (\bibinfo
  {year} {2020})}\BibitemShut {NoStop}%
\bibitem [{\citenamefont {Siddiqi}(2021)}]{siddiqi2021engineering}%
  \BibitemOpen
  \bibfield  {author} {\bibinfo {author} {\bibfnamefont {I.}~\bibnamefont
  {Siddiqi}},\ }\bibfield  {title} {\bibinfo {title} {Engineering
  high-coherence superconducting qubits},\ }\href
  {https://www.nature.com/articles/s41578-021-00370-4} {\bibfield  {journal}
  {\bibinfo  {journal} {Nature Reviews Materials}\ }\textbf {\bibinfo {volume}
  {6}},\ \bibinfo {pages} {875} (\bibinfo {year} {2021})}\BibitemShut {NoStop}%
\bibitem [{\citenamefont {Reed}\ \emph {et~al.}(2010)\citenamefont {Reed},
  \citenamefont {Johnson}, \citenamefont {Houck}, \citenamefont {DiCarlo},
  \citenamefont {Chow}, \citenamefont {Schuster}, \citenamefont {Frunzio},\
  and\ \citenamefont {Schoelkopf}}]{reed2010fast}%
  \BibitemOpen
  \bibfield  {author} {\bibinfo {author} {\bibfnamefont {M.~D.}\ \bibnamefont
  {Reed}}, \bibinfo {author} {\bibfnamefont {B.~R.}\ \bibnamefont {Johnson}},
  \bibinfo {author} {\bibfnamefont {A.~A.}\ \bibnamefont {Houck}}, \bibinfo
  {author} {\bibfnamefont {L.}~\bibnamefont {DiCarlo}}, \bibinfo {author}
  {\bibfnamefont {J.~M.}\ \bibnamefont {Chow}}, \bibinfo {author}
  {\bibfnamefont {D.~I.}\ \bibnamefont {Schuster}}, \bibinfo {author}
  {\bibfnamefont {L.}~\bibnamefont {Frunzio}},\ and\ \bibinfo {author}
  {\bibfnamefont {R.~J.}\ \bibnamefont {Schoelkopf}},\ }\bibfield  {title}
  {\bibinfo {title} {Fast reset and suppressing spontaneous emission of a
  superconducting qubit},\ }\href
  {https://pubs.aip.org/aip/apl/article/96/20/203110/119911/Fast-reset-and-suppressing-spontaneous-emission-of}
  {\bibfield  {journal} {\bibinfo  {journal} {Applied Physics Letters}\
  }\textbf {\bibinfo {volume} {96}},\ \bibinfo {pages} {203110} (\bibinfo
  {year} {2010})}\BibitemShut {NoStop}%
\bibitem [{\citenamefont {Paik}\ \emph {et~al.}(2011)\citenamefont {Paik},
  \citenamefont {Schuster}, \citenamefont {Bishop}, \citenamefont {Kirchmair},
  \citenamefont {Catelani}, \citenamefont {Sears}, \citenamefont {Johnson},
  \citenamefont {Reagor}, \citenamefont {Frunzio}, \citenamefont {Glazman}
  \emph {et~al.}}]{paik2011observation}%
  \BibitemOpen
  \bibfield  {author} {\bibinfo {author} {\bibfnamefont {H.}~\bibnamefont
  {Paik}}, \bibinfo {author} {\bibfnamefont {D.~I.}\ \bibnamefont {Schuster}},
  \bibinfo {author} {\bibfnamefont {L.~S.}\ \bibnamefont {Bishop}}, \bibinfo
  {author} {\bibfnamefont {G.}~\bibnamefont {Kirchmair}}, \bibinfo {author}
  {\bibfnamefont {G.}~\bibnamefont {Catelani}}, \bibinfo {author}
  {\bibfnamefont {A.~P.}\ \bibnamefont {Sears}}, \bibinfo {author}
  {\bibfnamefont {B.}~\bibnamefont {Johnson}}, \bibinfo {author} {\bibfnamefont
  {M.}~\bibnamefont {Reagor}}, \bibinfo {author} {\bibfnamefont
  {L.}~\bibnamefont {Frunzio}}, \bibinfo {author} {\bibfnamefont {L.~I.}\
  \bibnamefont {Glazman}}, \emph {et~al.},\ }\bibfield  {title} {\bibinfo
  {title} {Observation of high coherence in {Josephson} junction qubits
  measured in a three-dimensional circuit {QED} architecture},\ }\href
  {https://journals.aps.org/prl/abstract/10.1103/PhysRevLett.107.240501}
  {\bibfield  {journal} {\bibinfo  {journal} {Physical Review Letters}\
  }\textbf {\bibinfo {volume} {107}},\ \bibinfo {pages} {240501} (\bibinfo
  {year} {2011})}\BibitemShut {NoStop}%
\bibitem [{\citenamefont {Yan}\ \emph {et~al.}(2016)\citenamefont {Yan},
  \citenamefont {Gustavsson}, \citenamefont {Kamal}, \citenamefont {Birenbaum},
  \citenamefont {Sears}, \citenamefont {Hover}, \citenamefont {Gudmundsen},
  \citenamefont {Rosenberg}, \citenamefont {Samach}, \citenamefont {Weber}
  \emph {et~al.}}]{yan2016flux}%
  \BibitemOpen
  \bibfield  {author} {\bibinfo {author} {\bibfnamefont {F.}~\bibnamefont
  {Yan}}, \bibinfo {author} {\bibfnamefont {S.}~\bibnamefont {Gustavsson}},
  \bibinfo {author} {\bibfnamefont {A.}~\bibnamefont {Kamal}}, \bibinfo
  {author} {\bibfnamefont {J.}~\bibnamefont {Birenbaum}}, \bibinfo {author}
  {\bibfnamefont {A.~P.}\ \bibnamefont {Sears}}, \bibinfo {author}
  {\bibfnamefont {D.}~\bibnamefont {Hover}}, \bibinfo {author} {\bibfnamefont
  {T.~J.}\ \bibnamefont {Gudmundsen}}, \bibinfo {author} {\bibfnamefont
  {D.}~\bibnamefont {Rosenberg}}, \bibinfo {author} {\bibfnamefont
  {G.}~\bibnamefont {Samach}}, \bibinfo {author} {\bibfnamefont
  {S.}~\bibnamefont {Weber}}, \emph {et~al.},\ }\bibfield  {title} {\bibinfo
  {title} {The flux qubit revisited to enhance coherence and reproducibility},\
  }\href {https://www.nature.com/articles/ncomms12964} {\bibfield  {journal}
  {\bibinfo  {journal} {Nature communications}\ }\textbf {\bibinfo {volume}
  {7}},\ \bibinfo {pages} {12964} (\bibinfo {year} {2016})}\BibitemShut
  {NoStop}%
\bibitem [{\citenamefont {Nguyen}\ \emph {et~al.}(2019)\citenamefont {Nguyen},
  \citenamefont {Lin}, \citenamefont {Somoroff}, \citenamefont {Mencia},
  \citenamefont {Grabon},\ and\ \citenamefont {Manucharyan}}]{nguyen2019high}%
  \BibitemOpen
  \bibfield  {author} {\bibinfo {author} {\bibfnamefont {L.~B.}\ \bibnamefont
  {Nguyen}}, \bibinfo {author} {\bibfnamefont {Y.-H.}\ \bibnamefont {Lin}},
  \bibinfo {author} {\bibfnamefont {A.}~\bibnamefont {Somoroff}}, \bibinfo
  {author} {\bibfnamefont {R.}~\bibnamefont {Mencia}}, \bibinfo {author}
  {\bibfnamefont {N.}~\bibnamefont {Grabon}},\ and\ \bibinfo {author}
  {\bibfnamefont {V.~E.}\ \bibnamefont {Manucharyan}},\ }\bibfield  {title}
  {\bibinfo {title} {High-coherence fluxonium qubit},\ }\href
  {https://journals.aps.org/prx/abstract/10.1103/PhysRevX.9.041041} {\bibfield
  {journal} {\bibinfo  {journal} {Physical Review X}\ }\textbf {\bibinfo
  {volume} {9}},\ \bibinfo {pages} {041041} (\bibinfo {year}
  {2019})}\BibitemShut {NoStop}%
\bibitem [{\citenamefont {Gordon}\ \emph {et~al.}(2022)\citenamefont {Gordon},
  \citenamefont {Murray}, \citenamefont {Kurter}, \citenamefont {Sandberg},
  \citenamefont {Hall}, \citenamefont {Balakrishnan}, \citenamefont {Shelby},
  \citenamefont {Wacaser}, \citenamefont {Stabile}, \citenamefont {Sleight}
  \emph {et~al.}}]{gordon2022environmental}%
  \BibitemOpen
  \bibfield  {author} {\bibinfo {author} {\bibfnamefont {R.}~\bibnamefont
  {Gordon}}, \bibinfo {author} {\bibfnamefont {C.}~\bibnamefont {Murray}},
  \bibinfo {author} {\bibfnamefont {C.}~\bibnamefont {Kurter}}, \bibinfo
  {author} {\bibfnamefont {M.}~\bibnamefont {Sandberg}}, \bibinfo {author}
  {\bibfnamefont {S.}~\bibnamefont {Hall}}, \bibinfo {author} {\bibfnamefont
  {K.}~\bibnamefont {Balakrishnan}}, \bibinfo {author} {\bibfnamefont
  {R.}~\bibnamefont {Shelby}}, \bibinfo {author} {\bibfnamefont
  {B.}~\bibnamefont {Wacaser}}, \bibinfo {author} {\bibfnamefont
  {A.}~\bibnamefont {Stabile}}, \bibinfo {author} {\bibfnamefont
  {J.}~\bibnamefont {Sleight}}, \emph {et~al.},\ }\bibfield  {title} {\bibinfo
  {title} {Environmental radiation impact on lifetimes and quasiparticle
  tunneling rates of fixed-frequency transmon qubits},\ }\href
  {https://pubs.aip.org/aip/apl/article/120/7/074002/2833253/Environmental-radiation-impact-on-lifetimes-and}
  {\bibfield  {journal} {\bibinfo  {journal} {Applied Physics Letters}\
  }\textbf {\bibinfo {volume} {120}},\ \bibinfo {pages} {074002} (\bibinfo
  {year} {2022})}\BibitemShut {NoStop}%
\bibitem [{\citenamefont {Murray}(2021)}]{Murray2021-fv}%
  \BibitemOpen
  \bibfield  {author} {\bibinfo {author} {\bibfnamefont {C.~E.}\ \bibnamefont
  {Murray}},\ }\bibfield  {title} {\bibinfo {title} {Material matters in
  superconducting qubits},\ }\href
  {https://www.sciencedirect.com/science/article/pii/S0927796X21000413}
  {\bibfield  {journal} {\bibinfo  {journal} {Mater. Sci. Eng. R Rep.}\
  }\textbf {\bibinfo {volume} {146}},\ \bibinfo {pages} {100646} (\bibinfo
  {year} {2021})}\BibitemShut {NoStop}%
\bibitem [{\citenamefont {Place}\ \emph {et~al.}(2021)\citenamefont {Place},
  \citenamefont {Rodgers}, \citenamefont {Mundada}, \citenamefont {Smitham},
  \citenamefont {Fitzpatrick}, \citenamefont {Leng}, \citenamefont {Premkumar},
  \citenamefont {Bryon}, \citenamefont {Vrajitoarea}, \citenamefont {Sussman}
  \emph {et~al.}}]{place2021new}%
  \BibitemOpen
  \bibfield  {author} {\bibinfo {author} {\bibfnamefont {A.~P.}\ \bibnamefont
  {Place}}, \bibinfo {author} {\bibfnamefont {L.~V.}\ \bibnamefont {Rodgers}},
  \bibinfo {author} {\bibfnamefont {P.}~\bibnamefont {Mundada}}, \bibinfo
  {author} {\bibfnamefont {B.~M.}\ \bibnamefont {Smitham}}, \bibinfo {author}
  {\bibfnamefont {M.}~\bibnamefont {Fitzpatrick}}, \bibinfo {author}
  {\bibfnamefont {Z.}~\bibnamefont {Leng}}, \bibinfo {author} {\bibfnamefont
  {A.}~\bibnamefont {Premkumar}}, \bibinfo {author} {\bibfnamefont
  {J.}~\bibnamefont {Bryon}}, \bibinfo {author} {\bibfnamefont
  {A.}~\bibnamefont {Vrajitoarea}}, \bibinfo {author} {\bibfnamefont
  {S.}~\bibnamefont {Sussman}}, \emph {et~al.},\ }\bibfield  {title} {\bibinfo
  {title} {New material platform for superconducting transmon qubits with
  coherence times exceeding 0.3 milliseconds},\ }\href
  {https://www.nature.com/articles/s41467-021-22030-5} {\bibfield  {journal}
  {\bibinfo  {journal} {Nature communications}\ }\textbf {\bibinfo {volume}
  {12}},\ \bibinfo {pages} {1779} (\bibinfo {year} {2021})}\BibitemShut
  {NoStop}%
\bibitem [{\citenamefont {Wang}\ \emph {et~al.}(2022)\citenamefont {Wang},
  \citenamefont {Li}, \citenamefont {Xu}, \citenamefont {Li}, \citenamefont
  {Wang}, \citenamefont {Yang}, \citenamefont {Mi}, \citenamefont {Liang},
  \citenamefont {Su}, \citenamefont {Yang} \emph {et~al.}}]{wang2022towards}%
  \BibitemOpen
  \bibfield  {author} {\bibinfo {author} {\bibfnamefont {C.}~\bibnamefont
  {Wang}}, \bibinfo {author} {\bibfnamefont {X.}~\bibnamefont {Li}}, \bibinfo
  {author} {\bibfnamefont {H.}~\bibnamefont {Xu}}, \bibinfo {author}
  {\bibfnamefont {Z.}~\bibnamefont {Li}}, \bibinfo {author} {\bibfnamefont
  {J.}~\bibnamefont {Wang}}, \bibinfo {author} {\bibfnamefont {Z.}~\bibnamefont
  {Yang}}, \bibinfo {author} {\bibfnamefont {Z.}~\bibnamefont {Mi}}, \bibinfo
  {author} {\bibfnamefont {X.}~\bibnamefont {Liang}}, \bibinfo {author}
  {\bibfnamefont {T.}~\bibnamefont {Su}}, \bibinfo {author} {\bibfnamefont
  {C.}~\bibnamefont {Yang}}, \emph {et~al.},\ }\bibfield  {title} {\bibinfo
  {title} {Towards practical quantum computers: {Transmon} qubit with a
  lifetime approaching 0.5 milliseconds},\ }\href
  {https://www.nature.com/articles/s41534-021-00510-2} {\bibfield  {journal}
  {\bibinfo  {journal} {npj Quantum Information}\ }\textbf {\bibinfo {volume}
  {8}},\ \bibinfo {pages} {3} (\bibinfo {year} {2022})}\BibitemShut {NoStop}%
\bibitem [{\citenamefont {Deng}\ \emph {et~al.}(2023)\citenamefont {Deng},
  \citenamefont {Song}, \citenamefont {Gao}, \citenamefont {Xia}, \citenamefont
  {Bao}, \citenamefont {Jiang}, \citenamefont {Ku}, \citenamefont {Li},
  \citenamefont {Ma}, \citenamefont {Qin} \emph {et~al.}}]{deng2023titanium}%
  \BibitemOpen
  \bibfield  {author} {\bibinfo {author} {\bibfnamefont {H.}~\bibnamefont
  {Deng}}, \bibinfo {author} {\bibfnamefont {Z.}~\bibnamefont {Song}}, \bibinfo
  {author} {\bibfnamefont {R.}~\bibnamefont {Gao}}, \bibinfo {author}
  {\bibfnamefont {T.}~\bibnamefont {Xia}}, \bibinfo {author} {\bibfnamefont
  {F.}~\bibnamefont {Bao}}, \bibinfo {author} {\bibfnamefont {X.}~\bibnamefont
  {Jiang}}, \bibinfo {author} {\bibfnamefont {H.-S.}\ \bibnamefont {Ku}},
  \bibinfo {author} {\bibfnamefont {Z.}~\bibnamefont {Li}}, \bibinfo {author}
  {\bibfnamefont {X.}~\bibnamefont {Ma}}, \bibinfo {author} {\bibfnamefont
  {J.}~\bibnamefont {Qin}}, \emph {et~al.},\ }\bibfield  {title} {\bibinfo
  {title} {Titanium nitride film on sapphire substrate with low dielectric loss
  for superconducting qubits},\ }\href
  {https://journals.aps.org/prapplied/abstract/10.1103/PhysRevApplied.19.024013}
  {\bibfield  {journal} {\bibinfo  {journal} {Physical Review Applied}\
  }\textbf {\bibinfo {volume} {19}},\ \bibinfo {pages} {024013} (\bibinfo
  {year} {2023})}\BibitemShut {NoStop}%
\bibitem [{\citenamefont {Klimov}\ \emph {et~al.}(2018)\citenamefont {Klimov},
  \citenamefont {Kelly}, \citenamefont {Chen}, \citenamefont {Neeley},
  \citenamefont {Megrant}, \citenamefont {Burkett}, \citenamefont {Barends},
  \citenamefont {Arya}, \citenamefont {Chiaro}, \citenamefont {Chen} \emph
  {et~al.}}]{klimov2018fluctuations}%
  \BibitemOpen
  \bibfield  {author} {\bibinfo {author} {\bibfnamefont {P.}~\bibnamefont
  {Klimov}}, \bibinfo {author} {\bibfnamefont {J.}~\bibnamefont {Kelly}},
  \bibinfo {author} {\bibfnamefont {Z.}~\bibnamefont {Chen}}, \bibinfo {author}
  {\bibfnamefont {M.}~\bibnamefont {Neeley}}, \bibinfo {author} {\bibfnamefont
  {A.}~\bibnamefont {Megrant}}, \bibinfo {author} {\bibfnamefont
  {B.}~\bibnamefont {Burkett}}, \bibinfo {author} {\bibfnamefont
  {R.}~\bibnamefont {Barends}}, \bibinfo {author} {\bibfnamefont
  {K.}~\bibnamefont {Arya}}, \bibinfo {author} {\bibfnamefont {B.}~\bibnamefont
  {Chiaro}}, \bibinfo {author} {\bibfnamefont {Y.}~\bibnamefont {Chen}}, \emph
  {et~al.},\ }\bibfield  {title} {\bibinfo {title} {Fluctuations of
  energy-relaxation times in superconducting qubits},\ }\href
  {https://journals.aps.org/prl/abstract/10.1103/PhysRevLett.121.090502}
  {\bibfield  {journal} {\bibinfo  {journal} {Physical review letters}\
  }\textbf {\bibinfo {volume} {121}},\ \bibinfo {pages} {090502} (\bibinfo
  {year} {2018})}\BibitemShut {NoStop}%
\bibitem [{\citenamefont {Burnett}\ \emph {et~al.}(2019)\citenamefont
  {Burnett}, \citenamefont {Bengtsson}, \citenamefont {Scigliuzzo},
  \citenamefont {Niepce}, \citenamefont {Kudra}, \citenamefont {Delsing},\ and\
  \citenamefont {Bylander}}]{burnett2019decoherence}%
  \BibitemOpen
  \bibfield  {author} {\bibinfo {author} {\bibfnamefont {J.~J.}\ \bibnamefont
  {Burnett}}, \bibinfo {author} {\bibfnamefont {A.}~\bibnamefont {Bengtsson}},
  \bibinfo {author} {\bibfnamefont {M.}~\bibnamefont {Scigliuzzo}}, \bibinfo
  {author} {\bibfnamefont {D.}~\bibnamefont {Niepce}}, \bibinfo {author}
  {\bibfnamefont {M.}~\bibnamefont {Kudra}}, \bibinfo {author} {\bibfnamefont
  {P.}~\bibnamefont {Delsing}},\ and\ \bibinfo {author} {\bibfnamefont
  {J.}~\bibnamefont {Bylander}},\ }\bibfield  {title} {\bibinfo {title}
  {Decoherence benchmarking of superconducting qubits},\ }\href
  {https://www.nature.com/articles/s41534-019-0168-5} {\bibfield  {journal}
  {\bibinfo  {journal} {npj Quantum Information}\ }\textbf {\bibinfo {volume}
  {5}},\ \bibinfo {pages} {54} (\bibinfo {year} {2019})}\BibitemShut {NoStop}%
\bibitem [{\citenamefont {Wilen}\ \emph {et~al.}(2021)\citenamefont {Wilen},
  \citenamefont {Abdullah}, \citenamefont {Kurinsky}, \citenamefont {Stanford},
  \citenamefont {Cardani}, \citenamefont {d’Imperio}, \citenamefont {Tomei},
  \citenamefont {Faoro}, \citenamefont {Ioffe}, \citenamefont {Liu} \emph
  {et~al.}}]{wilen2021correlated}%
  \BibitemOpen
  \bibfield  {author} {\bibinfo {author} {\bibfnamefont {C.~D.}\ \bibnamefont
  {Wilen}}, \bibinfo {author} {\bibfnamefont {S.}~\bibnamefont {Abdullah}},
  \bibinfo {author} {\bibfnamefont {N.}~\bibnamefont {Kurinsky}}, \bibinfo
  {author} {\bibfnamefont {C.}~\bibnamefont {Stanford}}, \bibinfo {author}
  {\bibfnamefont {L.}~\bibnamefont {Cardani}}, \bibinfo {author} {\bibfnamefont
  {G.}~\bibnamefont {d’Imperio}}, \bibinfo {author} {\bibfnamefont
  {C.}~\bibnamefont {Tomei}}, \bibinfo {author} {\bibfnamefont
  {L.}~\bibnamefont {Faoro}}, \bibinfo {author} {\bibfnamefont
  {L.}~\bibnamefont {Ioffe}}, \bibinfo {author} {\bibfnamefont
  {C.}~\bibnamefont {Liu}}, \emph {et~al.},\ }\bibfield  {title} {\bibinfo
  {title} {Correlated charge noise and relaxation errors in superconducting
  qubits},\ }\href {https://www.nature.com/articles/s41586-021-03557-5}
  {\bibfield  {journal} {\bibinfo  {journal} {Nature}\ }\textbf {\bibinfo
  {volume} {594}},\ \bibinfo {pages} {369} (\bibinfo {year}
  {2021})}\BibitemShut {NoStop}%
\bibitem [{\citenamefont {McEwen}\ \emph {et~al.}(2022)\citenamefont {McEwen},
  \citenamefont {Faoro}, \citenamefont {Arya}, \citenamefont {Dunsworth},
  \citenamefont {Huang}, \citenamefont {Kim}, \citenamefont {Burkett},
  \citenamefont {Fowler}, \citenamefont {Arute}, \citenamefont {Bardin} \emph
  {et~al.}}]{mcewen2022resolving}%
  \BibitemOpen
  \bibfield  {author} {\bibinfo {author} {\bibfnamefont {M.}~\bibnamefont
  {McEwen}}, \bibinfo {author} {\bibfnamefont {L.}~\bibnamefont {Faoro}},
  \bibinfo {author} {\bibfnamefont {K.}~\bibnamefont {Arya}}, \bibinfo {author}
  {\bibfnamefont {A.}~\bibnamefont {Dunsworth}}, \bibinfo {author}
  {\bibfnamefont {T.}~\bibnamefont {Huang}}, \bibinfo {author} {\bibfnamefont
  {S.}~\bibnamefont {Kim}}, \bibinfo {author} {\bibfnamefont {B.}~\bibnamefont
  {Burkett}}, \bibinfo {author} {\bibfnamefont {A.}~\bibnamefont {Fowler}},
  \bibinfo {author} {\bibfnamefont {F.}~\bibnamefont {Arute}}, \bibinfo
  {author} {\bibfnamefont {J.~C.}\ \bibnamefont {Bardin}}, \emph {et~al.},\
  }\bibfield  {title} {\bibinfo {title} {Resolving catastrophic error bursts
  from cosmic rays in large arrays of superconducting qubits},\ }\href
  {https://www.nature.com/articles/s41567-021-01432-8} {\bibfield  {journal}
  {\bibinfo  {journal} {Nature Physics}\ }\textbf {\bibinfo {volume} {18}},\
  \bibinfo {pages} {107} (\bibinfo {year} {2022})}\BibitemShut {NoStop}%
\bibitem [{\citenamefont {Wang}\ \emph {et~al.}(2015)\citenamefont {Wang},
  \citenamefont {Axline}, \citenamefont {Gao}, \citenamefont {Brecht},
  \citenamefont {Chu}, \citenamefont {Frunzio}, \citenamefont {Devoret},\ and\
  \citenamefont {Schoelkopf}}]{wang2015surface}%
  \BibitemOpen
  \bibfield  {author} {\bibinfo {author} {\bibfnamefont {C.}~\bibnamefont
  {Wang}}, \bibinfo {author} {\bibfnamefont {C.}~\bibnamefont {Axline}},
  \bibinfo {author} {\bibfnamefont {Y.~Y.}\ \bibnamefont {Gao}}, \bibinfo
  {author} {\bibfnamefont {T.}~\bibnamefont {Brecht}}, \bibinfo {author}
  {\bibfnamefont {Y.}~\bibnamefont {Chu}}, \bibinfo {author} {\bibfnamefont
  {L.}~\bibnamefont {Frunzio}}, \bibinfo {author} {\bibfnamefont
  {M.}~\bibnamefont {Devoret}},\ and\ \bibinfo {author} {\bibfnamefont {R.~J.}\
  \bibnamefont {Schoelkopf}},\ }\bibfield  {title} {\bibinfo {title} {Surface
  participation and dielectric loss in superconducting qubits},\ }\href
  {https://pubs.aip.org/aip/apl/article/107/16/162601/593971/Surface-participation-and-dielectric-loss-in}
  {\bibfield  {journal} {\bibinfo  {journal} {Applied Physics Letters}\
  }\textbf {\bibinfo {volume} {107}},\ \bibinfo {pages} {162601} (\bibinfo
  {year} {2015})}\BibitemShut {NoStop}%
\bibitem [{\citenamefont {Martinis}\ \emph {et~al.}(2005)\citenamefont
  {Martinis}, \citenamefont {Cooper}, \citenamefont {McDermott}, \citenamefont
  {Steffen}, \citenamefont {Ansmann}, \citenamefont {Osborn}, \citenamefont
  {Cicak}, \citenamefont {Oh}, \citenamefont {Pappas}, \citenamefont {Simmonds}
  \emph {et~al.}}]{martinis2005decoherence}%
  \BibitemOpen
  \bibfield  {author} {\bibinfo {author} {\bibfnamefont {J.~M.}\ \bibnamefont
  {Martinis}}, \bibinfo {author} {\bibfnamefont {K.~B.}\ \bibnamefont
  {Cooper}}, \bibinfo {author} {\bibfnamefont {R.}~\bibnamefont {McDermott}},
  \bibinfo {author} {\bibfnamefont {M.}~\bibnamefont {Steffen}}, \bibinfo
  {author} {\bibfnamefont {M.}~\bibnamefont {Ansmann}}, \bibinfo {author}
  {\bibfnamefont {K.}~\bibnamefont {Osborn}}, \bibinfo {author} {\bibfnamefont
  {K.}~\bibnamefont {Cicak}}, \bibinfo {author} {\bibfnamefont
  {S.}~\bibnamefont {Oh}}, \bibinfo {author} {\bibfnamefont {D.~P.}\
  \bibnamefont {Pappas}}, \bibinfo {author} {\bibfnamefont {R.~W.}\
  \bibnamefont {Simmonds}}, \emph {et~al.},\ }\bibfield  {title} {\bibinfo
  {title} {Decoherence in josephson qubits from dielectric loss},\ }\href
  {https://journals.aps.org/prl/abstract/10.1103/PhysRevLett.95.210503}
  {\bibfield  {journal} {\bibinfo  {journal} {Physical review letters}\
  }\textbf {\bibinfo {volume} {95}},\ \bibinfo {pages} {210503} (\bibinfo
  {year} {2005})}\BibitemShut {NoStop}%
\bibitem [{\citenamefont {Grabovskij}\ \emph {et~al.}(2012)\citenamefont
  {Grabovskij}, \citenamefont {Peichl}, \citenamefont {Lisenfeld},
  \citenamefont {Weiss},\ and\ \citenamefont {Ustinov}}]{grabovskij2012strain}%
  \BibitemOpen
  \bibfield  {author} {\bibinfo {author} {\bibfnamefont {G.~J.}\ \bibnamefont
  {Grabovskij}}, \bibinfo {author} {\bibfnamefont {T.}~\bibnamefont {Peichl}},
  \bibinfo {author} {\bibfnamefont {J.}~\bibnamefont {Lisenfeld}}, \bibinfo
  {author} {\bibfnamefont {G.}~\bibnamefont {Weiss}},\ and\ \bibinfo {author}
  {\bibfnamefont {A.~V.}\ \bibnamefont {Ustinov}},\ }\bibfield  {title}
  {\bibinfo {title} {Strain tuning of individual atomic tunneling systems
  detected by a superconducting qubit},\ }\href
  {https://www.science.org/doi/full/10.1126/science.1226487} {\bibfield
  {journal} {\bibinfo  {journal} {Science}\ }\textbf {\bibinfo {volume}
  {338}},\ \bibinfo {pages} {232} (\bibinfo {year} {2012})}\BibitemShut
  {NoStop}%
\bibitem [{\citenamefont {Lisenfeld}\ \emph {et~al.}(2019)\citenamefont
  {Lisenfeld}, \citenamefont {Bilmes}, \citenamefont {Megrant}, \citenamefont
  {Barends}, \citenamefont {Kelly}, \citenamefont {Klimov}, \citenamefont
  {Weiss}, \citenamefont {Martinis},\ and\ \citenamefont
  {Ustinov}}]{lisenfeld2019electric}%
  \BibitemOpen
  \bibfield  {author} {\bibinfo {author} {\bibfnamefont {J.}~\bibnamefont
  {Lisenfeld}}, \bibinfo {author} {\bibfnamefont {A.}~\bibnamefont {Bilmes}},
  \bibinfo {author} {\bibfnamefont {A.}~\bibnamefont {Megrant}}, \bibinfo
  {author} {\bibfnamefont {R.}~\bibnamefont {Barends}}, \bibinfo {author}
  {\bibfnamefont {J.}~\bibnamefont {Kelly}}, \bibinfo {author} {\bibfnamefont
  {P.}~\bibnamefont {Klimov}}, \bibinfo {author} {\bibfnamefont
  {G.}~\bibnamefont {Weiss}}, \bibinfo {author} {\bibfnamefont {J.~M.}\
  \bibnamefont {Martinis}},\ and\ \bibinfo {author} {\bibfnamefont {A.~V.}\
  \bibnamefont {Ustinov}},\ }\bibfield  {title} {\bibinfo {title} {Electric
  field spectroscopy of material defects in transmon qubits},\ }\href
  {https://www.nature.com/articles/s41534-019-0224-1} {\bibfield  {journal}
  {\bibinfo  {journal} {npj Quantum Information}\ }\textbf {\bibinfo {volume}
  {5}},\ \bibinfo {pages} {105} (\bibinfo {year} {2019})}\BibitemShut {NoStop}%
\bibitem [{\citenamefont {Wang}\ \emph {et~al.}(2014)\citenamefont {Wang},
  \citenamefont {Gao}, \citenamefont {Pop}, \citenamefont {Vool}, \citenamefont
  {Axline}, \citenamefont {Brecht}, \citenamefont {Heeres}, \citenamefont
  {Frunzio}, \citenamefont {Devoret}, \citenamefont {Catelani} \emph
  {et~al.}}]{wang2014measurement}%
  \BibitemOpen
  \bibfield  {author} {\bibinfo {author} {\bibfnamefont {C.}~\bibnamefont
  {Wang}}, \bibinfo {author} {\bibfnamefont {Y.~Y.}\ \bibnamefont {Gao}},
  \bibinfo {author} {\bibfnamefont {I.~M.}\ \bibnamefont {Pop}}, \bibinfo
  {author} {\bibfnamefont {U.}~\bibnamefont {Vool}}, \bibinfo {author}
  {\bibfnamefont {C.}~\bibnamefont {Axline}}, \bibinfo {author} {\bibfnamefont
  {T.}~\bibnamefont {Brecht}}, \bibinfo {author} {\bibfnamefont {R.~W.}\
  \bibnamefont {Heeres}}, \bibinfo {author} {\bibfnamefont {L.}~\bibnamefont
  {Frunzio}}, \bibinfo {author} {\bibfnamefont {M.~H.}\ \bibnamefont
  {Devoret}}, \bibinfo {author} {\bibfnamefont {G.}~\bibnamefont {Catelani}},
  \emph {et~al.},\ }\bibfield  {title} {\bibinfo {title} {Measurement and
  control of quasiparticle dynamics in a superconducting qubit},\ }\href
  {https://www.nature.com/articles/ncomms6836} {\bibfield  {journal} {\bibinfo
  {journal} {Nature communications}\ }\textbf {\bibinfo {volume} {5}},\
  \bibinfo {pages} {5836} (\bibinfo {year} {2014})}\BibitemShut {NoStop}%
\bibitem [{\citenamefont {Serniak}\ \emph {et~al.}(2018)\citenamefont
  {Serniak}, \citenamefont {Hays}, \citenamefont {De~Lange}, \citenamefont
  {Diamond}, \citenamefont {Shankar}, \citenamefont {Burkhart}, \citenamefont
  {Frunzio}, \citenamefont {Houzet},\ and\ \citenamefont
  {Devoret}}]{serniak2018hot}%
  \BibitemOpen
  \bibfield  {author} {\bibinfo {author} {\bibfnamefont {K.}~\bibnamefont
  {Serniak}}, \bibinfo {author} {\bibfnamefont {M.}~\bibnamefont {Hays}},
  \bibinfo {author} {\bibfnamefont {G.}~\bibnamefont {De~Lange}}, \bibinfo
  {author} {\bibfnamefont {S.}~\bibnamefont {Diamond}}, \bibinfo {author}
  {\bibfnamefont {S.}~\bibnamefont {Shankar}}, \bibinfo {author} {\bibfnamefont
  {L.}~\bibnamefont {Burkhart}}, \bibinfo {author} {\bibfnamefont
  {L.}~\bibnamefont {Frunzio}}, \bibinfo {author} {\bibfnamefont
  {M.}~\bibnamefont {Houzet}},\ and\ \bibinfo {author} {\bibfnamefont
  {M.}~\bibnamefont {Devoret}},\ }\bibfield  {title} {\bibinfo {title} {Hot
  nonequilibrium quasiparticles in transmon qubits},\ }\href
  {https://journals.aps.org/prl/abstract/10.1103/PhysRevLett.121.157701}
  {\bibfield  {journal} {\bibinfo  {journal} {Physical review letters}\
  }\textbf {\bibinfo {volume} {121}},\ \bibinfo {pages} {157701} (\bibinfo
  {year} {2018})}\BibitemShut {NoStop}%
\bibitem [{\citenamefont {Cardani}\ \emph {et~al.}(2021)\citenamefont
  {Cardani}, \citenamefont {Valenti}, \citenamefont {Casali}, \citenamefont
  {Catelani}, \citenamefont {Charpentier}, \citenamefont {Clemenza},
  \citenamefont {Colantoni}, \citenamefont {Cruciani}, \citenamefont
  {D’Imperio}, \citenamefont {Gironi} \emph {et~al.}}]{cardani2021reducing}%
  \BibitemOpen
  \bibfield  {author} {\bibinfo {author} {\bibfnamefont {L.}~\bibnamefont
  {Cardani}}, \bibinfo {author} {\bibfnamefont {F.}~\bibnamefont {Valenti}},
  \bibinfo {author} {\bibfnamefont {N.}~\bibnamefont {Casali}}, \bibinfo
  {author} {\bibfnamefont {G.}~\bibnamefont {Catelani}}, \bibinfo {author}
  {\bibfnamefont {T.}~\bibnamefont {Charpentier}}, \bibinfo {author}
  {\bibfnamefont {M.}~\bibnamefont {Clemenza}}, \bibinfo {author}
  {\bibfnamefont {I.}~\bibnamefont {Colantoni}}, \bibinfo {author}
  {\bibfnamefont {A.}~\bibnamefont {Cruciani}}, \bibinfo {author}
  {\bibfnamefont {G.}~\bibnamefont {D’Imperio}}, \bibinfo {author}
  {\bibfnamefont {L.}~\bibnamefont {Gironi}}, \emph {et~al.},\ }\bibfield
  {title} {\bibinfo {title} {Reducing the impact of radioactivity on quantum
  circuits in a deep-underground facility},\ }\href
  {https://www.nature.com/articles/s41467-021-23032-z} {\bibfield  {journal}
  {\bibinfo  {journal} {Nature communications}\ }\textbf {\bibinfo {volume}
  {12}},\ \bibinfo {pages} {2733} (\bibinfo {year} {2021})}\BibitemShut
  {NoStop}%
\bibitem [{\citenamefont {Veps{\"a}l{\"a}inen}\ \emph
  {et~al.}(2020)\citenamefont {Veps{\"a}l{\"a}inen}, \citenamefont {Karamlou},
  \citenamefont {Orrell}, \citenamefont {Dogra}, \citenamefont {Loer},
  \citenamefont {Vasconcelos}, \citenamefont {Kim}, \citenamefont {Melville},
  \citenamefont {Niedzielski}, \citenamefont {Yoder} \emph
  {et~al.}}]{vepsalainen2020impact}%
  \BibitemOpen
  \bibfield  {author} {\bibinfo {author} {\bibfnamefont {A.~P.}\ \bibnamefont
  {Veps{\"a}l{\"a}inen}}, \bibinfo {author} {\bibfnamefont {A.~H.}\
  \bibnamefont {Karamlou}}, \bibinfo {author} {\bibfnamefont {J.~L.}\
  \bibnamefont {Orrell}}, \bibinfo {author} {\bibfnamefont {A.~S.}\
  \bibnamefont {Dogra}}, \bibinfo {author} {\bibfnamefont {B.}~\bibnamefont
  {Loer}}, \bibinfo {author} {\bibfnamefont {F.}~\bibnamefont {Vasconcelos}},
  \bibinfo {author} {\bibfnamefont {D.~K.}\ \bibnamefont {Kim}}, \bibinfo
  {author} {\bibfnamefont {A.~J.}\ \bibnamefont {Melville}}, \bibinfo {author}
  {\bibfnamefont {B.~M.}\ \bibnamefont {Niedzielski}}, \bibinfo {author}
  {\bibfnamefont {J.~L.}\ \bibnamefont {Yoder}}, \emph {et~al.},\ }\bibfield
  {title} {\bibinfo {title} {Impact of ionizing radiation on superconducting
  qubit coherence},\ }\href {https://www.nature.com/articles/s41586-020-2619-8}
  {\bibfield  {journal} {\bibinfo  {journal} {Nature}\ }\textbf {\bibinfo
  {volume} {584}},\ \bibinfo {pages} {551} (\bibinfo {year}
  {2020})}\BibitemShut {NoStop}%
\bibitem [{\citenamefont {Thorbeck}\ \emph {et~al.}(2022)\citenamefont
  {Thorbeck}, \citenamefont {Eddins}, \citenamefont {Lauer}, \citenamefont
  {McClure},\ and\ \citenamefont {Carroll}}]{thorbeck_tls_2022}%
  \BibitemOpen
  \bibfield  {author} {\bibinfo {author} {\bibfnamefont {T.}~\bibnamefont
  {Thorbeck}}, \bibinfo {author} {\bibfnamefont {A.}~\bibnamefont {Eddins}},
  \bibinfo {author} {\bibfnamefont {I.}~\bibnamefont {Lauer}}, \bibinfo
  {author} {\bibfnamefont {D.~T.}\ \bibnamefont {McClure}},\ and\ \bibinfo
  {author} {\bibfnamefont {M.}~\bibnamefont {Carroll}},\ }\href
  {https://arxiv.org/abs/2210.04780} {\bibinfo {title} {{TLS} {Dynamics} in a
  {Superconducting} {Qubit} {Due} to {Background} {Ionizing} {Radiation}}}
  (\bibinfo {year} {2022}),\ \bibinfo {note} {arXiv:2210.04780 [cond-mat,
  physics:quant-ph]}\BibitemShut {NoStop}%
\bibitem [{\citenamefont {Olivieri}\ \emph {et~al.}(2017)\citenamefont
  {Olivieri}, \citenamefont {Billard}, \citenamefont {De~Jesus}, \citenamefont
  {Juillard},\ and\ \citenamefont {Leder}}]{olivieri2017vibrations}%
  \BibitemOpen
  \bibfield  {author} {\bibinfo {author} {\bibfnamefont {E.}~\bibnamefont
  {Olivieri}}, \bibinfo {author} {\bibfnamefont {J.}~\bibnamefont {Billard}},
  \bibinfo {author} {\bibfnamefont {M.}~\bibnamefont {De~Jesus}}, \bibinfo
  {author} {\bibfnamefont {A.}~\bibnamefont {Juillard}},\ and\ \bibinfo
  {author} {\bibfnamefont {A.}~\bibnamefont {Leder}},\ }\bibfield  {title}
  {\bibinfo {title} {Vibrations on pulse tube based dry dilution refrigerators
  for low noise measurements},\ }\href
  {https://www.sciencedirect.com/science/article/pii/S0168900217303984}
  {\bibfield  {journal} {\bibinfo  {journal} {Nuclear Instruments and Methods
  in Physics Research Section A: Accelerators, Spectrometers, Detectors and
  Associated Equipment}\ }\textbf {\bibinfo {volume} {858}},\ \bibinfo {pages}
  {73} (\bibinfo {year} {2017})}\BibitemShut {NoStop}%
\bibitem [{\citenamefont {M{\"u}ller}\ \emph {et~al.}(2019)\citenamefont
  {M{\"u}ller}, \citenamefont {Cole},\ and\ \citenamefont
  {Lisenfeld}}]{muller2019towards}%
  \BibitemOpen
  \bibfield  {author} {\bibinfo {author} {\bibfnamefont {C.}~\bibnamefont
  {M{\"u}ller}}, \bibinfo {author} {\bibfnamefont {J.~H.}\ \bibnamefont
  {Cole}},\ and\ \bibinfo {author} {\bibfnamefont {J.}~\bibnamefont
  {Lisenfeld}},\ }\bibfield  {title} {\bibinfo {title} {Towards understanding
  two-level-systems in amorphous solids: insights from quantum circuits},\
  }\href {https://iopscience.iop.org/article/10.1088/1361-6633/ab3a7e/meta}
  {\bibfield  {journal} {\bibinfo  {journal} {Reports on Progress in Physics}\
  }\textbf {\bibinfo {volume} {82}},\ \bibinfo {pages} {124501} (\bibinfo
  {year} {2019})}\BibitemShut {NoStop}%
\bibitem [{\citenamefont {Glazman}\ and\ \citenamefont
  {Catelani}(2021)}]{glazman2021bogoliubov}%
  \BibitemOpen
  \bibfield  {author} {\bibinfo {author} {\bibfnamefont {L.}~\bibnamefont
  {Glazman}}\ and\ \bibinfo {author} {\bibfnamefont {G.}~\bibnamefont
  {Catelani}},\ }\bibfield  {title} {\bibinfo {title} {Bogoliubov
  quasiparticles in superconducting qubits},\ }\href
  {https://www.scipost.org/10.21468/SciPostPhysLectNotes.31?acad_field_slug=physics}
  {\bibfield  {journal} {\bibinfo  {journal} {SciPost Physics Lecture Notes}\
  ,\ \bibinfo {pages} {031}} (\bibinfo {year} {2021})}\BibitemShut {NoStop}%
\bibitem [{\citenamefont {Rosen}\ \emph {et~al.}(2019)\citenamefont {Rosen},
  \citenamefont {Horsley}, \citenamefont {Harrison}, \citenamefont {Holland},
  \citenamefont {Chang}, \citenamefont {Bond},\ and\ \citenamefont
  {DuBois}}]{rosen2019protecting}%
  \BibitemOpen
  \bibfield  {author} {\bibinfo {author} {\bibfnamefont {Y.~J.}\ \bibnamefont
  {Rosen}}, \bibinfo {author} {\bibfnamefont {M.~A.}\ \bibnamefont {Horsley}},
  \bibinfo {author} {\bibfnamefont {S.~E.}\ \bibnamefont {Harrison}}, \bibinfo
  {author} {\bibfnamefont {E.~T.}\ \bibnamefont {Holland}}, \bibinfo {author}
  {\bibfnamefont {A.~S.}\ \bibnamefont {Chang}}, \bibinfo {author}
  {\bibfnamefont {T.}~\bibnamefont {Bond}},\ and\ \bibinfo {author}
  {\bibfnamefont {J.~L.}\ \bibnamefont {DuBois}},\ }\bibfield  {title}
  {\bibinfo {title} {Protecting superconducting qubits from phonon mediated
  decay},\ }\href
  {https://pubs.aip.org/aip/apl/article/114/20/202601/36775/Protecting-superconducting-qubits-from-phonon}
  {\bibfield  {journal} {\bibinfo  {journal} {Applied Physics Letters}\
  }\textbf {\bibinfo {volume} {114}},\ \bibinfo {pages} {202601} (\bibinfo
  {year} {2019})}\BibitemShut {NoStop}%
\bibitem [{\citenamefont {Pirro}\ \emph {et~al.}(2000)\citenamefont {Pirro},
  \citenamefont {Alessandrello}, \citenamefont {Brofferio}, \citenamefont
  {Bucci}, \citenamefont {Cremonesi}, \citenamefont {Coccia}, \citenamefont
  {Fiorini}, \citenamefont {Fafone}, \citenamefont {Giuliani}, \citenamefont
  {Nucciotti}, \citenamefont {Pavan}, \citenamefont {Pessina}, \citenamefont
  {Previtali}, \citenamefont {Vanzini},\ and\ \citenamefont
  {Zanotti}}]{PIRRO2000331}%
  \BibitemOpen
  \bibfield  {author} {\bibinfo {author} {\bibfnamefont {S.}~\bibnamefont
  {Pirro}}, \bibinfo {author} {\bibfnamefont {A.}~\bibnamefont
  {Alessandrello}}, \bibinfo {author} {\bibfnamefont {C.}~\bibnamefont
  {Brofferio}}, \bibinfo {author} {\bibfnamefont {C.}~\bibnamefont {Bucci}},
  \bibinfo {author} {\bibfnamefont {O.}~\bibnamefont {Cremonesi}}, \bibinfo
  {author} {\bibfnamefont {E.}~\bibnamefont {Coccia}}, \bibinfo {author}
  {\bibfnamefont {E.}~\bibnamefont {Fiorini}}, \bibinfo {author} {\bibfnamefont
  {V.}~\bibnamefont {Fafone}}, \bibinfo {author} {\bibfnamefont
  {A.}~\bibnamefont {Giuliani}}, \bibinfo {author} {\bibfnamefont
  {A.}~\bibnamefont {Nucciotti}}, \bibinfo {author} {\bibfnamefont
  {M.}~\bibnamefont {Pavan}}, \bibinfo {author} {\bibfnamefont
  {G.}~\bibnamefont {Pessina}}, \bibinfo {author} {\bibfnamefont
  {E.}~\bibnamefont {Previtali}}, \bibinfo {author} {\bibfnamefont
  {M.}~\bibnamefont {Vanzini}},\ and\ \bibinfo {author} {\bibfnamefont
  {L.}~\bibnamefont {Zanotti}},\ }\bibfield  {title} {\bibinfo {title}
  {Vibrational and thermal noise reduction for cryogenic detectors},\ }\href
  {https://www.sciencedirect.com/science/article/pii/S0168900299013765}
  {\bibfield  {journal} {\bibinfo  {journal} {Nuclear Instruments and Methods
  in Physics Research Section A: Accelerators, Spectrometers, Detectors and
  Associated Equipment}\ }\textbf {\bibinfo {volume} {444}},\ \bibinfo {pages}
  {331} (\bibinfo {year} {2000})}\BibitemShut {NoStop}%
\bibitem [{\citenamefont {Maisonobe}\ \emph {et~al.}(2018)\citenamefont
  {Maisonobe}, \citenamefont {Billard}, \citenamefont {De~Jesus}, \citenamefont
  {Juillard}, \citenamefont {Misiak}, \citenamefont {Olivieri}, \citenamefont
  {Sayah},\ and\ \citenamefont {Vagneron}}]{maisonobe2018vibration}%
  \BibitemOpen
  \bibfield  {author} {\bibinfo {author} {\bibfnamefont {R.}~\bibnamefont
  {Maisonobe}}, \bibinfo {author} {\bibfnamefont {J.}~\bibnamefont {Billard}},
  \bibinfo {author} {\bibfnamefont {M.}~\bibnamefont {De~Jesus}}, \bibinfo
  {author} {\bibfnamefont {A.}~\bibnamefont {Juillard}}, \bibinfo {author}
  {\bibfnamefont {D.}~\bibnamefont {Misiak}}, \bibinfo {author} {\bibfnamefont
  {E.}~\bibnamefont {Olivieri}}, \bibinfo {author} {\bibfnamefont
  {S.}~\bibnamefont {Sayah}},\ and\ \bibinfo {author} {\bibfnamefont
  {L.}~\bibnamefont {Vagneron}},\ }\bibfield  {title} {\bibinfo {title}
  {Vibration decoupling system for massive bolometers in dry cryostats},\
  }\href
  {https://iopscience.iop.org/article/10.1088/1748-0221/13/08/T08009/meta}
  {\bibfield  {journal} {\bibinfo  {journal} {Journal of Instrumentation}\
  }\textbf {\bibinfo {volume} {13}}\bibinfo  {number} { (08)},\ \bibinfo
  {pages} {T08009}}\BibitemShut {NoStop}%
\bibitem [{\citenamefont {Kalra}\ \emph {et~al.}(2016)\citenamefont {Kalra},
  \citenamefont {Laucht}, \citenamefont {Dehollain}, \citenamefont {Bar},
  \citenamefont {Freer}, \citenamefont {Simmons}, \citenamefont {Muhonen},\
  and\ \citenamefont {Morello}}]{Kalra2016-ey}%
  \BibitemOpen
\bibfield  {number} {  }\bibfield  {author} {\bibinfo {author} {\bibfnamefont
  {R.}~\bibnamefont {Kalra}}, \bibinfo {author} {\bibfnamefont
  {A.}~\bibnamefont {Laucht}}, \bibinfo {author} {\bibfnamefont {J.~P.}\
  \bibnamefont {Dehollain}}, \bibinfo {author} {\bibfnamefont {D.}~\bibnamefont
  {Bar}}, \bibinfo {author} {\bibfnamefont {S.}~\bibnamefont {Freer}}, \bibinfo
  {author} {\bibfnamefont {S.}~\bibnamefont {Simmons}}, \bibinfo {author}
  {\bibfnamefont {J.~T.}\ \bibnamefont {Muhonen}},\ and\ \bibinfo {author}
  {\bibfnamefont {A.}~\bibnamefont {Morello}},\ }\bibfield  {title} {\bibinfo
  {title} {Vibration-induced electrical noise in a cryogen-free dilution
  refrigerator: {Characterization}, mitigation, and impact on qubit
  coherence},\ }\href
  {https://pubs.aip.org/aip/rsi/article/87/7/073905/357812/Vibration-induced-electrical-noise-in-a-cryogen}
  {\bibfield  {journal} {\bibinfo  {journal} {Rev. Sci. Instrum.}\ }\textbf
  {\bibinfo {volume} {87}},\ \bibinfo {pages} {073905} (\bibinfo {year}
  {2016})}\BibitemShut {NoStop}%
\bibitem [{\citenamefont {Cao}(2022)}]{cao2022vibration}%
  \BibitemOpen
  \bibfield  {author} {\bibinfo {author} {\bibfnamefont {H.}~\bibnamefont
  {Cao}},\ }\bibfield  {title} {\bibinfo {title} {Vibration control for
  mechanical cryocoolers},\ }\href
  {https://www.sciencedirect.com/science/article/pii/S0011227522001771}
  {\bibfield  {journal} {\bibinfo  {journal} {Cryogenics}\ ,\ \bibinfo {pages}
  {103595}} (\bibinfo {year} {2022})}\BibitemShut {NoStop}%
\bibitem [{\citenamefont {Uhlig}(2023)}]{uhlig2023dry}%
  \BibitemOpen
  \bibfield  {author} {\bibinfo {author} {\bibfnamefont {K.}~\bibnamefont
  {Uhlig}},\ }\bibfield  {title} {\bibinfo {title} {Dry dilution refrigerator
  with pulse tube shutoff option},\ }\href
  {https://www.sciencedirect.com/science/article/pii/S0011227523000231}
  {\bibfield  {journal} {\bibinfo  {journal} {Cryogenics}\ ,\ \bibinfo {pages}
  {103649}} (\bibinfo {year} {2023})}\BibitemShut {NoStop}%
\bibitem [{\citenamefont {Osman}\ \emph {et~al.}(2021)\citenamefont {Osman},
  \citenamefont {Simon}, \citenamefont {Bengtsson}, \citenamefont {Kosen},
  \citenamefont {Krantz}, \citenamefont {P.~Lozano}, \citenamefont
  {Scigliuzzo}, \citenamefont {Delsing}, \citenamefont {Bylander},\ and\
  \citenamefont {Fadavi~Roudsari}}]{osman2021simplified}%
  \BibitemOpen
  \bibfield  {author} {\bibinfo {author} {\bibfnamefont {A.}~\bibnamefont
  {Osman}}, \bibinfo {author} {\bibfnamefont {J.}~\bibnamefont {Simon}},
  \bibinfo {author} {\bibfnamefont {A.}~\bibnamefont {Bengtsson}}, \bibinfo
  {author} {\bibfnamefont {S.}~\bibnamefont {Kosen}}, \bibinfo {author}
  {\bibfnamefont {P.}~\bibnamefont {Krantz}}, \bibinfo {author} {\bibfnamefont
  {D.}~\bibnamefont {P.~Lozano}}, \bibinfo {author} {\bibfnamefont
  {M.}~\bibnamefont {Scigliuzzo}}, \bibinfo {author} {\bibfnamefont
  {P.}~\bibnamefont {Delsing}}, \bibinfo {author} {\bibfnamefont
  {J.}~\bibnamefont {Bylander}},\ and\ \bibinfo {author} {\bibfnamefont
  {A.}~\bibnamefont {Fadavi~Roudsari}},\ }\bibfield  {title} {\bibinfo {title}
  {Simplified {Josephson}-junction fabrication process for reproducibly
  high-performance superconducting qubits},\ }\href
  {https://pubs.aip.org/aip/apl/article/118/6/064002/40060/Simplified-Josephson-junction-fabrication-process}
  {\bibfield  {journal} {\bibinfo  {journal} {Applied Physics Letters}\
  }\textbf {\bibinfo {volume} {118}},\ \bibinfo {pages} {064002} (\bibinfo
  {year} {2021})}\BibitemShut {NoStop}%
\bibitem [{\citenamefont {Jeffrey}\ \emph {et~al.}(2014)\citenamefont
  {Jeffrey}, \citenamefont {Sank}, \citenamefont {Mutus}, \citenamefont
  {White}, \citenamefont {Kelly}, \citenamefont {Barends}, \citenamefont
  {Chen}, \citenamefont {Chen}, \citenamefont {Chiaro}, \citenamefont
  {Dunsworth} \emph {et~al.}}]{jeffrey2014fast}%
  \BibitemOpen
  \bibfield  {author} {\bibinfo {author} {\bibfnamefont {E.}~\bibnamefont
  {Jeffrey}}, \bibinfo {author} {\bibfnamefont {D.}~\bibnamefont {Sank}},
  \bibinfo {author} {\bibfnamefont {J.}~\bibnamefont {Mutus}}, \bibinfo
  {author} {\bibfnamefont {T.}~\bibnamefont {White}}, \bibinfo {author}
  {\bibfnamefont {J.}~\bibnamefont {Kelly}}, \bibinfo {author} {\bibfnamefont
  {R.}~\bibnamefont {Barends}}, \bibinfo {author} {\bibfnamefont
  {Y.}~\bibnamefont {Chen}}, \bibinfo {author} {\bibfnamefont {Z.}~\bibnamefont
  {Chen}}, \bibinfo {author} {\bibfnamefont {B.}~\bibnamefont {Chiaro}},
  \bibinfo {author} {\bibfnamefont {A.}~\bibnamefont {Dunsworth}}, \emph
  {et~al.},\ }\bibfield  {title} {\bibinfo {title} {Fast accurate state
  measurement with superconducting qubits},\ }\href
  {https://journals.aps.org/prl/abstract/10.1103/PhysRevLett.112.190504}
  {\bibfield  {journal} {\bibinfo  {journal} {Physical review letters}\
  }\textbf {\bibinfo {volume} {112}},\ \bibinfo {pages} {190504} (\bibinfo
  {year} {2014})}\BibitemShut {NoStop}%
\bibitem [{\citenamefont {Macklin}\ \emph {et~al.}(2015)\citenamefont
  {Macklin}, \citenamefont {O’brien}, \citenamefont {Hover}, \citenamefont
  {Schwartz}, \citenamefont {Bolkhovsky}, \citenamefont {Zhang}, \citenamefont
  {Oliver},\ and\ \citenamefont {Siddiqi}}]{macklin2015near}%
  \BibitemOpen
  \bibfield  {author} {\bibinfo {author} {\bibfnamefont {C.}~\bibnamefont
  {Macklin}}, \bibinfo {author} {\bibfnamefont {K.}~\bibnamefont {O’brien}},
  \bibinfo {author} {\bibfnamefont {D.}~\bibnamefont {Hover}}, \bibinfo
  {author} {\bibfnamefont {M.}~\bibnamefont {Schwartz}}, \bibinfo {author}
  {\bibfnamefont {V.}~\bibnamefont {Bolkhovsky}}, \bibinfo {author}
  {\bibfnamefont {X.}~\bibnamefont {Zhang}}, \bibinfo {author} {\bibfnamefont
  {W.}~\bibnamefont {Oliver}},\ and\ \bibinfo {author} {\bibfnamefont
  {I.}~\bibnamefont {Siddiqi}},\ }\bibfield  {title} {\bibinfo {title} {A
  near--quantum-limited {Josephson} traveling-wave parametric amplifier},\
  }\href {https://www.science.org/doi/full/10.1126/science.aaa8525} {\bibfield
  {journal} {\bibinfo  {journal} {Science}\ }\textbf {\bibinfo {volume}
  {350}},\ \bibinfo {pages} {307} (\bibinfo {year} {2015})}\BibitemShut
  {NoStop}%
\bibitem [{\citenamefont {Vijay}\ \emph {et~al.}(2011)\citenamefont {Vijay},
  \citenamefont {Slichter},\ and\ \citenamefont
  {Siddiqi}}]{vijay2011observation}%
  \BibitemOpen
  \bibfield  {author} {\bibinfo {author} {\bibfnamefont {R.}~\bibnamefont
  {Vijay}}, \bibinfo {author} {\bibfnamefont {D.}~\bibnamefont {Slichter}},\
  and\ \bibinfo {author} {\bibfnamefont {I.}~\bibnamefont {Siddiqi}},\
  }\bibfield  {title} {\bibinfo {title} {Observation of quantum jumps in a
  superconducting artificial atom},\ }\href
  {https://journals.aps.org/prl/abstract/10.1103/PhysRevLett.106.110502}
  {\bibfield  {journal} {\bibinfo  {journal} {Physical review letters}\
  }\textbf {\bibinfo {volume} {106}},\ \bibinfo {pages} {110502} (\bibinfo
  {year} {2011})}\BibitemShut {NoStop}%
\bibitem [{\citenamefont {Youssefi}\ \emph {et~al.}(2022)\citenamefont
  {Youssefi}, \citenamefont {Kono}, \citenamefont {Chegnizadeh},\ and\
  \citenamefont {Kippenberg}}]{youssefi2022squeezed}%
  \BibitemOpen
  \bibfield  {author} {\bibinfo {author} {\bibfnamefont {A.}~\bibnamefont
  {Youssefi}}, \bibinfo {author} {\bibfnamefont {S.}~\bibnamefont {Kono}},
  \bibinfo {author} {\bibfnamefont {M.}~\bibnamefont {Chegnizadeh}},\ and\
  \bibinfo {author} {\bibfnamefont {T.~J.}\ \bibnamefont {Kippenberg}},\
  }\bibfield  {title} {\bibinfo {title} {A squeezed mechanical oscillator with
  milli-second quantum decoherence},\ }\href {https://arxiv.org/abs/2208.13082}
  {\bibfield  {journal} {\bibinfo  {journal} {arXiv preprint arXiv:2208.13082}\
  } (\bibinfo {year} {2022})}\BibitemShut {NoStop}%
\bibitem [{\citenamefont {C{\'o}rcoles}\ \emph {et~al.}(2011)\citenamefont
  {C{\'o}rcoles}, \citenamefont {Chow}, \citenamefont {Gambetta}, \citenamefont
  {Rigetti}, \citenamefont {Rozen}, \citenamefont {Keefe}, \citenamefont
  {Beth~Rothwell}, \citenamefont {Ketchen},\ and\ \citenamefont
  {Steffen}}]{corcoles2011protecting}%
  \BibitemOpen
  \bibfield  {author} {\bibinfo {author} {\bibfnamefont {A.~D.}\ \bibnamefont
  {C{\'o}rcoles}}, \bibinfo {author} {\bibfnamefont {J.~M.}\ \bibnamefont
  {Chow}}, \bibinfo {author} {\bibfnamefont {J.~M.}\ \bibnamefont {Gambetta}},
  \bibinfo {author} {\bibfnamefont {C.}~\bibnamefont {Rigetti}}, \bibinfo
  {author} {\bibfnamefont {J.~R.}\ \bibnamefont {Rozen}}, \bibinfo {author}
  {\bibfnamefont {G.~A.}\ \bibnamefont {Keefe}}, \bibinfo {author}
  {\bibfnamefont {M.}~\bibnamefont {Beth~Rothwell}}, \bibinfo {author}
  {\bibfnamefont {M.~B.}\ \bibnamefont {Ketchen}},\ and\ \bibinfo {author}
  {\bibfnamefont {M.}~\bibnamefont {Steffen}},\ }\bibfield  {title} {\bibinfo
  {title} {Protecting superconducting qubits from radiation},\ }\href
  {https://pubs.aip.org/aip/apl/article/99/18/181906/341405/Protecting-superconducting-qubits-from-radiation}
  {\bibfield  {journal} {\bibinfo  {journal} {Applied Physics Letters}\
  }\textbf {\bibinfo {volume} {99}},\ \bibinfo {pages} {181906} (\bibinfo
  {year} {2011})}\BibitemShut {NoStop}%
\bibitem [{\citenamefont {Martinis}(2021)}]{martinis2021saving}%
  \BibitemOpen
  \bibfield  {author} {\bibinfo {author} {\bibfnamefont {J.~M.}\ \bibnamefont
  {Martinis}},\ }\bibfield  {title} {\bibinfo {title} {Saving superconducting
  quantum processors from decay and correlated errors generated by gamma and
  cosmic rays},\ }\href {https://www.nature.com/articles/s41534-021-00431-0}
  {\bibfield  {journal} {\bibinfo  {journal} {npj Quantum Information}\
  }\textbf {\bibinfo {volume} {7}},\ \bibinfo {pages} {90} (\bibinfo {year}
  {2021})}\BibitemShut {NoStop}%
\bibitem [{\citenamefont {Lisenfeld}\ \emph {et~al.}(2015)\citenamefont
  {Lisenfeld}, \citenamefont {Grabovskij}, \citenamefont {M{\"u}ller},
  \citenamefont {Cole}, \citenamefont {Weiss},\ and\ \citenamefont
  {Ustinov}}]{lisenfeld2015observation}%
  \BibitemOpen
  \bibfield  {author} {\bibinfo {author} {\bibfnamefont {J.}~\bibnamefont
  {Lisenfeld}}, \bibinfo {author} {\bibfnamefont {G.~J.}\ \bibnamefont
  {Grabovskij}}, \bibinfo {author} {\bibfnamefont {C.}~\bibnamefont
  {M{\"u}ller}}, \bibinfo {author} {\bibfnamefont {J.~H.}\ \bibnamefont
  {Cole}}, \bibinfo {author} {\bibfnamefont {G.}~\bibnamefont {Weiss}},\ and\
  \bibinfo {author} {\bibfnamefont {A.~V.}\ \bibnamefont {Ustinov}},\
  }\bibfield  {title} {\bibinfo {title} {Observation of directly interacting
  coherent two-level systems in an amorphous material},\ }\href
  {https://www.nature.com/articles/ncomms7182} {\bibfield  {journal} {\bibinfo
  {journal} {Nature communications}\ }\textbf {\bibinfo {volume} {6}},\
  \bibinfo {pages} {6182} (\bibinfo {year} {2015})}\BibitemShut {NoStop}%
\bibitem [{\citenamefont {Andersson}\ \emph {et~al.}(2021)\citenamefont
  {Andersson}, \citenamefont {Bilobran}, \citenamefont {Scigliuzzo},
  \citenamefont {de~Lima}, \citenamefont {Cole},\ and\ \citenamefont
  {Delsing}}]{andersson2021acoustic}%
  \BibitemOpen
  \bibfield  {author} {\bibinfo {author} {\bibfnamefont {G.}~\bibnamefont
  {Andersson}}, \bibinfo {author} {\bibfnamefont {A.}~\bibnamefont {Bilobran}},
  \bibinfo {author} {\bibfnamefont {M.}~\bibnamefont {Scigliuzzo}}, \bibinfo
  {author} {\bibfnamefont {M.}~\bibnamefont {de~Lima}}, \bibinfo {author}
  {\bibfnamefont {J.}~\bibnamefont {Cole}},\ and\ \bibinfo {author}
  {\bibfnamefont {P.}~\bibnamefont {Delsing}},\ }\bibfield  {title} {\bibinfo
  {title} {Acoustic spectral hole-burning in a two-level system ensemble},\
  }\href {https://www.nature.com/articles/s41534-020-00348-0} {\bibfield
  {journal} {\bibinfo  {journal} {npj Quantum Information}\ }\textbf {\bibinfo
  {volume} {7}},\ \bibinfo {pages} {1} (\bibinfo {year} {2021})}\BibitemShut
  {NoStop}%
\bibitem [{\citenamefont {Kono}\ \emph {et~al.}(2020)\citenamefont {Kono},
  \citenamefont {Koshino}, \citenamefont {Lachance-Quirion}, \citenamefont
  {Van~Loo}, \citenamefont {Tabuchi}, \citenamefont {Noguchi},\ and\
  \citenamefont {Nakamura}}]{kono2020breaking}%
  \BibitemOpen
  \bibfield  {author} {\bibinfo {author} {\bibfnamefont {S.}~\bibnamefont
  {Kono}}, \bibinfo {author} {\bibfnamefont {K.}~\bibnamefont {Koshino}},
  \bibinfo {author} {\bibfnamefont {D.}~\bibnamefont {Lachance-Quirion}},
  \bibinfo {author} {\bibfnamefont {A.~F.}\ \bibnamefont {Van~Loo}}, \bibinfo
  {author} {\bibfnamefont {Y.}~\bibnamefont {Tabuchi}}, \bibinfo {author}
  {\bibfnamefont {A.}~\bibnamefont {Noguchi}},\ and\ \bibinfo {author}
  {\bibfnamefont {Y.}~\bibnamefont {Nakamura}},\ }\bibfield  {title} {\bibinfo
  {title} {Breaking the trade-off between fast control and long lifetime of a
  superconducting qubit},\ }\href
  {https://www.nature.com/articles/s41467-020-17511-y} {\bibfield  {journal}
  {\bibinfo  {journal} {Nature Communications}\ }\textbf {\bibinfo {volume}
  {11}},\ \bibinfo {pages} {3683} (\bibinfo {year} {2020})}\BibitemShut
  {NoStop}%
\bibitem [{\citenamefont {Lienhard}\ \emph {et~al.}(2019)\citenamefont
  {Lienhard}, \citenamefont {Braum{\"u}ller}, \citenamefont {Woods},
  \citenamefont {Rosenberg}, \citenamefont {Calusine}, \citenamefont {Weber},
  \citenamefont {Veps{\"a}l{\"a}inen}, \citenamefont {O'Brien}, \citenamefont
  {Orlando}, \citenamefont {Gustavsson} \emph
  {et~al.}}]{lienhard2019microwave}%
  \BibitemOpen
  \bibfield  {author} {\bibinfo {author} {\bibfnamefont {B.}~\bibnamefont
  {Lienhard}}, \bibinfo {author} {\bibfnamefont {J.}~\bibnamefont
  {Braum{\"u}ller}}, \bibinfo {author} {\bibfnamefont {W.}~\bibnamefont
  {Woods}}, \bibinfo {author} {\bibfnamefont {D.}~\bibnamefont {Rosenberg}},
  \bibinfo {author} {\bibfnamefont {G.}~\bibnamefont {Calusine}}, \bibinfo
  {author} {\bibfnamefont {S.}~\bibnamefont {Weber}}, \bibinfo {author}
  {\bibfnamefont {A.}~\bibnamefont {Veps{\"a}l{\"a}inen}}, \bibinfo {author}
  {\bibfnamefont {K.}~\bibnamefont {O'Brien}}, \bibinfo {author} {\bibfnamefont
  {T.~P.}\ \bibnamefont {Orlando}}, \bibinfo {author} {\bibfnamefont
  {S.}~\bibnamefont {Gustavsson}}, \emph {et~al.},\ }\bibfield  {title}
  {\bibinfo {title} {Microwave packaging for superconducting qubits},\ }in\
  \href {https://ieeexplore.ieee.org/abstract/document/8701119} {\emph
  {\bibinfo {booktitle} {2019 IEEE MTT-S International Microwave Symposium
  (IMS)}}}\ (\bibinfo {organization} {IEEE},\ \bibinfo {year} {2019})\ pp.\
  \bibinfo {pages} {275--278}\BibitemShut {NoStop}%
\bibitem [{\citenamefont {Kreikebaum}(2020)}]{kreikebaum2020superconducting}%
  \BibitemOpen
  \bibfield  {author} {\bibinfo {author} {\bibfnamefont {J.~M.}\ \bibnamefont
  {Kreikebaum}},\ }\href@noop {} {\emph {\bibinfo {title} {Superconducting
  Qubit Enabled Single Microwave Photon Detection}}}\ (\bibinfo  {publisher}
  {University of California, Berkeley},\ \bibinfo {year} {2020})\BibitemShut
  {NoStop}%
\bibitem [{\citenamefont {Nersisyan}\ \emph {et~al.}(2019)\citenamefont
  {Nersisyan}, \citenamefont {Poletto}, \citenamefont {Alidoust}, \citenamefont
  {Manenti}, \citenamefont {Renzas}, \citenamefont {Bui}, \citenamefont {Vu},
  \citenamefont {Whyland}, \citenamefont {Mohan}, \citenamefont {Sete} \emph
  {et~al.}}]{nersisyan2019manufacturing}%
  \BibitemOpen
  \bibfield  {author} {\bibinfo {author} {\bibfnamefont {A.}~\bibnamefont
  {Nersisyan}}, \bibinfo {author} {\bibfnamefont {S.}~\bibnamefont {Poletto}},
  \bibinfo {author} {\bibfnamefont {N.}~\bibnamefont {Alidoust}}, \bibinfo
  {author} {\bibfnamefont {R.}~\bibnamefont {Manenti}}, \bibinfo {author}
  {\bibfnamefont {R.}~\bibnamefont {Renzas}}, \bibinfo {author} {\bibfnamefont
  {C.-V.}\ \bibnamefont {Bui}}, \bibinfo {author} {\bibfnamefont
  {K.}~\bibnamefont {Vu}}, \bibinfo {author} {\bibfnamefont {T.}~\bibnamefont
  {Whyland}}, \bibinfo {author} {\bibfnamefont {Y.}~\bibnamefont {Mohan}},
  \bibinfo {author} {\bibfnamefont {E.~A.}\ \bibnamefont {Sete}}, \emph
  {et~al.},\ }\bibfield  {title} {\bibinfo {title} {Manufacturing low
  dissipation superconducting quantum processors},\ }in\ \href
  {https://ieeexplore.ieee.org/abstract/document/8993458} {\emph {\bibinfo
  {booktitle} {2019 IEEE international electron devices meeting (IEDM)}}}\
  (\bibinfo {organization} {IEEE},\ \bibinfo {year} {2019})\ pp.\ \bibinfo
  {pages} {31--1}\BibitemShut {NoStop}%
\bibitem [{\citenamefont {{Van Vliet}}\ and\ \citenamefont
  {Handel}(1982)}]{VANVLIET1982261}%
  \BibitemOpen
  \bibfield  {author} {\bibinfo {author} {\bibfnamefont {C.~M.}\ \bibnamefont
  {{Van Vliet}}}\ and\ \bibinfo {author} {\bibfnamefont {P.~H.}\ \bibnamefont
  {Handel}},\ }\bibfield  {title} {\bibinfo {title} {A new transform theorem
  for stochastic processes with special application to counting statistics},\
  }\href {https://www.sciencedirect.com/science/article/pii/037843718290019X}
  {\bibfield  {journal} {\bibinfo  {journal} {Physica A: Statistical Mechanics
  and its Applications}\ }\textbf {\bibinfo {volume} {113}},\ \bibinfo {pages}
  {261} (\bibinfo {year} {1982})}\BibitemShut {NoStop}%
\bibitem [{\citenamefont {Boissonneault}\ \emph {et~al.}(2009)\citenamefont
  {Boissonneault}, \citenamefont {Gambetta},\ and\ \citenamefont
  {Blais}}]{boissonneault2009dispersive}%
  \BibitemOpen
  \bibfield  {author} {\bibinfo {author} {\bibfnamefont {M.}~\bibnamefont
  {Boissonneault}}, \bibinfo {author} {\bibfnamefont {J.~M.}\ \bibnamefont
  {Gambetta}},\ and\ \bibinfo {author} {\bibfnamefont {A.}~\bibnamefont
  {Blais}},\ }\bibfield  {title} {\bibinfo {title} {Dispersive regime of
  circuit {QED}: Photon-dependent qubit dephasing and relaxation rates},\
  }\href {https://journals.aps.org/pra/abstract/10.1103/PhysRevA.79.013819}
  {\bibfield  {journal} {\bibinfo  {journal} {Physical Review A}\ }\textbf
  {\bibinfo {volume} {79}},\ \bibinfo {pages} {013819} (\bibinfo {year}
  {2009})}\BibitemShut {NoStop}%
\end{thebibliography}%

\bigskip

\end{document}